\documentclass[twocolumn,preprintnumbers,amsmath,amssymb,showkeys,mathtools,letterpaper]{revtex4-1}

\usepackage{graphicx}
\usepackage{upgreek}
\usepackage{amsmath}
\usepackage{amssymb}
\usepackage{dcolumn}
\usepackage{chngpage}

\usepackage{graphicx}
\usepackage{dcolumn}
\usepackage{bm}
\usepackage{amssymb} 
\usepackage{subfigure}
\usepackage{color}
\usepackage{epstopdf}
\usepackage{tabularx}
\newcolumntype{C}{>{\centering\arraybackslash}X}
\usepackage{array,booktabs}

\begin{document}

\preprint{{\it J. Chem. Phys.}}

\title{Role of Internal Motions and Molecular Geometry on the NMR Relaxation of Hydrocarbons}
\author{P. M. Singer}
\author{D. Asthagiri}
\author{Z. Chen}
\author{Arjun V. Parambathu}
\author{G. J. Hirasaki}
\author{W. G. Chapman}
\affiliation{Department of Chemical and Biomolecular Engineering, Rice University, 6100 Main St., Houston, TX 77005, USA}

\date{\today}

\keywords{In{\it tra}molecular relaxation, In{\it ter}molecular relaxation, BPP theory, Rigid molecules, Cross-relaxation}

\begin{abstract}

The role of internal motions and molecular geometry on $^1$H NMR relaxation times $T_{1,2}$ in hydrocarbons is investigated using MD (molecular dynamics) simulations of the autocorrelation functions for in{\it tra}molecular $G_R(t)$ and in{\it ter}molecular $G_T(t)$ $^1$H-$^1$H dipole-dipole interactions arising from rotational ($R$) and translational ($T$) diffusion, respectively. We show that molecules with increased molecular symmetry such as neopentane, benzene, and isooctane show better agreement with traditional hard-sphere models than their corresponding straight-chain $n$-alkane, and furthermore that spherically-symmetric neopentane agrees well with the Stokes-Einstein theory. The influence of internal motions on the dynamics and $T_{1,2}$ relaxation of $n$-alkanes are investigated by simulating rigid $n$-alkanes and comparing with flexible (i.e. non-rigid) $n$-alkanes. Internal motions cause the rotational and translational correlation-times $\tau_{R,T}$ to get significantly shorter and the relaxation times $T_{1,2}$ to get significantly longer, especially for longer-chain $n$-alkanes. Site-by-site simulations of $^1$H's along the chains indicate significant variations in $\tau_{R,T}$ and $T_{1,2}$ across the chain, especially for longer-chain $n$-alkanes. The extent of the stretched (i.e. multi-exponential) decay in the autocorrelation functions $G_{R,T}(t)$ are quantified using inverse Laplace transforms, for both rigid and flexible molecules, and on a site-by-site bases. Comparison of $T_{1,2}$ measurements with the site-by-site simulations indicate that cross-relaxation (partially) averages-out the variations in $\tau_{R,T}$ and $T_{1,2}$ across the chain of long-chain $n$-alkanes. This work also has implications on the role of nano-pore confinement on the NMR relaxation of fluids in the organic-matter pores of kerogen and bitumen.

\end{abstract}


\maketitle

\section{Introduction}\label{sc:Intro}

Recent studies have shown that MD (molecular dynamics) simulations can successfully predict $^1$H NMR (nuclear magnetic resonance) relaxation times $T_{1,2}$ and diffusion coefficients of bulk hydrocarbons and water, without any adjustable parameters in the interpretation of the simulation data \cite{singer:jmr2017}. Besides validating the forcefields used in the simulations, the MD simulations reveal new insight about the NMR relaxation from in{\it tra}molecular versus in{\it ter}molecular $^1$H-$^1$H dipole-dipole interactions in fluids, which are not easily accessible experimentally \cite{woessner:jcp1964}. For instance, the simulations quantify the relative strength of the two relaxation mechanisms, indicating that in{\it tra}molecular increasingly dominates over in{\it ter}molecular relaxation with increasing molecular chain-length (i.e. increasing carbon number). This validates the common practice of only considering in{\it tra}molecular dipole-dipole interactions for simulations of macromolecules such as proteins \cite{kowalewski:book} or polymers in the short-time regime \cite{henritzi:ssnmr2013}.

However, as also reported in \cite{singer:jmr2017}, the {\it functional forms} of the simulated autocorrelation functions for in{\it tra}molecular $G_{R}(t)$ (i.e. rotational) and in{\it ter}molecular $G_{T}(t)$ (i.e. translational) $^1$H-$^1$H dipole-dipole interactions show significant deviations from the traditional hard-sphere models by Bloembergen, Purcell, Pound (BPP) \cite{bloembergen:pr1948} and Torrey \cite{torrey:pr1953}, respectively. In the case of in{\it tra}molecular interactions, the BPP model predicts a single-exponential decay for $G_{R}(t)$ with rotational-correlation time $\tau_R$ however, the simulations clearly indicate an increasingly ``stretched" (i.e. multi-exponential) decay with increasing chain-length. For in{\it ter}molecular interactions, the Torrey model predicts a specific functional form for $G_{T}(t)$ (and an associated translational-correlation time $\tau_T$), however, the simulations also indicate a ``stretched" decay at large chain-lengths. Another prediction from the hard-sphere models \cite{bloembergen:pr1948,torrey:pr1953} is that the ratio of translational-diffusion correlation-time $\tau_D$ (where $\tau_D = \frac{5}{2} \tau_T$) \cite{cowan:book} to rotational-correlation time $\tau_R$ should be $\tau_D/\tau_R = 9$ \cite{abragam:book}, which we show in this report is the case for spherically-symmetric neopentane. However, the simulations clearly show that $\tau_D/\tau_R \ll 9$ at large chain-lengths \cite{singer:jmr2017}. These findings indicate that the Stokes-Einstein radius for rotational and translational motion are comparable for short-chain and spherical alkanes, but increasingly diverge with increasing chain length, indicating limitations of ascribing a single ``radius" for chain molecules. 

While the molecular dynamics of long-chain $n$-alkanes are expected to deviate from hard-sphere predictions, the underlying theory for the deviations remain elusive and difficult to verify experimentally. Simplified models for the autocorrelation function of non-spherical and non-rigid molecules have been successfully developed in the past, such as those by Woessner and others which describe spin-relaxation processes in two-proton systems undergoing anisotropic reorientation \cite{woessner:jcp1962,woessner:jcp1962b,huntress:jcp1968} and internal motions \cite{woessner:jcp1965}, as well as three-proton systems \cite{hubbard:jcp1969}. These anisotropic reorientation models have been successful for highly symmetric molecules such as benzene, where in-plane and perpendicular-to-plane rotational diffusion-coefficients (and rotational correlation-times) can be determined from NMR measurements and MD simulations \cite{laaksonen:jcp1998,witt:jpca2000}. In the case of in{\it tra}molecular (i.e. rotational) relaxation for molecules of lower symmetry, such theories result in a multi-exponential decay for the autocorrelation function $G_{R}(t)$, with a corresponding distribution in correlation times. Generalizations of such models for the autocorrelation function later arose which take a ``model free" or heuristic approach to the internal motions, such as the Lipari-Szabo model \cite{lipari:jacs1982,lipari:jacs1982b} often used in polymers \cite{kariyo:macro2008a} and proteins \cite{frueh:pnmrs2002}. However, as the internal motions of the molecule become more complex, one is invariably forced to develop even more heuristic models of the autocorrelation function and its associated spectral density \cite{beckmann:prep1988,bakhmutov:book}. Such heuristic approaches include a variety of distribution functions for the underlying correlation times, including the Cole-Davidson distribution function \cite{bormuth:macro2013,henritzi:ssnmr2013}, the generalized gamma function \cite{kariyo:macro2008a}, the Kohlrausch-Williams-Watts \cite{henritzi:ssnmr2013} functions, and the Singer-Hirasaki function \cite{singer:EF2018}, each chosen to fit the observations, but without theoretical justification. In the case of bulk water (which exhibits hydrogen-bonding), $G_{R}(t)$ is stretched to a similar extent as $n$-pentane \cite{singer:jmr2017}, which may be a result of large, discrete angular-jumps \cite{calero:jpcb2015} superimposed on a continuous-time rotational-diffusion process \cite{madhavi:jpcb2017}.

In this report, we study the influence of internal motions and molecular geometry on the molecular dynamics of hydrocarbons by simulating the autocorrelation functions $G_{R,T}(t)$ of rigid $n$-alkanes (i.e. without internal motions), compared with flexible $n$-alkanes (i.e. with internal motions). While such rigid $n$-alkanes do not exist in nature, they present an ideal testing ground for simulating the influence of internal motions on correlation times and relaxation times. We also report site-by-site simulations $G_{R,T}(t)$ for the $^1$H's across the chain, thereby quantifying the underlying distribution in correlation times $\tau_{R,T}$ and relaxation times $T_{1,2}$ across the molecule. The simulations of both the rigid molecules and the site-by-site $^1$H's reveal key insights about the functional forms of $G_{R,T}(t)$ as a function of chain length, without invoking any heuristic models. The site-by-site simulations are also compared with $T_{1,2}$ measurements in the case of $n$-decane and $n$-heptadecane, which show that cross-relaxation \cite{campbell:jmr1973,kalk:jmr1976} partially (in the case of $n$-heptadecane) averages out the underlying variations in $\tau_{R,T}$ and $T_{1,2}$. Such comparisons between measurement and site-by-site simulation could in principle be used as a new method for determining the cross-relaxation rates at low magnetic-fields, provided the proper theoretical framework is developed.

Besides addressing the fundamental science of molecular dynamics of bulk fluids, the present work also opens up new opportunities for investigating the effects of nanometer confinement on fluids, such as the light (i.e. low-viscosity) hydrocarbons confined in the organic-matter pores of the kerogen and bitumen typically found in organic-rich shale \cite{passey:spe2010,loucks:aapg2012}. There is increasing evidence that the NMR surface relaxation of the light hydrocarbons confined in such organic nano-pores is dominated by in{\it tra}molecular \cite{singer:petro2016} and in{\it ter}molecular \cite{washburn:jmr2017} $^1$H-$^1$H dipole-dipole interactions, as opposed to surface paramagnetism. Provided the molecular dynamics of the bulk fluid is well understood, the MD simulations of $^1$H-$^1$H dipole-dipole interactions can then in principle be used to characterize the complex NMR response of fluids confined in organic nano-pores \cite{ozen:petro2013,rylander:spe2013,jiang:spwla2013,singer:sca2013,washburn:jmr2013,kausik:sca2014,daigle:urtec2014,washburn:cmr2014,korb:jpcc2014,nicot:petro2015,lessenger:SPWLA2015,birdwell:ef2015,fleury:jpse2016,kausik:petro2016,singer:petro2016,sun:spwla2016,sondergeld:spwla2016,yang:ef2016,chen:petro2017,valori:ef2017,washburn:jmr2017}, which as of yet is not well understood. The effects of nano-confinement on fluids in organic shale can also affect the phase behavior and partitioning of components between the matrix and production fractures as a result of the strong interaction between the fluid molecules and pore surface. It is clear that the fundamental understanding of such complex systems would significantly improve by integrating NMR measurements, MD simulations, and molecular DFT (density functional theory) \cite{liu:lang2017} techniques. 

The rest of the article is organized as follows. Section \ref{sc:Method} presents the methodology including: the hydrocarbons investigated and labeled in Section \ref{ssc:Molecules}, the MD simulation background in Section \ref{ssc:MD}, and the autocorrelation function and NMR relaxation background in Section \ref{ssc:NMR}. Section \ref{sc:Results} presents the results and discussions including: the symmetric molecules and hard-sphere models in Section \ref{ssc:Sphere}, the internal motions in flexible versus rigid molecules in Section \ref{ssc:Rigid}, the site-by-site simulations and distribution in correlation times in Section \ref{ssc:Site}, including the cross-relaxation effects and comparison with measurements in Section \ref{sssc:Cross}. Conclusions are presented in Section \ref{sc:Conc}.

\section{Methodology} \label{sc:Method}

This section introduces the hydrocarbons under investigation, and presents a brief review of the MD simulations, the definition and properties of the autocorrelation functions, and, the derivation of the NMR relaxation times. Further information about the methodology can be found in \cite{singer:jmr2017}.

\subsection{Hydrocarbons investigated and labeled} \label{ssc:Molecules}

 \begin{figure}[!ht]
	\begin{center}
		\includegraphics[width=1\columnwidth]{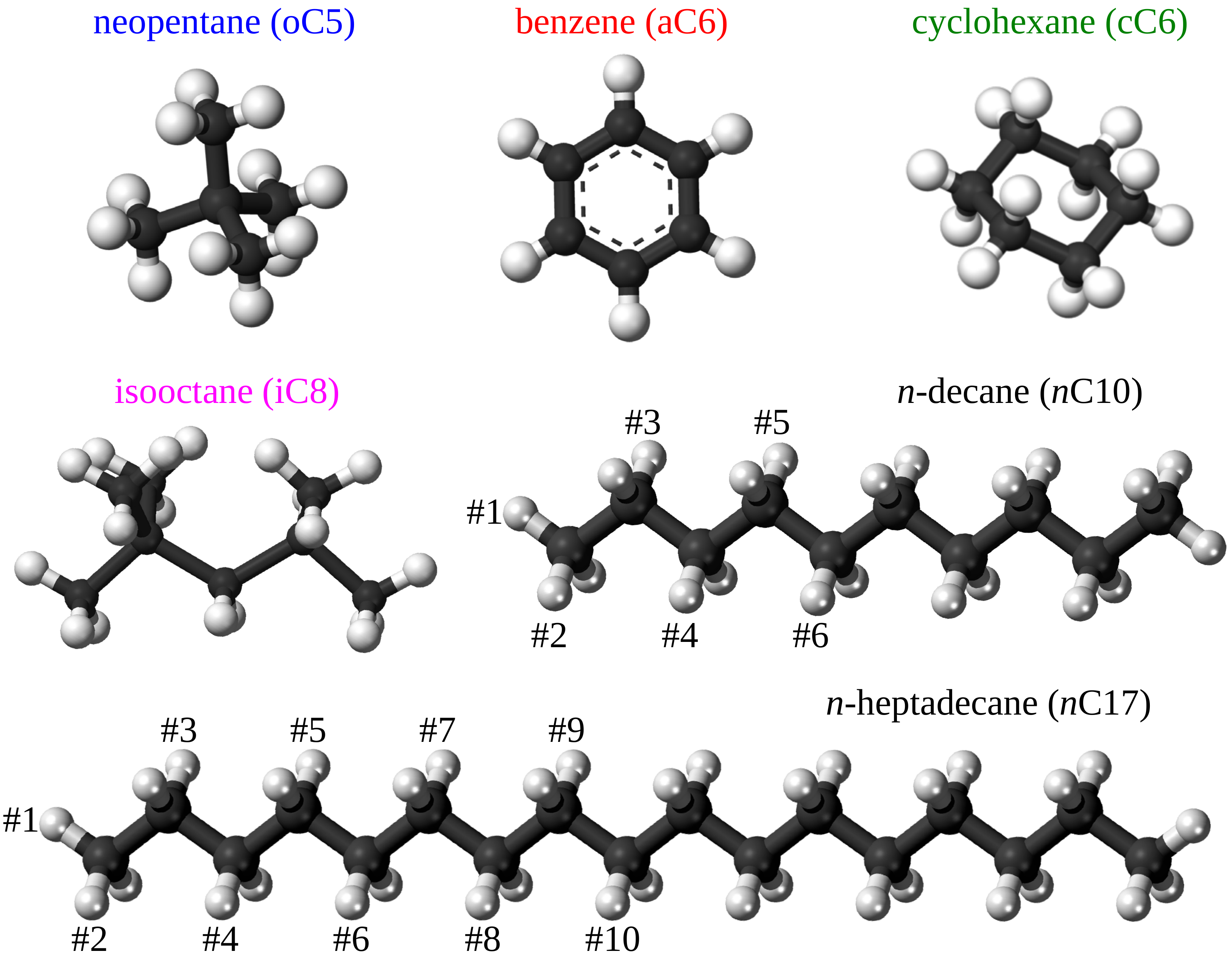} 
	\end{center}
	\caption{Ball-and-stick illustrations of the hydrocarbons simulated in this report, including (in order of increasing molar mass): (1) neopentane C$_5$H$_{12}$ (2,2-dimethylpropane) or ``oC5" for short, (2) benzene C$_6$H$_{6}$ or ``aC6" for short, (3) cyclohexane C$_6$H$_{12}$ or ``cC6" for short, (4) isooctane C$_8$H$_{18}$ (2,2,4-trimethylpentane) or ``iC8" for short, (5) $n$-decane $n$-C$_{10}$H$_{22}$ or ``$n$C10" for short, and (6) $n$-heptadecane $n$-C$_{17}$H$_{36}$ or ``$n$C17" for short. $n$-decane and $n$-heptadecane also show labels for the inequivalent $^1$H's along the chain. The other $n$-alkanes (not shown) are labeled $n$C$\#$ for carbon number C$\#$.}
	\label{fg:Molecules}
\end{figure} 

The hydrocarbons simulated in this report are shown in Fig. \ref{fg:Molecules}. Neopentane (C$_5$H$_{12}$) refers to the isomer 2,2-dimethylpropane, and is labeled ``oC5" for short; it is spherically symmetric, and all $^1$H's are topologically equivalent. Benzene (C$_6$H$_{6}$) is labeled as ``aC6" for short; it is planar symmetric, and all $^1$H's are equivalent. Cyclohexane (C$_6$H$_{12}$) is labeled as ``cC6" for short; it contains two inequivalent $^1$H's sites, namely axial and equatorial. Isooctane (C$_8$H$_{18}$) refers to the isomer 2,2,4-trimethylpentane, and is labeled ``iC8" for short; it is more spherical than the normal isomer $n$-octane (or ``$n$C8 for short), with many inequivalent $^1$H's. The other $n$-alkanes (not shown) are labeled $n$C$\#$ for carbon number C$\#$.

The site-by-site simulations are conducted on $n$-decane ($n$-C$_{10}$H$_{22}$) labeled as ``$n$C10" for short, and $n$-heptadecane ($n$-C$_{17}$H$_{36}$) labeled as ``$n$C17" for short. In the case of $n$-decane, sites $\#2 \leftrightarrow\#$6 have a degeneracy of 4, while $\#1$ has a degeneracy of 2. In the case of $n$-heptadecane, sites $\#2\leftrightarrow\#$9 have a degeneracy of 4, while $\#1$ and $\#10$ have a degeneracy of 2. These degeneracies are used to compute the weighted average of the autocorrelation functions. The site-by-site simulations are also conducted on rigid $n$-decane, with the same labels as flexible $n$-decane.

\subsection{Molecular dynamics simulation} \label{ssc:MD}

The MD simulations of all the flexible hydrocarbons were performed using NAMD \cite{namd} version 2.11. The bulk alkanes were modeled using the CHARMM General Force field (CGenFF) \cite{cgenff}. 
The protocol for setting-up the initial simulation configuration was exactly as before \cite{singer:jmr2017}. We use the $n$-alkane simulation trajectory from the earlier work for the analysis noted below. 
For the cyclic alkanes and isooctane, as before, we created the initial simulation system by packing $N$ copies of the molecule into a cube of volume $L^3$ using the Packmol program \cite{packmol}. The volume was chosen 
such that the number density $N/V$ corresponds to the experimentally determined number density at 293.15~K. The simulation approach for these systems using NAMD was as before \cite{singer:jmr2017}. 

All the rigid body simulations are performed within LAMMPS \cite{plimpton:jcop1995}. For the rigid $n$-alkanes, we took the initial configuration from our earlier work, i.e.\ the configuration obtained after the Packmol packing procedure. At this stage, by construction all the $n$-alkanes are in the fully 
extended (all-trans) configuration. Then using the CHARMM-to-LAMMPS tool from the Enhanced Monte Carlo package \cite{veld:macro2003}, we prepared the molecular configuration and forcefield information into a 
LAMMPS input data file. To remove potential steric overlaps from the packing procedure, within LAMMPS we first run 30 steps of constant energy molecular dynamics using the {\sc nve/limit 0.1} option. (The {\sc limit} 
option enforces the maximum distance a particle can move and allows the simulation to proceed despite possible overlaps.) Initial velocities for the MD simulation are obtained with a Gaussian distribution and are adjusted to give a temperature of 293.15~K. We find that 30 steps is more than sufficient to remove any potential steric overlaps and also avoid distorting the geometry of the alkanes from their original all-trans configuration. 
After this dynamics step, we run rigid body molecular dynamics simulations using the {\sc rigid/nve/small molecule} option within LAMMPS. The system is thermostated using a Langevin thermostat with a damping
coefficient of 5~ps$^{-1}$. The cutoff within LAMMPS was exactly as it was in our earlier NAMD runs for the flexible molecules. Specifically,  the Lennard-Jones interactions were terminated at 14.00~{\AA} 
by smoothly switching to zero starting at 13.00 {\AA}. The time step for integrating the equations of motion is 1~fs and the equilibration phase was over 1~ns.  In the subsequent production phase, we remove the 
thermostat. The production phase lasted at least 1~ns and configurations are saved every 100~steps to obtain at least $10^4$ frames for analysis. As before \cite{singer:jmr2017}, the average temperature in the rigid body simulations is within $\pm$3~K of the target temperature of 293.15~K ($20^{\rm o}$C).

\subsection{Autocorrelation function and NMR relaxation} \label{ssc:NMR}

The theory of NMR relaxation in liquids is well known \cite{bloembergen:pr1948,torrey:pr1953,abragam:book,mcconnell:book,cowan:book,kimmich:book}, and the expressions for the autocorrelation functions for {\it isotropic} in{\it tra}molecular $G_R(t)$ and in{\it ter}molecular $G_T(t)$ $^1$H-$^1$H dipole-dipole interactions are derived in Ref. \cite{singer:jmr2017}. Cross-correlation effects are neglected, which is generally justified for isotropic systems \cite{kalk:jmr1976}. We use the convention based on the text by McConnell \cite{mcconnell:book}, with $G_{R,T}(t)$ in units of s$^{-2}$: 
\begin{multline}
G_{R,T}(t) =
\frac{3}{16} \! \left(\frac{\mu_0}{4\pi}\right)^2 \! \hbar^2 \gamma^4 \frac{1}{N_{R,T}} \times \\ \sum\limits_{i \neq j}^{N_{R,T}} 
\left< \frac{(3\cos^{2}\!\theta_{ij}\!(t+\tau)-1)}{r_{ij}^3\!\left(t+\tau\right)}  \frac{(3\cos^{2}\!\theta_{ij}\!(\tau)-1)}{r_{ij}^3\!(\tau)} \right>_{\!\! \tau}.
\label{eq:GmRT}
\end{multline}
The correlation times $\tau_{R,T}$ are derived from the following expressions \cite{cowan:book}:
\begin{equation}
\tau_{R,T} = \frac{1}{G_{R,T}(0)}\int_{0}^{\infty}\!G_{R,T}(t)\, dt.
\label{eq:TauRT}
\end{equation}
In other words, $\tau_{R,T}$ is defined as the normalized area under the autocorrelation functions $G_{R,T}(t)$. The areas in this report are determined using Simpson's rule, as opposed to the trapezoidal rule used previously \cite{singer:jmr2017}. All flexible $n$-alkane data in Sections \ref{ssc:Sphere} and \ref{ssc:Rigid} are taken from \cite{singer:jmr2017}, and interpreted using Simpson's rule.

\begin{figure}[!ht]
	\begin{center}
		\includegraphics[width=0.9\columnwidth]{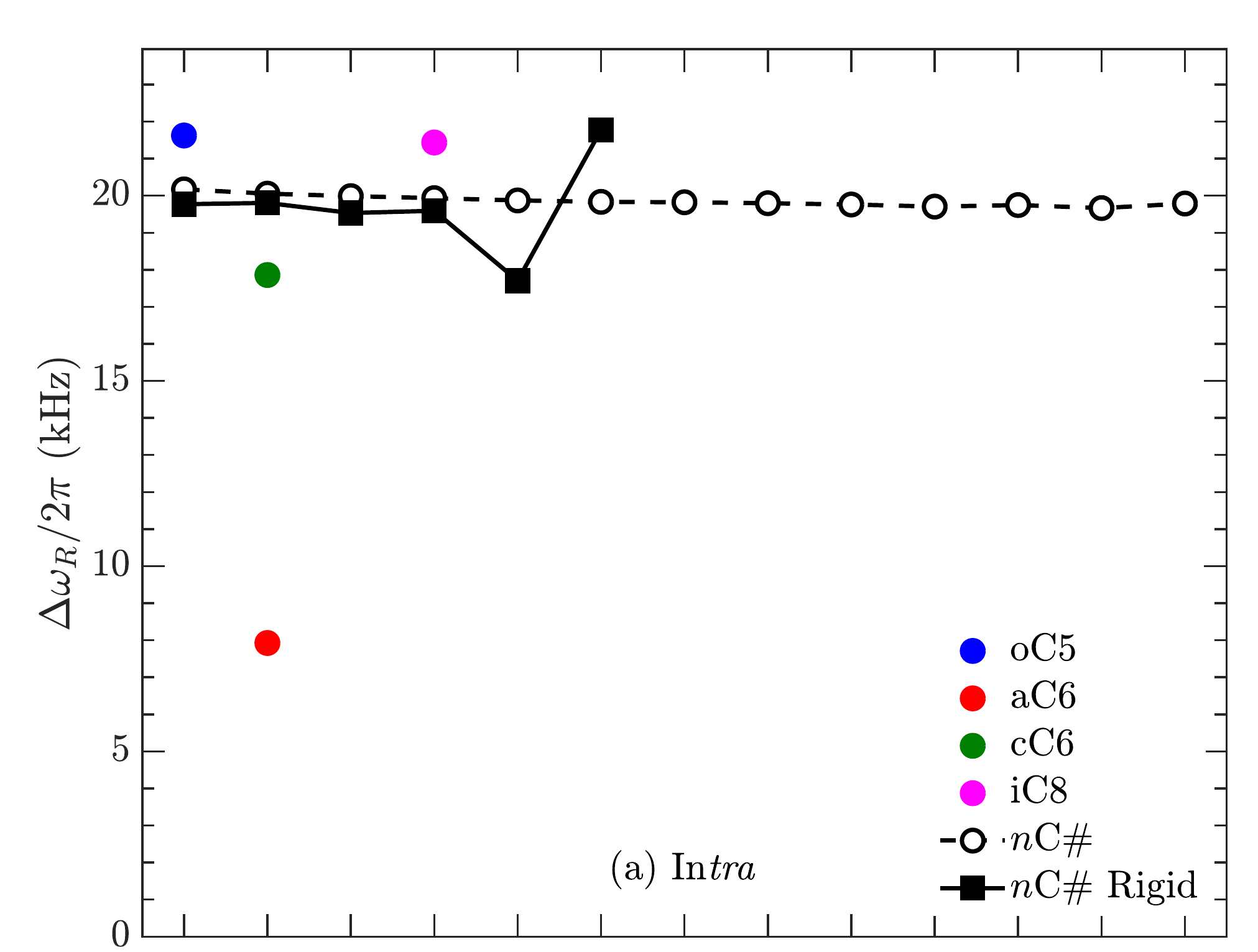}  
		\includegraphics[width=0.9\columnwidth]{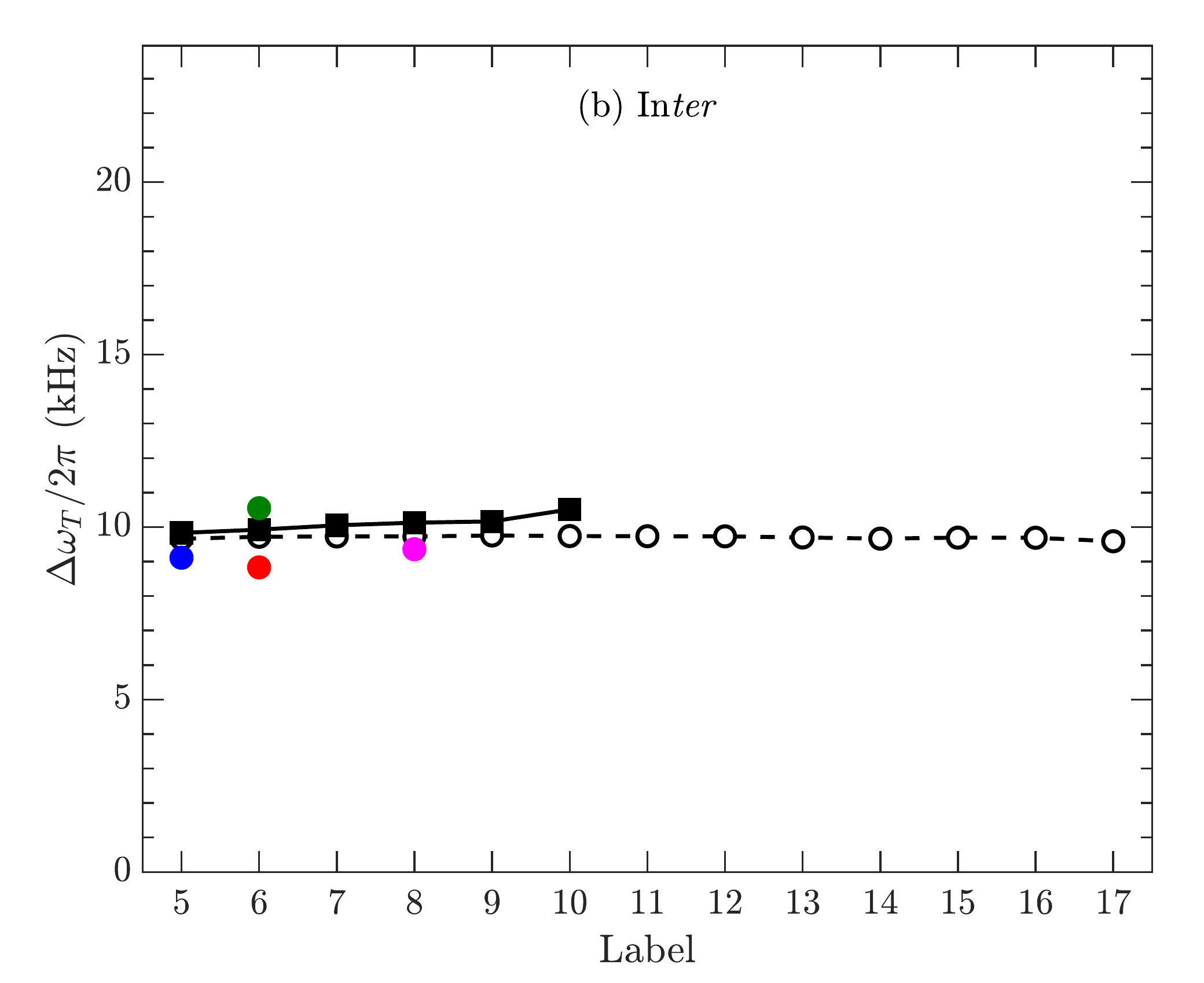} 
	\end{center}
	\caption{MD simulation results for the dipolar strength of (a) in{\it tra}molecular $\Delta\omega_R$ and (b) in{\it ter}molecular $\Delta\omega_T$ interactions, using Eq. \ref{eq:Dipolar} and reported in units of kHz. The hydrocarbon short-hand labels are defined in Fig. \ref{fg:Molecules}, and are plotted in order of increasing carbon number C$\#$. $n$C$\#$ refers to $n$-alkanes of carbon number C$\#$. ``Rigid" refers to rigid hydrocarbons, while all others are flexible. The flexible $n$C$\#$ data are taken from \cite{singer:jmr2017}.} \label{fg:Dipolar}
\end{figure}

Besides the correlation time, the other parameter of interest is the ``second-moment" $\Delta\omega_{R,T}^2$ given by the following expression \cite{cowan:book}:
\begin{equation}
G_{R,T}(0) = \frac{1}{3}\Delta\omega_{R,T}^2.
\label{eq:G0dipolar}
\end{equation}
The dipolar strength $\Delta\omega_{R,T}$ of the dipole-dipole interaction is derived from Eqs. \ref{eq:GmRT} and \ref{eq:G0dipolar} to yield the following expression:
\begin{multline}
\Delta\omega_{R,T} = \\ \sqrt{\frac{9}{16} \! \left(\frac{\mu_0}{4\pi}\right)^2 \! \hbar^2 \gamma^4  \frac{1}{N_{R,T}}\sum\limits_{i \neq j}^{N_{R,T}} 
\left< \frac{(3\cos^{2}\!\theta_{ij}\!(\tau)-1)^2}{r_{ij}^6\!\left(\tau\right)}  \right>_{\!\! \tau}}.
\label{eq:Dipolar}
\end{multline}
The simulated $\Delta\omega_{R,T}$ are shown in Fig. \ref{fg:Dipolar}, in units of kHz. The data indicate that $\Delta\omega_{R} \simeq 2 \Delta\omega_{T}$ in all cases, except the case of benzene where $\Delta\omega_{R} \simeq \Delta\omega_{T}$. The case of benzene can be understood from Eq. \ref{eq:Dipolar} since (a) the $^1$H coordination number (i.e. the number of nearest neighbors) $n_H$ is smaller and (b) the nearest neighbor $^1$H-$^1$H distance $r_H$ is larger, therefore $\Delta\omega_{R} \propto \sqrt{n_H}/r_H^3$ is smaller. In all cases, $\Delta\omega_{T}$ is roughly the same given that the $^1$H spin density is roughly the same. Note that all cases reported here are in the motional averaging regime $\Delta\omega_{R,T}\, \tau_{R,T} \ll 1$, which is typical of low-viscosity liquids.

The spectral densities $J_{R,T}(\omega)$ (in units of s$^{-1}$) of the local magnetic-field fluctuations can be determined from the Fourier transform of the autocorrelation function:
\begin{equation}
J_{R,T}(\omega) = 2\int_{0}^{\infty}\!G_{R,T}(t)\cos\left(\omega t\right) dt,
\label{eq:FourierRTcos}
\end{equation}
where $G_{R,T}(t)$ is real and even. In the fast motion regime applicable here (i.e. $\omega_0 \, \tau_{R,T} \ll 1$, where $\omega_0$ is the Larmor frequency for $^1$H), the spectral density follows $J_{R,T}(\omega_0) = J_{R,T}(0) $, and the following can be derived:
\begin{equation}
\frac{1}{T_{1,R,T}} = \frac{1}{T_{2,R,T}} = 5J_{R,T}(0) = \frac{10}{3} \Delta\omega_{R,T}^2 \tau_{R,T}.
\label{eq:T12RTmotional}
\end{equation}
In other words, $T_{1,R} = T_{2,R}$ and $T_{1,T} = T_{2,T}$, which is characteristic for low viscosity fluids in the fast motion regime. The total relaxation rates are then equal to the sum of in{\it tra}molecular and in{\it ter}molecular rates:
\begin{equation}
\frac{1}{T_1} = \frac{1}{T_2} =\frac{10}{3} \Delta\omega_R^2  \tau_R +  \frac{10}{3} \Delta\omega_T^2 \tau_T,
\label{eq:T12motional}
\end{equation}
which is the final expression used to predict the NMR relaxation time $T_{1} = T_{2}$ from simulation results.

In the case of symmetric molecules such as neopentane, benzene, and methane (CH$_4$), there is an additional contribution to $^1$H NMR relaxation from the spin-rotation interaction \cite{hubbard:pr1963}. In the case of methane, relaxation from the spin-rotation interaction dominates at low densities (i.e., in the gas phase), whereas relaxation from the $^1$H-$^1$H dipole-dipole interaction dominates at high densities (i.e. in the liquid phase) \cite{lo:SPE2002}. Simulations of the spin-rotation interaction for these molecules will be reported elsewhere.

In order to quantify the departure of $G_R(t)$ from single-exponential decay, we fit $G_R(t)$ to a sum of multi-exponential decays and determine the underlying distribution in correlation times $\tau$. More specifically, we perform an inversion of the following Laplace transform \cite{venkataramanan:ieee2002,song:jmr2002}: 
\begin{eqnarray}
\begin{aligned}
G_{R,T}(t) &= \int\! P_{R,T}(\tau) \exp\!\left(\!-\frac{t}{\tau}\right) d\tau, \\
G_{R,T}(0) &= \int\! P_{R,T}(\tau)\, d\tau,
\end{aligned} \label{eq:ILT}
\end{eqnarray}
where $P_{R,T}(\tau)$ (in units of s$^{-3}$) is the probability distribution function derived from the inversion. Also listed is the integral of $P_{R,T}(\tau)$, which is equal to $G_{R,T}(0)$.

In the case of the BBP hard-sphere model, $P_{R}(\tau)$ is a delta-function at $\tau_R$, i.e. $P_{R}(\tau) = G_{R}(0) \, \delta(\tau - \tau_R)$. However, as shown by the simulations below, $G_R(t)$ is always stretched (i.e. multi-exponential) to some degree, therefore $P_{R}(\tau)$ has a finite distribution. The decomposition of $G_{R,T}(t)$ into a sum of exponential decays is common practice in heuristic models of complex molecules \cite{beckmann:prep1988,bakhmutov:book}, where the more complex the molecule dynamics, the more terms are required. This justifies our general approach of decomposing $G_{R,T}(t)$ into a ``model free" sum of exponential decays in Eq. \ref{eq:ILT}, for the purposes of quantifying the departure from hard-sphere models.

The $P_{R,T}(\tau)$ distributions were determined by using the discrete form of Eq. \ref{eq:ILT}, using 150 logarithmically-spaced $\tau$ bins ranging from 10$^{-2}$ ps $\leq \tau \leq$ 10$^{3}$ ps, and a fixed regularization parameter $\alpha = 10^{-1}$ \cite{venkataramanan:ieee2002,song:jmr2002}. The resulting $P_{R,T}(\tau)$ distributions were then analyzed using the following :
\begin{eqnarray}
\begin{aligned}
\mu_{R,T} &= \frac{1}{G_{R,T}(0) }\int\! P_{R,T}\!\left(\tau\right) \ln(\tau) \,d\tau, \\
\sigma^2_{R,T} &= \frac{1}{G_{R,T}(0) }\int\! P_{R,T}\!\left(\tau\right) \left(\ln(\tau) - \mu_{R,T} \right)^2 d\tau.
\end{aligned} \label{eq:CvRT}
\end{eqnarray}
$\sigma_{R,T}$ is the standard deviation and $\mu_{R,T}$ is the mean of the variable $\ln(\tau)$. A natural logarithm in $\tau$ is used as the variable since the underlying $P_{R,T}(\tau)$ distributions are discrete and evenly spaced in $\ln(\tau)$. 

In the case of $n$-decane and $n$-heptadecane, the MD simulation were compared with $^1$H $T_{2}$ measurements. The $n$-decane ($\geq $ 99\% purity) and $n$-heptadecane (99\% purity) were obtained from Sigma-Aldrich. The $n$-alkanes were de-oxygenated by bubbling high-purity N$_2$ gas overnight and sealing the vial, thereby removing paramagnetic O$_2$ in solution. The NMR measurements were acquired at ambient conditions ($\simeq$ 25 $^{\rm o}$C) using a GeoSpec2 from Oxford Instruments in an 18 mm probe, at a Larmor frequency of $\omega_0/2\pi = 2.3$ MHz for $^1$H (where $\omega_0 = \gamma B_0$, for magnetic-field strength $B_0$ and gyro-magnetic ratio $\gamma/2\pi = 42.57$ MHz/T for $^1$H). A CPMG (Carr-Purcell-Meiboom-Gill) echo train was used with an echo spacing of $T_E = $ 0.2 ms, and averaged up to a signal-to-noise ratio of SNR $\simeq$ 10$^3$. The $1/T_2$ distributions were determined using a 1D inverse Laplace transform \cite{venkataramanan:ieee2002,song:jmr2002}, with a regularization parameter of $\alpha \simeq $ 10$^{-3}$ corresponding to the noise floor. The width of the $1/T_2$ distributions was found to be the same at a slightly elevated temperature of 30 $^{\rm o}$C, which for the case of $n$-heptadecane rules out any potential influence from the nearby phase-transition at $\simeq$ 20 $^{\rm o}$C.

\section{Results and Discussions}\label{sc:Results}

\subsection{Symmetric molecules and hard-sphere models} \label{ssc:Sphere}

The hard-sphere models developed by BPP \cite{bloembergen:pr1948} and Torrey \cite{torrey:pr1953} have been the building blocks for the interpretation of NMR relaxation in liquids. In this section, we show how the case of the more spherical isomers neopentane and isooctane approach the BPP and Torrey theories, and in particular neopentane shows good agreement with the Stokes-Einstein theory. Such agreement validates the hard-sphere theories in a manner never reported before, and at the same time further validates our MD simulation methodology \cite{singer:jmr2017}.

The in{\it tra}molecular dipole-dipole interaction \cite{bloembergen:pr1948} is traditionally derived using the rotational diffusion equation of rank 2 for hard-spheres of radius $R_R$, and is parameterized by the rotational diffusion coefficient $D_R$. The parameter $D_R$ is simply related to rotational correlation time $\tau_R$, which is defined as the average time it takes the molecule to rotate by 1 radian. The in{\it ter}molecular dipole-dipole interaction \cite{torrey:pr1953} is traditionally derived using the Brownian motion model where the diffusion propagator is derived for hard-spheres of radius $R_T$, and is parameterized by the translational diffusion coefficient $D_T$. The parameter $D_T$ is simply related to the translational-diffusion correlation-time $\tau_D$, which is defined as the average time it takes the molecule to diffuse by one hard-core diameter $2R_T$ \cite{mcconnell:book}. 

\begin{figure}[!ht]
	\begin{center}
		\includegraphics[width=0.9\columnwidth]{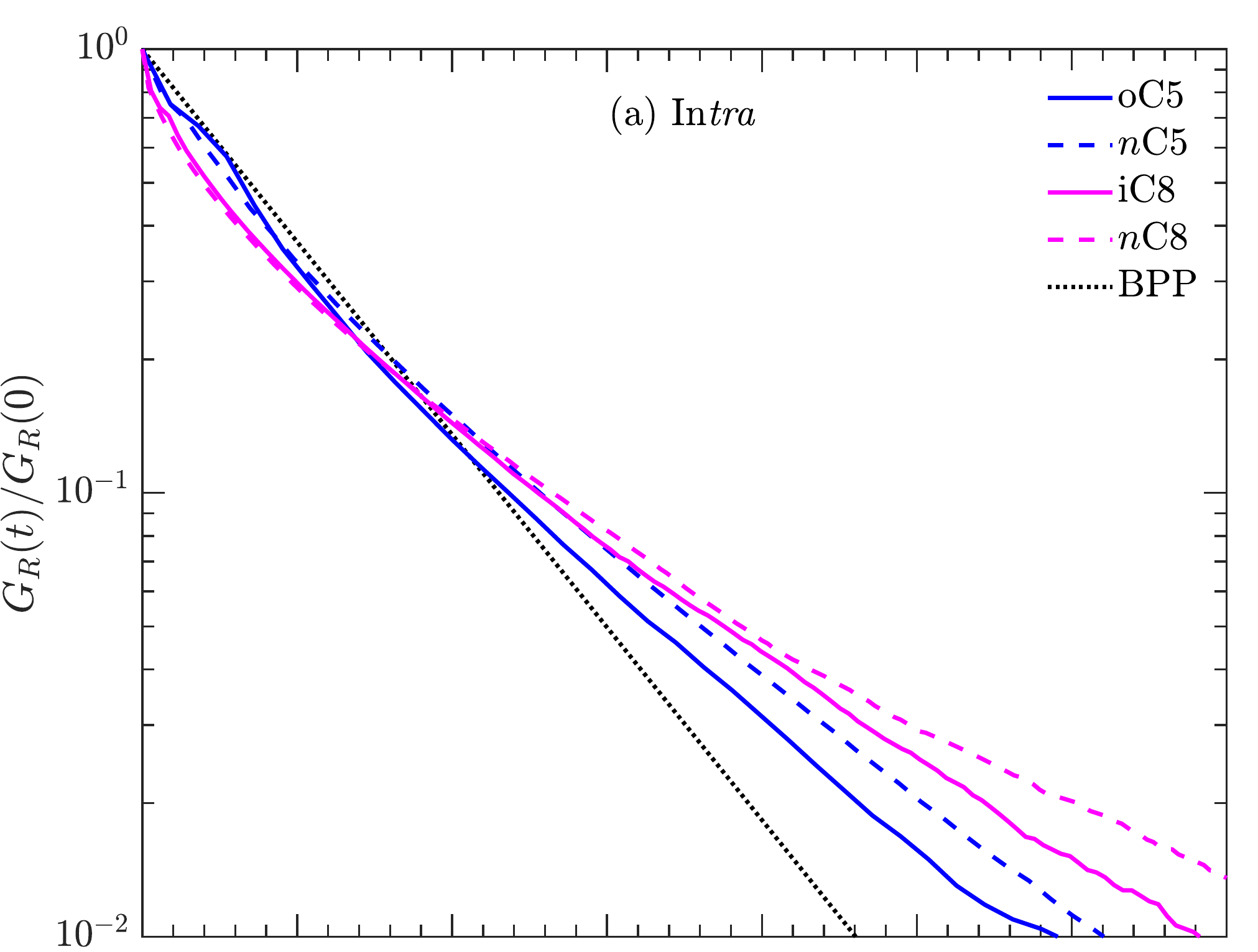}  
		\includegraphics[width=0.9\columnwidth]{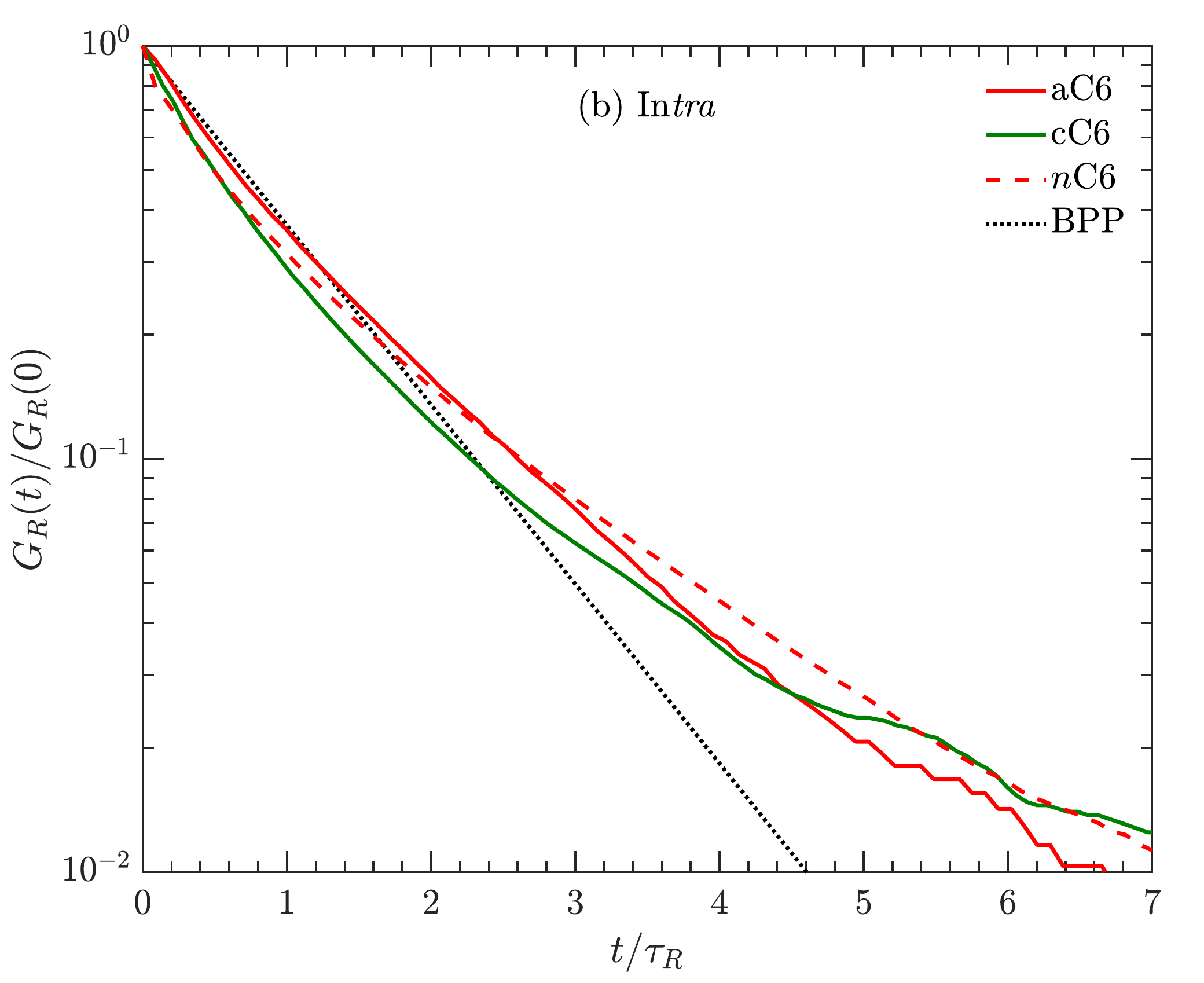} 
	\end{center}
	\caption{MD simulations of the autocorrelation function $G_R(t)$ for in{\it tra}molecular interactions using Eq. \ref{eq:GmRT}, for selected hydrocarbons defined in Fig. \ref{fg:Molecules}, split into (a) and (b) for clarity. The $y$-axis has been normalized by zero time value $G_{R}(0)$, and the $x$-axis has been normalized by correlation time $\tau_{R}$ (Eq. \ref{eq:TauRT}). The prediction from the BPP \cite{bloembergen:pr1948} hard-sphere model (Eq.~\ref{eq:Gmodel}) is shown by the black dotted line.} 
	\label{fg:GtIntra}
\end{figure}

\begin{figure}[!ht]
	\begin{center}
		\includegraphics[width=0.9\columnwidth]{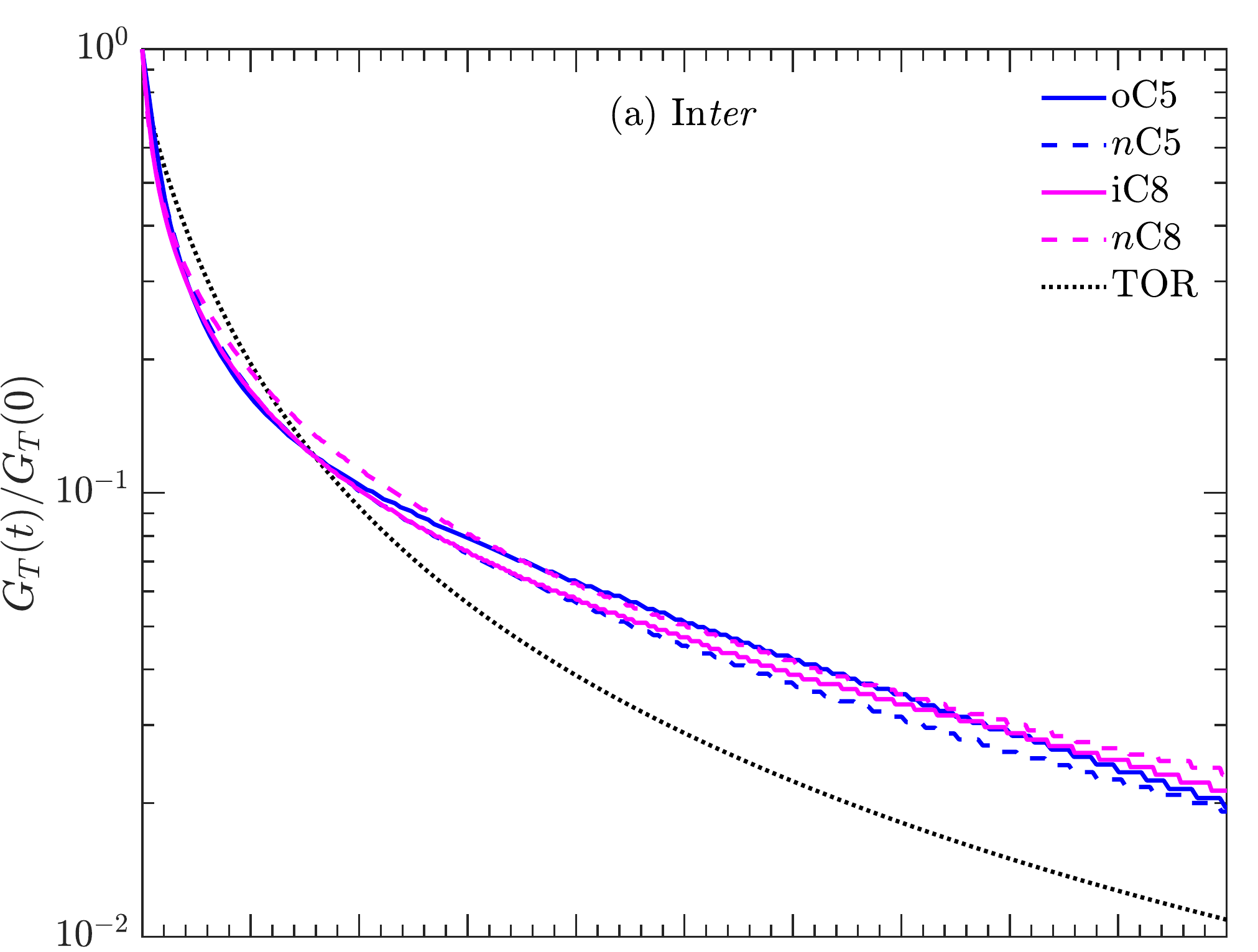}  
		\includegraphics[width=0.9\columnwidth]{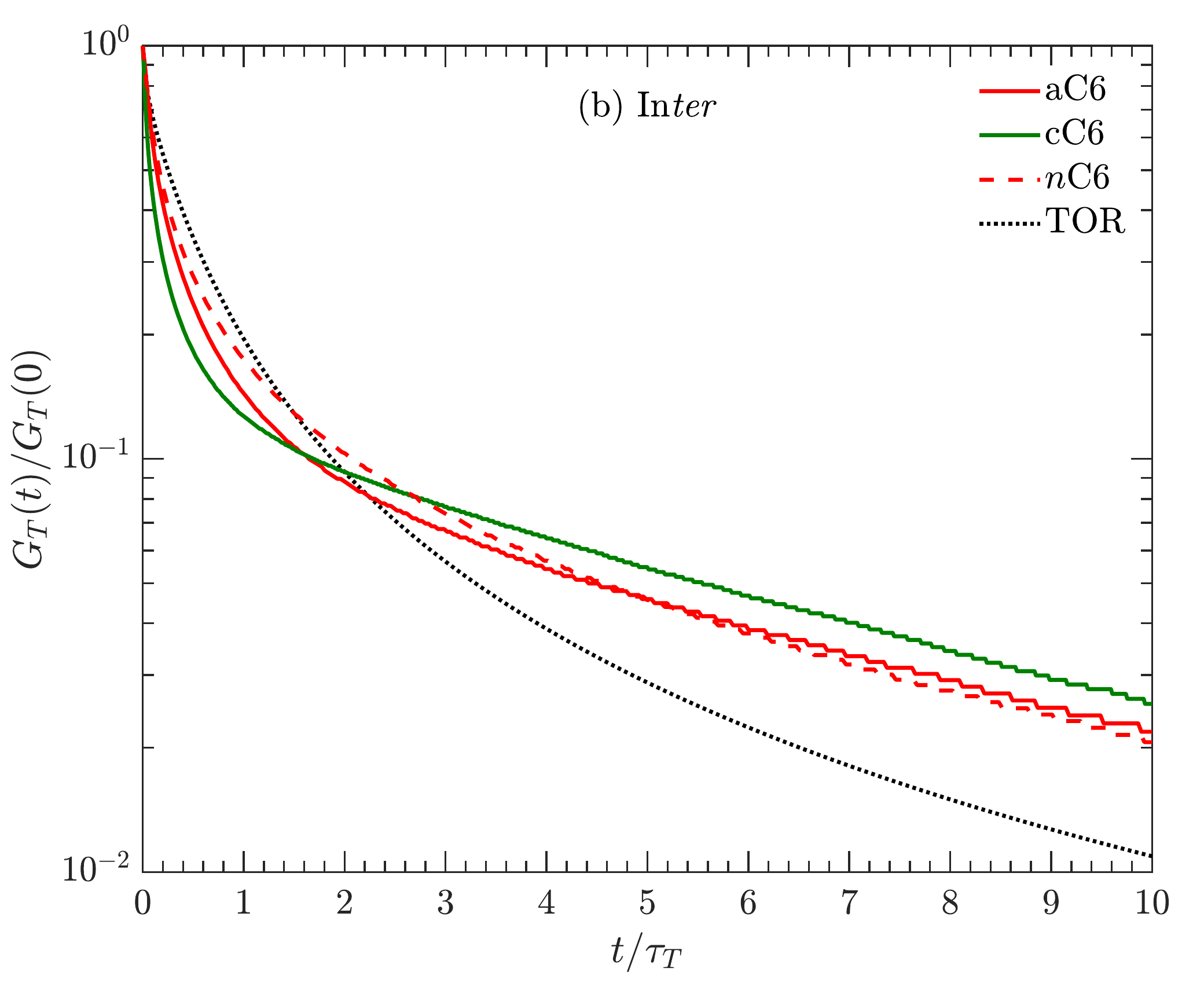} 
	\end{center}
	\caption{MD simulations of the autocorrelation function $G_T(t)$ for in{\it ter}molecular interactions using Eq. \ref{eq:GmRT}, for selected hydrocarbons defined in Fig. \ref{fg:Molecules}, split into (a) and (b) for clarity. The $y$-axis has been normalized by zero time value $G_{T}(0)$, and the $x$-axis has been normalized by correlation time $\tau_{T}$ (Eq. \ref{eq:TauRT}). The prediction from the Torrey \cite{torrey:pr1953} (``TOR") hard-sphere model (Eq.~\ref{eq:Gmodel}) is shown by the black dotted line.} 
	\label{fg:GtInter}
\end{figure}

For hard-spheres, the diffusion coefficients (and correlation times) can be related to the bulk properties of the fluid, namely the viscosity $\eta$ and absolute temperature $T$, using the traditional Stokes-Einstein relation \cite{abragam:book}:
\begin{eqnarray}
\begin{aligned}
\tau_R &= \frac{1}{6D_R} = \frac{4\pi }{3k_B} R_R^3 \frac{\eta}{T}, \\
\tau_D &=  \frac{2R_T^2}{D_T} = \frac{12\pi}{k_B} R_T^3 \frac{\eta}{T}= \frac{5}{2}\tau_T.
\end{aligned}  \label{eq:TauModel}
\end{eqnarray}
Note that in order to relate the Stokes-Einstein translational-diffusion correlation time $\tau_D$ to the NMR derived translational correlation time $\tau_T$ (Eq. \ref{eq:TauRT}), a factor of $\frac{5}{2}$ is required, i.e. $\tau_D = \frac{5}{2}\tau_T$ \cite{cowan:book}. An expression for the two autocorrelation functions are then derived:
\begin{eqnarray}
\begin{aligned}
G_R(t) &= G_R(0) \exp\!\left(\!-\frac{t}{\tau_R}\right), \\
G_T(t) &= G_T(0) \int_0^{\infty}\!3 \frac{J_{3/2}^2(x)}{x}\exp\!\left(\!-x^2\frac{t}{\frac{5}{2}\tau_T}\right) dx,
\end{aligned}  \label{eq:Gmodel}
\end{eqnarray}
where both expressions reduce to $G_{R,T}(0)$ (Eq. \ref{eq:G0dipolar}) at $t = 0$. 

\begin{figure}[!ht]
	\begin{center}
		\includegraphics[width=0.9\columnwidth]{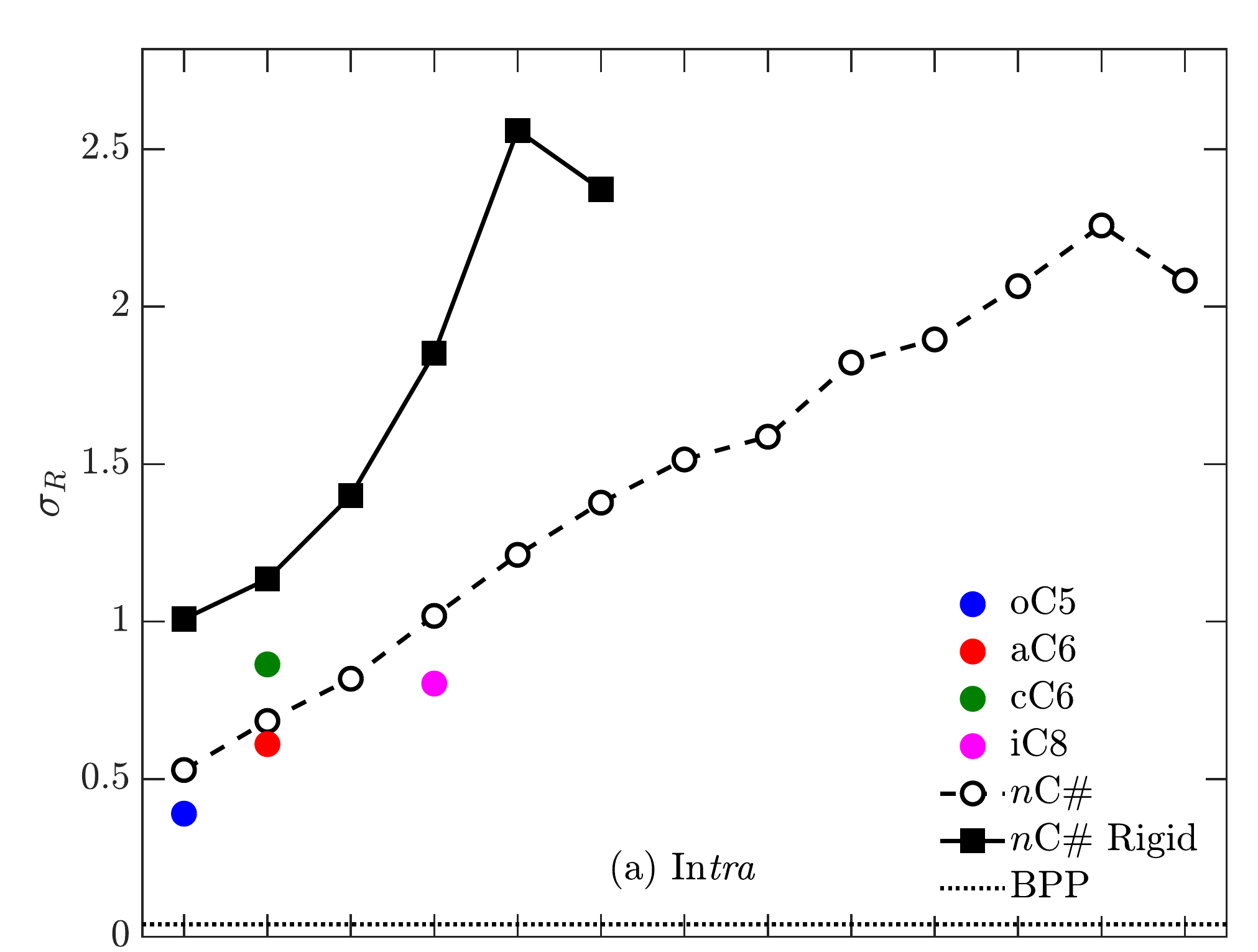}  
		\includegraphics[width=0.9\columnwidth]{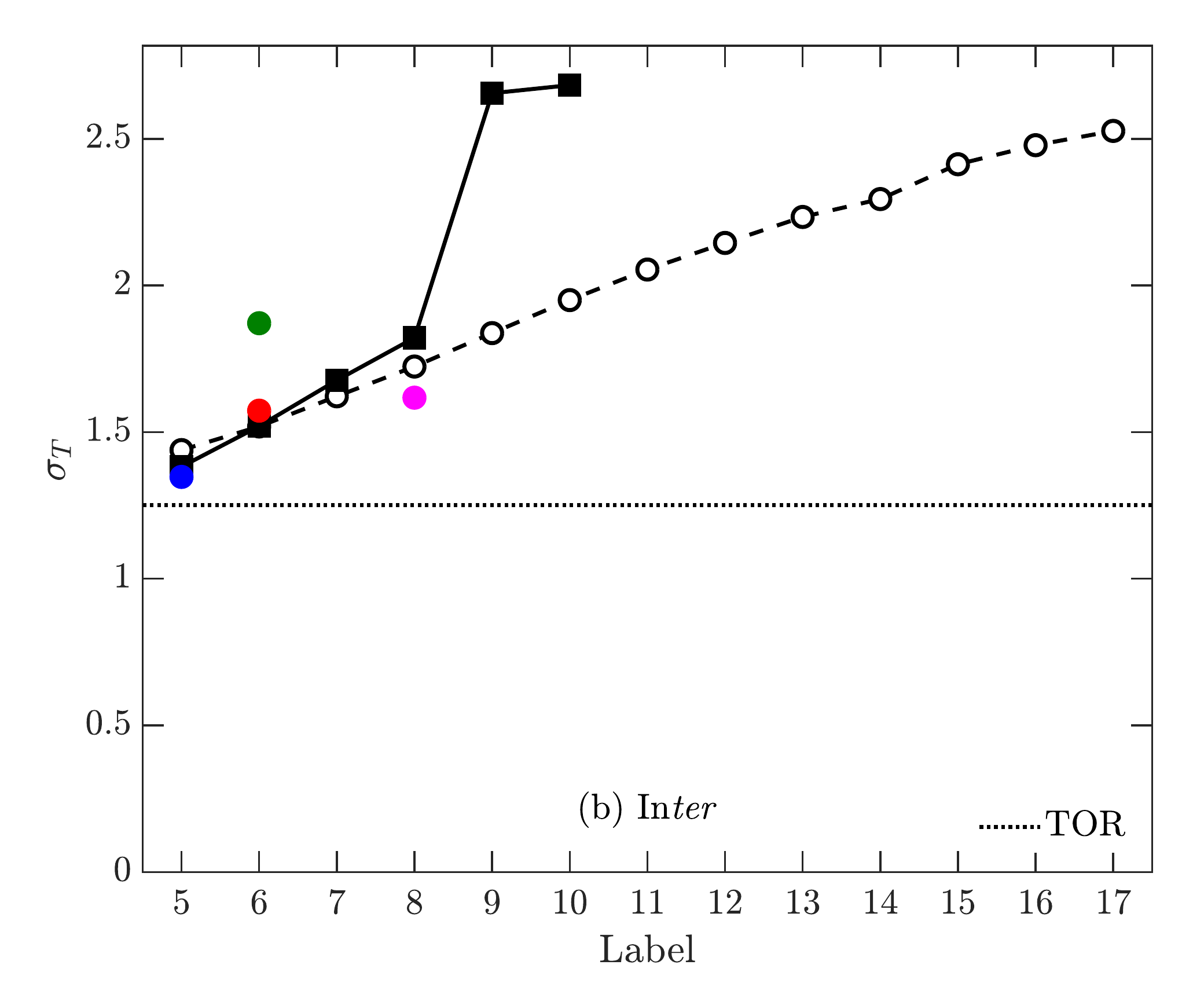} 
	\end{center}
	\caption{MD simulation results for the standard deviation of (a) in{\it tra}molecular $\sigma_R$ and (b) in{\it ter}molecular $\sigma_T$ interactions, using Eq. \ref{eq:CvRT} determined from the $P_{R,T}(\tau)$ distributions (Eq. \ref{eq:ILT}) shown in Appendix A. Same hydrocarbon labels as Fig. \ref{fg:Dipolar}. Also shown are the BPP model which predicts $\sigma_R = 0$ (or $\simeq$ 0.038 in our computation), and the Torrey (TOR) model which predicts $\sigma_T \simeq$ 1.25.} \label{fg:CvRT_HS}
\end{figure}

The BPP hard-sphere model for the in{\it tra}molecular autocorrelation $G_{R}(t)$ is plotted in Fig. \ref{fg:GtIntra}(a) as the dotted black lines. The hard-sphere model is a single-exponential decay (Eq. \ref{eq:Gmodel}), which is a straight-line decay on the semilog-$y$ plot. The in{\it tra}molecular $G_R(t)$ appear to be more ``stretched" (i.e. multi-exponential) in nature than the hard-sphere model, however neopentane (solid line) is clearly closer to a hard-sphere, relative to the linear isomer $n$-pentane (dashed line). Similarly, the isooctane (solid line) is closer to the hard-sphere prediction, relative to the linear isomer $n$-octane (dashed line). These observations are quantified in Fig. \ref{fg:CvRT_HS}, where the standard deviation $\sigma_{R,T}$ are plotted for the same dataset as Fig. \ref{fg:Dipolar}. A more stretched (i.e. multi-exponential) decay in $G_{R}(t)$ yields a larger distribution $P_{R}(\tau)$ in correlation times $\tau$ (Eq. \ref{eq:ILT}), which yields a larger $\sigma_{R,T}$ (Eq. \ref{eq:CvRT}). The BPP hard-sphere model $G_{R}(t)$ in Eq. \ref{eq:Gmodel}, on the other hand, predicts a single exponential decay, i.e. $P_{R}(\tau) = G_{R}(0) \, \delta(\tau - \tau_R)$, corresponding to $\sigma_R = 0$ (which is limited to $\sigma_R \simeq$ 0.038 in our case due to regularization effects of the inverse Laplace transform \cite{venkataramanan:ieee2002,song:jmr2002}). The $P_{R,T}(\tau)$ distributions from the inverse Laplace transform (Eq. \ref{eq:ILT}) used to compute $\sigma_{R,T}$ are shown in Appendix A, including the hard-sphere model. The data in Fig. \ref{fg:CvRT_HS}(a) indicate that $\sigma_R$ drops by $\simeq $ 27 $\%$ for neopentane compared to $n$-pentane; likewise, $\sigma_R$ drops by $\simeq $ 22 $\%$ for isooctane compared to $n$-octane. This quantifies the observation in Fig. \ref{fg:GtIntra} that the more spherical the molecule, the closer to single-exponential decay in $G_{R}(t)$.

Meanwhile, the Torrey hard-sphere model ``TOR" for the in{\it ter}molecular autocorrelation $G_{T}(t)$ is plotted in Fig. \ref{fg:GtInter}(a) as the dotted black lines. The effect of molecular symmetry does not have a significant effect on the functional form of $G_{T}(t)$, and remains equally departed from the hard-sphere model. More specifically, the same analysis of  $G_{T}(t)$ in Fig. \ref{fg:GtInter}(b) indicates roughly similar $\sigma_T$ values with molecular symmetry. Note however that the Torrey hard-sphere model (TOR) has a multi-exponential distribution characterized by $\sigma_T\simeq$ 1.25. This is due to the fact that $G_{T}(t)$ in Eq. \ref{eq:Gmodel} is not a single exponential decay, even though it contains only one correlation time $\tau_T$ (see Apendix A for more details). What is remarkable is that the Torrey hard-sphere prediction $\sigma_T\simeq$ 1.25 is consistent with neopentane, and even to some extent $n$-pentane.

\begin{figure}[!ht]
	\begin{center}
		\includegraphics[width=0.9\columnwidth]{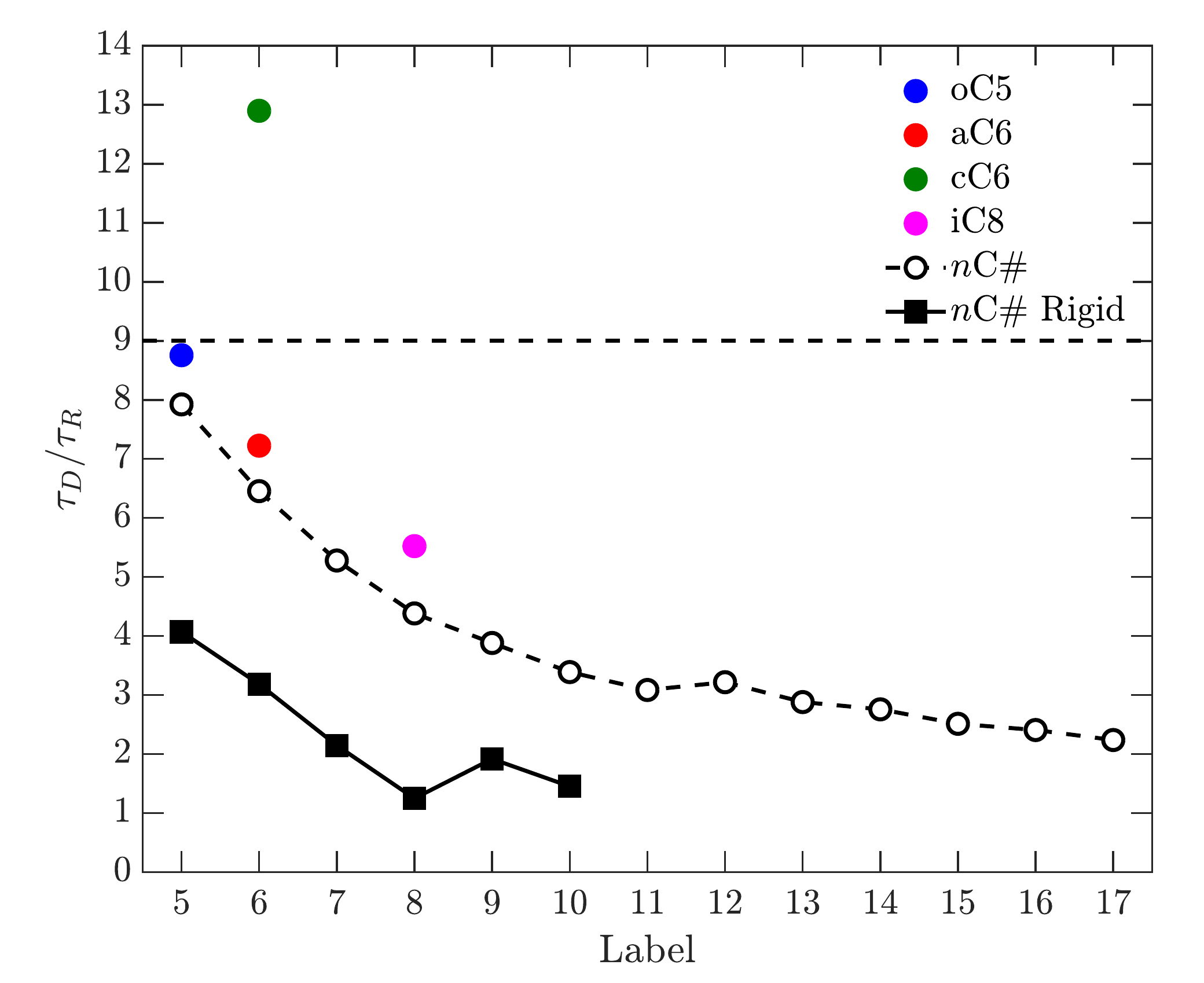}  
	\end{center}
	\caption{MD simulation result for ratio $\tau_D/\tau_R$ of  translational-diffusion correlation-time ($\tau_D = \frac{5}{2}\tau_T$) to rotational-correlation time ($\tau_R$), along with predicted value $\tau_D/\tau_R = 9$ \cite{abragam:book} from Stokes-Einstein relation (Eq. \ref{eq:TauModel}). Same hydrocarbon labels as Fig. \ref{fg:Dipolar}.} \label{fg:TauRatio}
\end{figure} 

Another way to quantify deviations from the hard-sphere model is to compute the ratio ${\tau_D}/{\tau_R}$, which according to Eq. \ref{eq:Gmodel} should be ${\tau_D}/{\tau_R} = 9$ for hard-spheres \cite{abragam:book} where the Stokes-Einstein radius for rotation $R_R$ equals the Stokes-Einstein radius for translation $R_T$. The ratio plotted in Fig. \ref{fg:TauRatio} clearly shows that lower $n$-alkanes tend towards the hard-spheres, while the higher $n$-alkanes increasingly depart from hard-spheres. Similarly, the ratio for isooctane is found to be closer to a hard-sphere than $n$-octane, as expected. Furthermore, in the case of neopentane, the ratio is found to be ${\tau_D}/{\tau_R} = 8.76$, which is very close to the hard-sphere prediction of 9. This is a remarkable finding which validates the hard-sphere models by BPP, Torrey, and Stokes-Einstein in a manner never reported before. 

A similar trend is found for the planar-symmetric molecule benzene, where Fig. \ref{fg:GtIntra}(b) indicates that benzene (solid line) is closer to the BPP prediction (i.e. straighter) than $n$-hexane (dashed line). More specifically, as shown in Fig. \ref{fg:CvRT_HS}(a), $\sigma_R$ is found to be $\simeq$ 10 $\%$ lower for benzene compared to $n$-hexane. Likewise, $\tau_D/\tau_R$ for benzene in Fig. \ref{fg:TauRatio} is closer to the hard-sphere prediction of 9 than $n$-hexane. This indicates that planar-symmetric molecules are more ``spherical" than linear chains of the same carbon number, at least when molecular dynamics are concerned.

On the other hand, Fig. \ref{fg:GtIntra}(b) indicates that cyclohexane (which contains $^1$H's in both axial and equatorial positions) is further away from the BPP prediction (i.e. more stretched) than $n$-hexane. More specifically, as shown in Fig. \ref{fg:CvRT_HS}(a), $\sigma_R$ is found to be $\simeq$ 15 $\%$ higher for cyclohexane compared to $n$-hexane. Likewise $\tau_D/\tau_R$ for cyclohexane in Fig. \ref{fg:TauRatio} is further away from the hard-sphere prediction of 9 than $n$-hexane. This indicates that the out of plane $^1$H sites in cyclohexane result in larger deviations from hard-spheres than linear chains of the same carbon number.

\begin{figure}[!ht]
	\begin{center}
		\includegraphics[width=0.9\columnwidth]{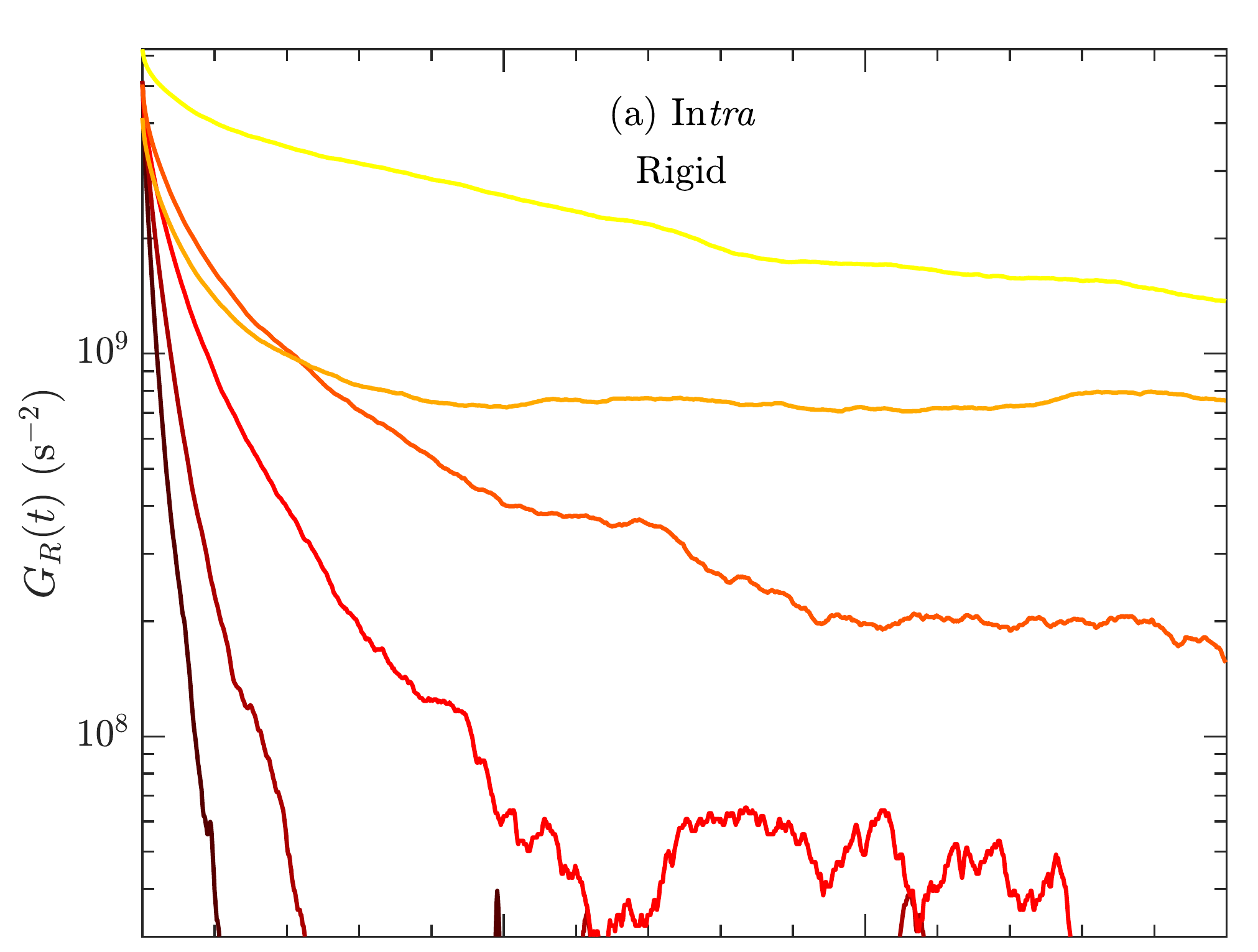} 
		\includegraphics[width=0.9\columnwidth]{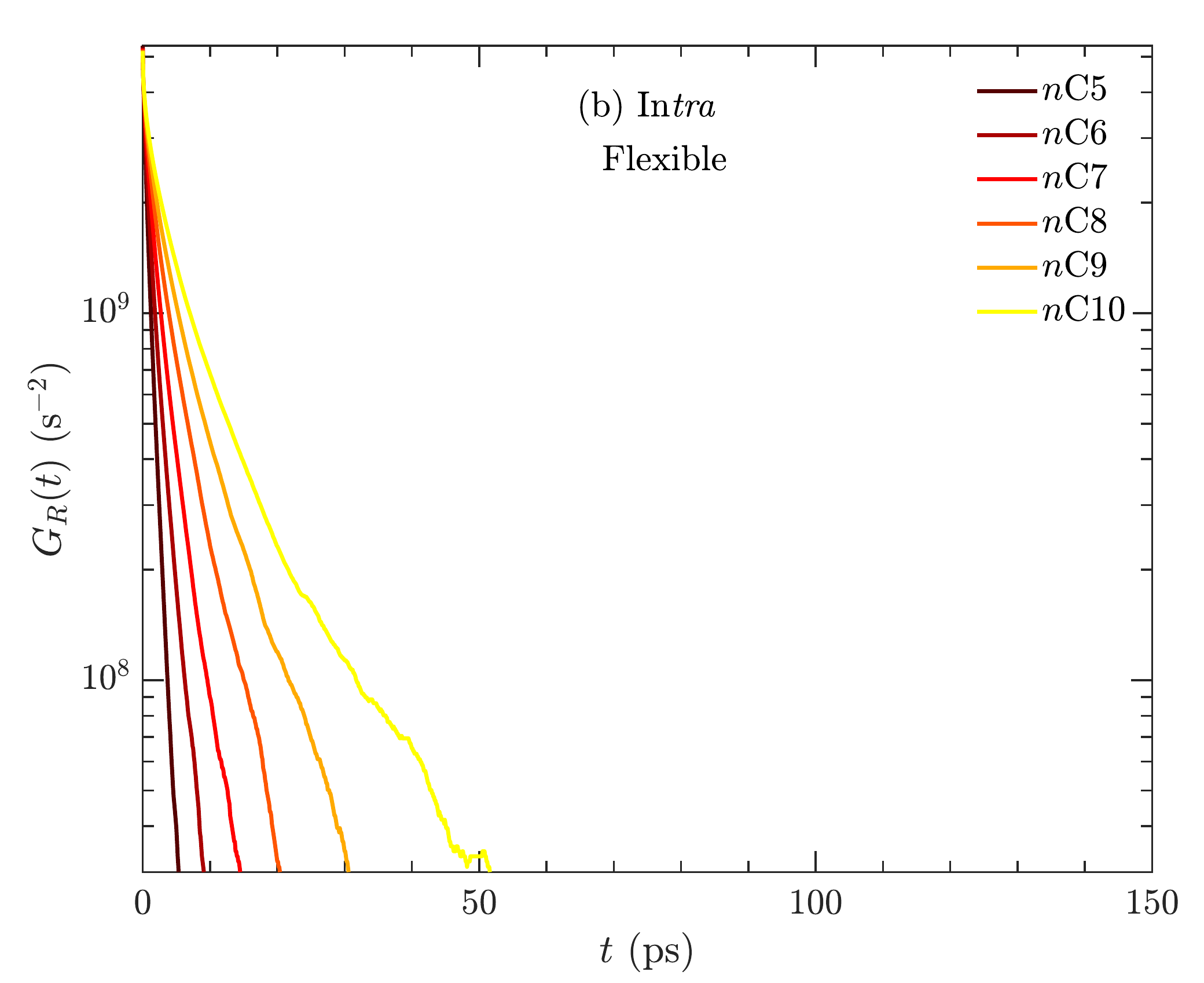}
	\end{center}
	\caption{MD simulations of the autocorrelation function $G_R(t)$ for in{\it tra}molecular interactions, for (a) rigid and (b) flexible hydrocarbons, from $n$-pentane ($n$C5) to $n$-decane ($n$C10).} 
	\label{fg:GtRigidIntra}
\end{figure}

\subsection{Internal motions in flexible versus rigid molecules} \label{ssc:Rigid}

As discussed above, neopentane is a spherical molecule with relatively stiff (i.e. rigid) bonds, and is therefore expected to agree with the above mentioned BPP, Torrey, and Stokes-Einstein theory of hard spheres. We now turn our attention to the effects of internal motions in the long-chain $n$-alkanes, which we study here by simulating completely rigid $n$-alkanes. While rigid molecules do not exist in nature, they provide an ideal testing ground for quantifying the effects of rigidity on the molecular dynamics and the NMR relaxation. For convenience, all data in this section and the previous section are listed in the supplementary material.

\begin{figure}[!ht]
	\begin{center}
		\includegraphics[width=0.9\columnwidth]{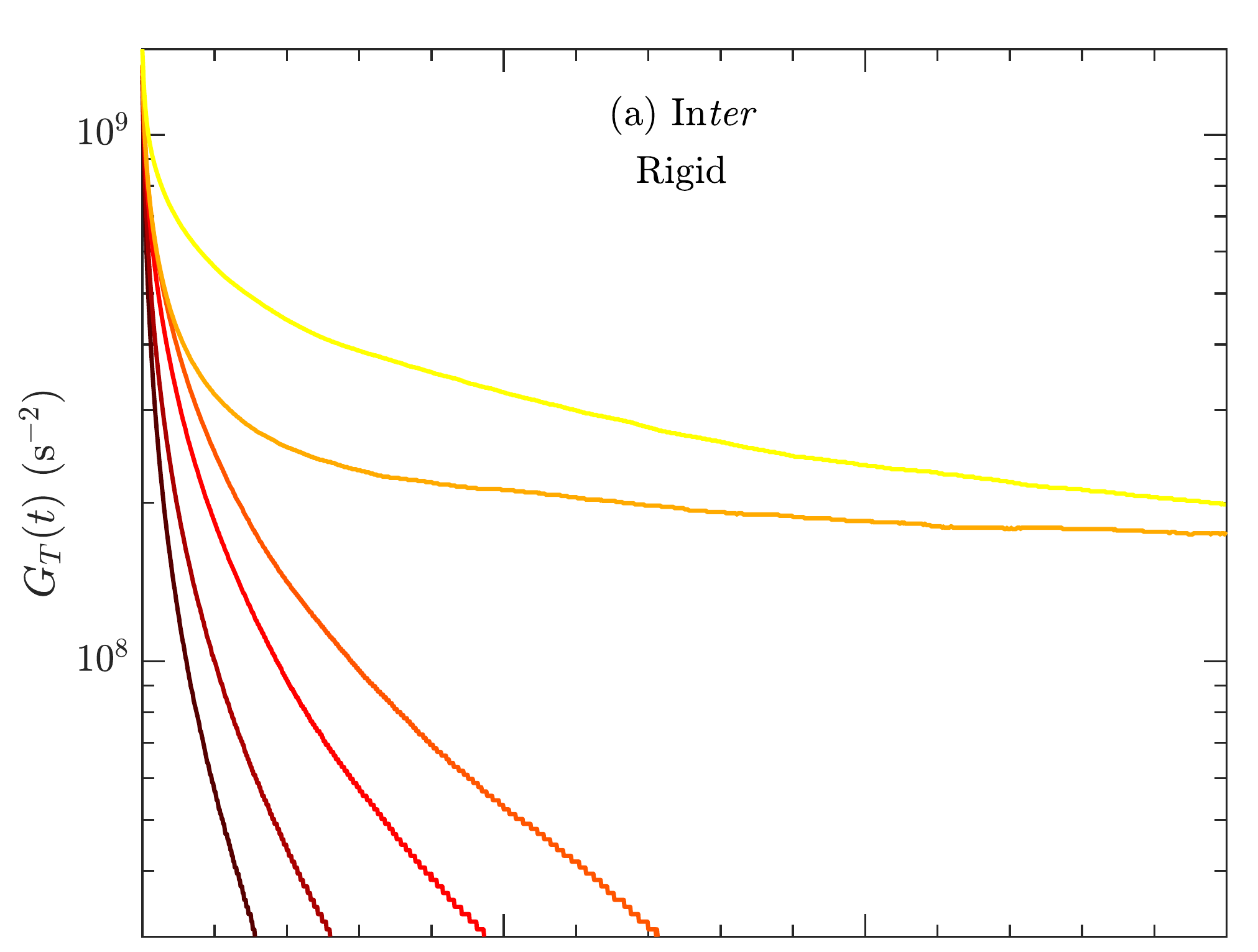} 
		\includegraphics[width=0.9\columnwidth]{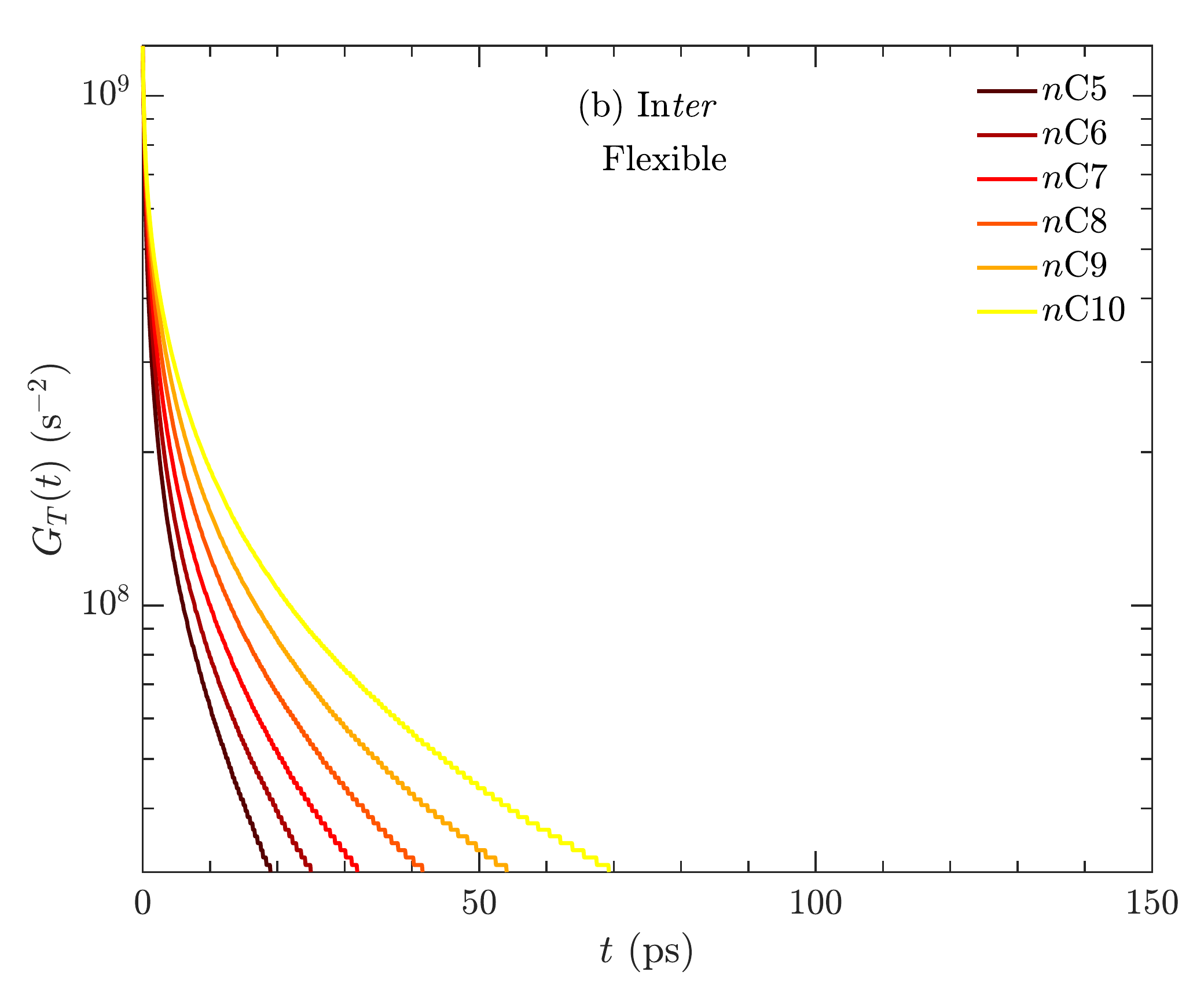}
	\end{center}
	\caption{MD simulations of the autocorrelation function $G_T(t)$ for in{\it ter}molecular interactions, for (a) rigid and (b) flexible hydrocarbons, from $n$-pentane ($n$C5) to $n$-decane ($n$C10).} 
	\label{fg:GtRigidInter}
\end{figure}

The characterization of internal motions in hydrocarbons originates from Woessner's theories \cite{woessner:jcp1965}, who postulated that internal motions would decrease the correlation times $\tau_{R,T}$ and therefore increase the relaxation times $T_{1,2}$ in liquids. As here for $n$-decane, we find that internal motions increase $\tau_{R}$ by a factor $\simeq$ 12, resulting in a factor $\simeq$ 13 decrease in $T_{1,2}$. Furthermore, it was postulated in \cite{woessner:jcp1965} that internal motions would cause the in{\it tra}molecular contribution to relaxation to decrease relative to the in{\it ter}molecular contribution. As shown here for $n$-alkanes, we find that internal motions decrease the in{\it tra}molecular contribution relative to the in{\it ter}molecular contribution by a factor $\simeq$ 2, on average. In other words, we can test the original theories using MD simulations, without invoking any heuristic models.

\begin{figure}[!ht]
	\begin{center}
		\includegraphics[width=0.9\columnwidth]{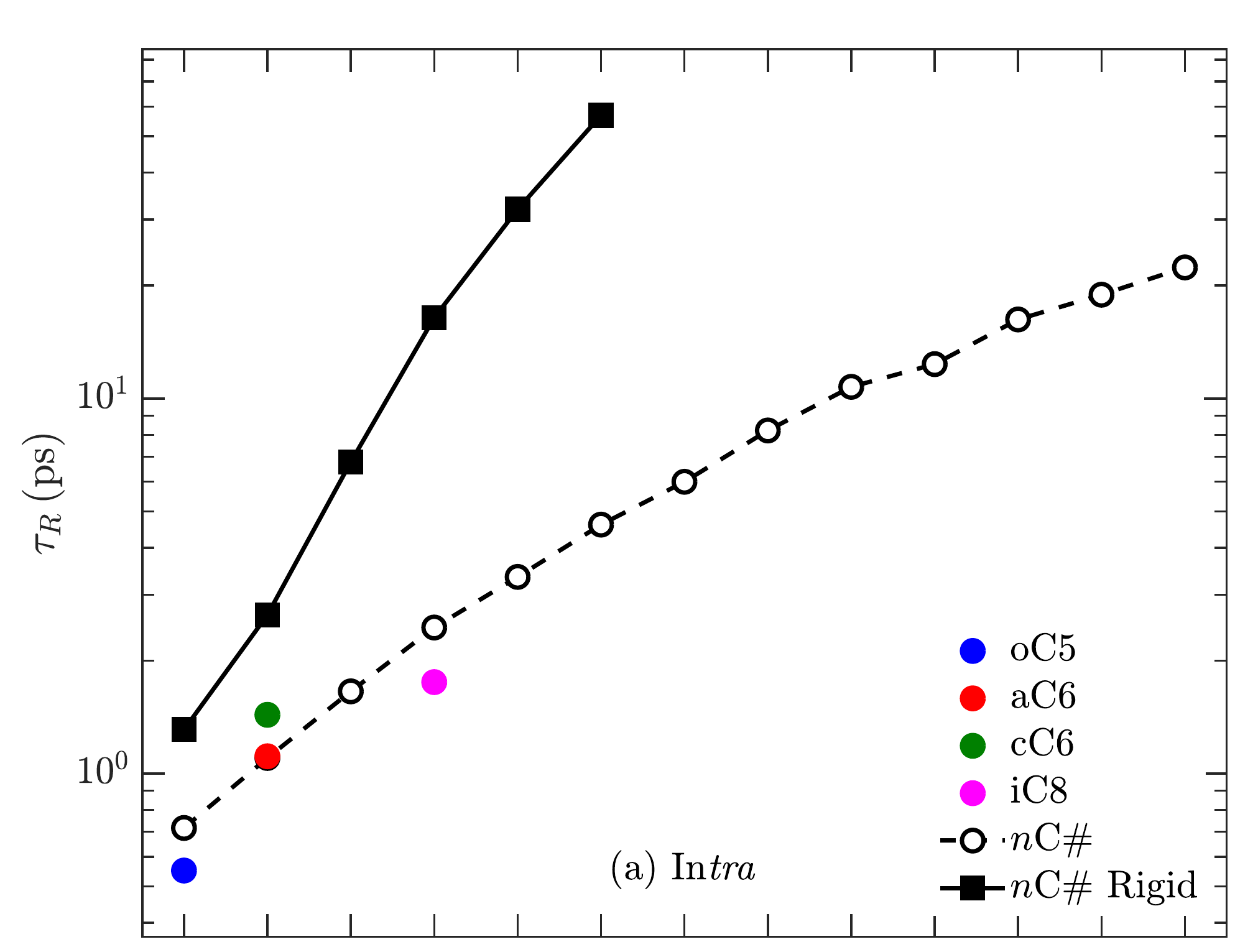}  
		\includegraphics[width=0.9\columnwidth]{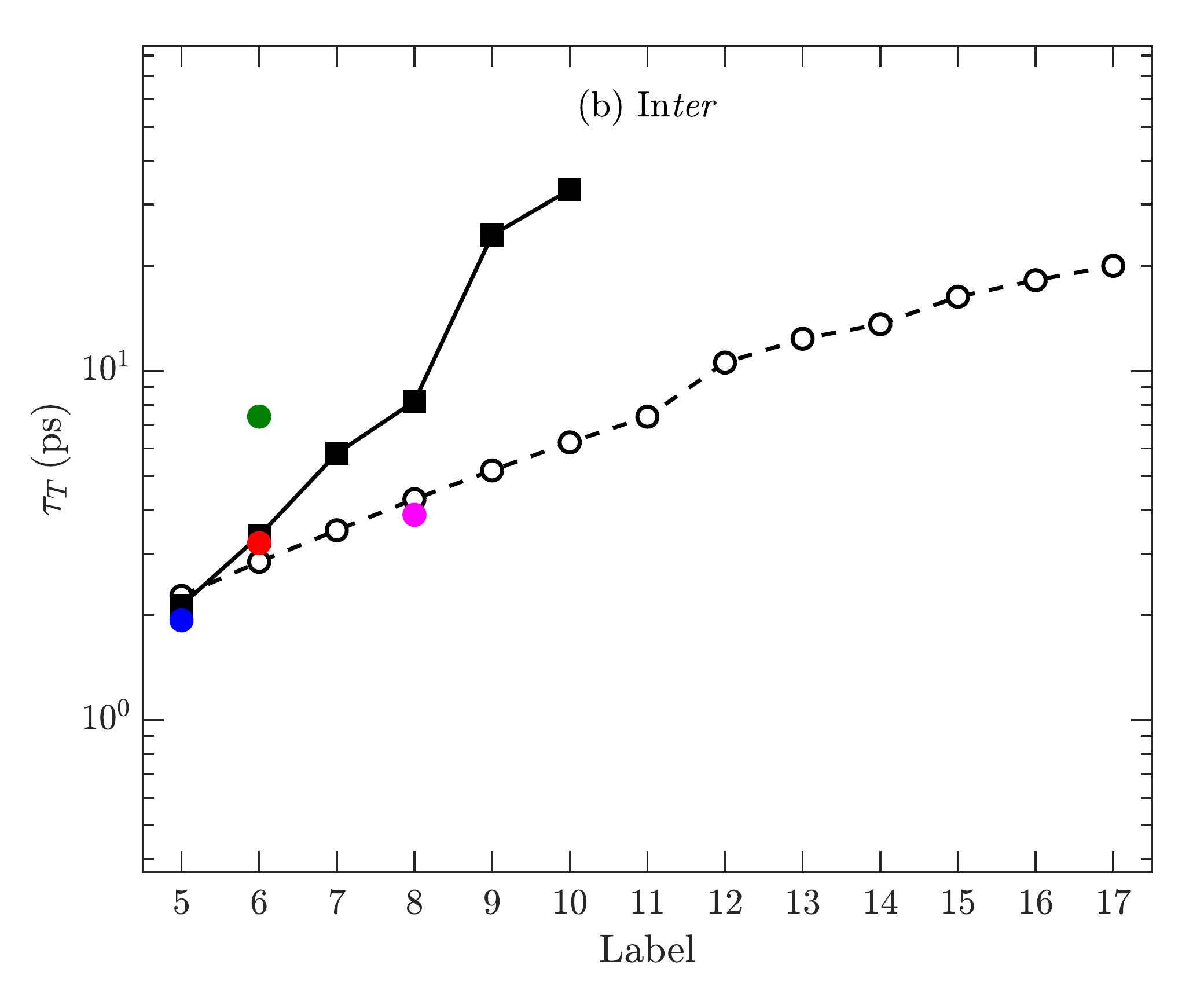}  
	\end{center}
	\caption{MD simulation results for correlation times from (a) in{\it tra}molecular $\tau_R$ and (b) in{\it ter}molecular $\tau_T$ interactions, using Eq. \ref{eq:TauRT}. Same hydrocarbon labels as Fig. \ref{fg:Dipolar}.} \label{fg:TauAlkane}
\end{figure}

\begin{figure}[!ht]
	\begin{center}
		\includegraphics[width=0.9\columnwidth]{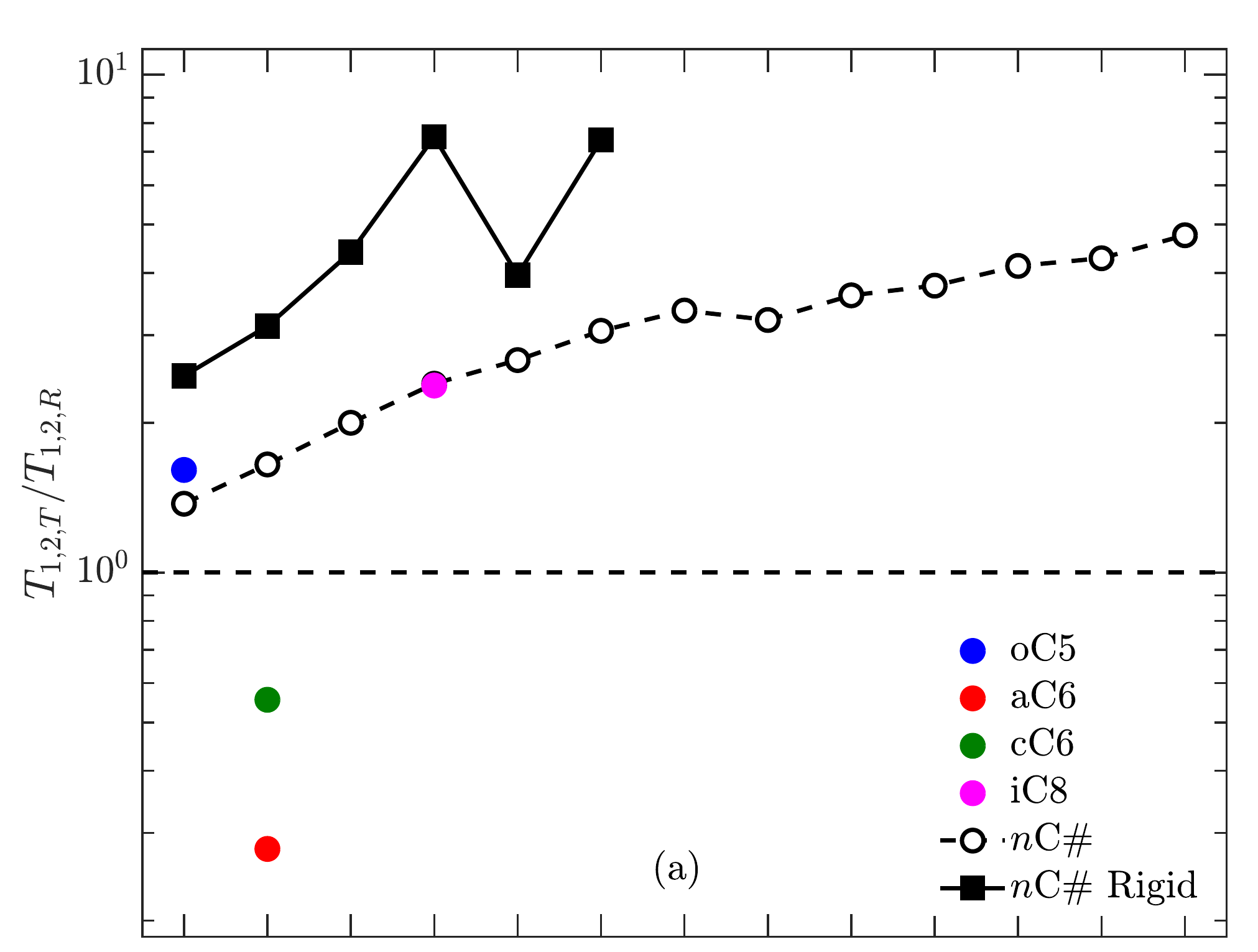}  
		\includegraphics[width=0.9\columnwidth]{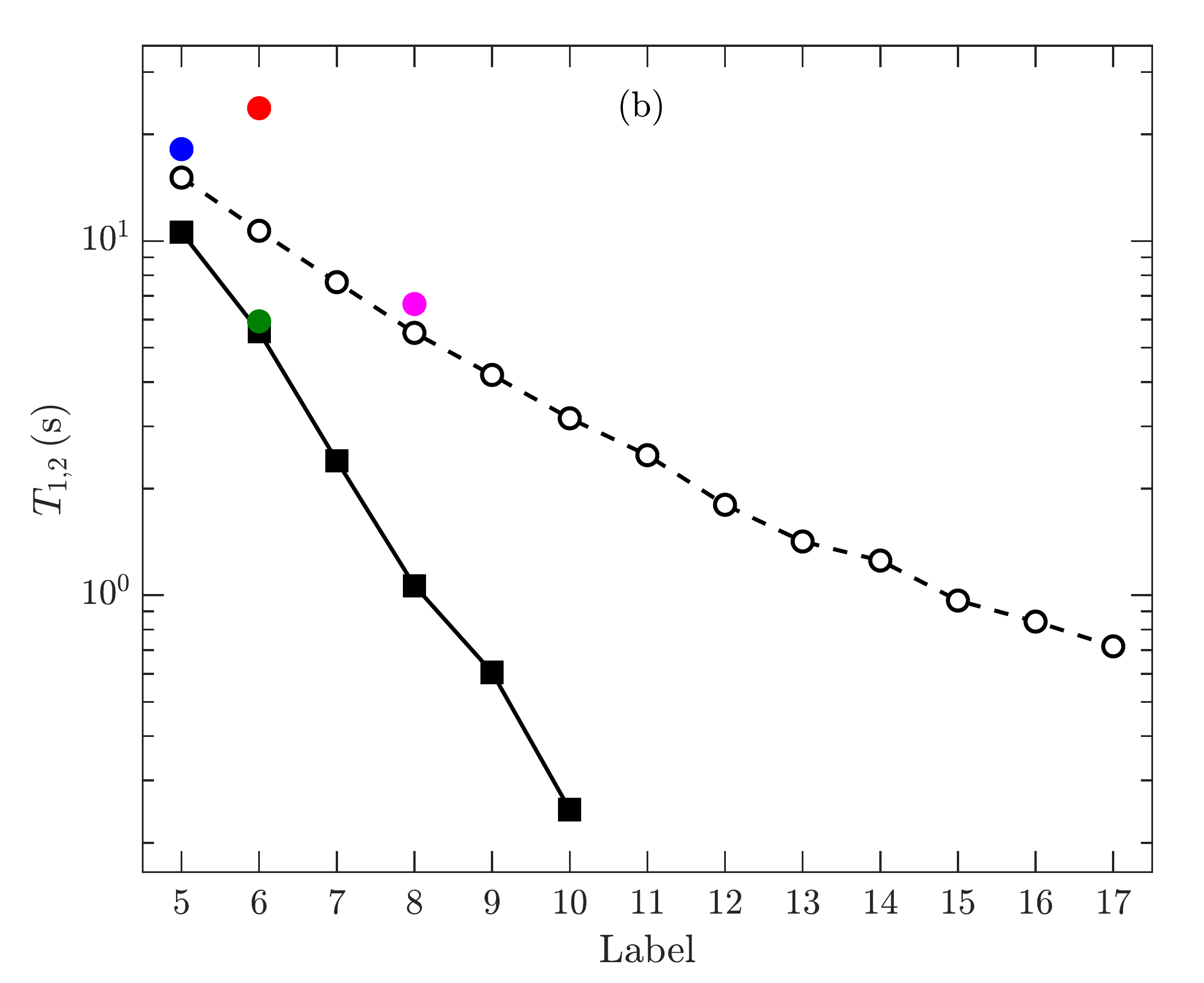} 
	\end{center}
	\caption{MD simulation results for (a) the ratio of relaxation times $T_{1,2,T}/T_{1,2,R}$ using Eq. \ref{eq:T12RTmotional}, where $T_{1,2,T}/T_{1,2,R} \gg 1$ indicates that in{\it tra}molecular dominates over in{\it ter}molecular, and (b) the total relaxation time $T_{1,2,}$ using Eq. \ref{eq:T12motional}. Same hydrocarbon labels as Fig. \ref{fg:Dipolar}.} \label{fg:T1Alkanes}
\end{figure}

The effects of internal motions on in{\it tra}molecular autocorrelation functions $G_R(t)$ for $n$C5$\leftrightarrow$$n$C10 are shown in Fig. \ref{fg:GtRigidIntra}, presented in such a way as to allow for direct comparison between (a) rigid molecules and (b) flexible molecules. The internal motions clearly decrease the correlation times $\tau_R$, as shown when going from (a) rigid to (b) flexible molecules. Likewise, the in{\it ter}molecular autocorrelation functions $G_T(t)$ for $n$C5$\leftrightarrow$$n$C10 are shown in Fig. \ref{fg:GtRigidInter}, and show a similar decrease in correlation times $\tau_T$ in going from (a) rigid to (b) flexible. What is also significant is that the stretched decay for in{\it tra}molecular $G_{R}(t)$ in Fig. \ref{fg:GtRigidIntra} persists for rigid molecules, indicating that molecular geometry plays a crucial role in the functional-form for $G_{R}(t)$. If internal motions were the only cause of the stretched decay, then the functional forms for rigid $G_{R}(t)$ (which do not have internal motions) would be straighter (i.e. closer to single exponential). In fact the opposite is found, namely the standard deviation $\sigma_R$ is a factor $\simeq$ 2 larger for rigid molecules compared to flexible molecules, as shown in Fig. \ref{fg:CvRT_HS}(a). This suggests that internal motions have a tendency to narrow the underlying distribution $P_{R}(\tau)$ (Eq. \ref{eq:ILT}) in correlation times $\tau$. 

Meanwhile, the functional forms for in{\it ter}molecular $G_{T}(t)$ of rigid molecules in Fig. \ref{fg:GtRigidInter} continues to deviate from the Torrey model, again indicating that molecular geometry also plays a crucial role in the functional-form for $G_{T}(t)$. As shown in Fig. \ref{fg:CvRT_HS}(b), the standard deviation $\sigma_T$ is a factor $\simeq$ 2 larger for rigid $n$C9 and $n$C10 compared to flexible $n$C9 and $n$C10. This suggests that internal motions have a tendency to narrow the underlying distribution $P_{T}(\tau)$ (Eq. \ref{eq:ILT}) in correlation times, for $n$C9 and above.

In order to quantify these findings, the same analysis for the dipolar strengths $\Delta\omega_{R,T}$ (Eq. \ref{eq:Dipolar}) and correlation times $\tau_{R,T}$ (Eq. \ref{eq:TauRT}) is applied to the rigid molecules. In the case of $\Delta\omega_{R,T}$, Fig. \ref{fg:Dipolar} indicates that rigidity does not significantly effect the dipolar strength. In the case of $\Delta\omega_{R}$ this is expected since on average over time, the nearest neighbor $^1$H's are the same distance apart for both rigid and flexible, i.e. the equilibrium positions are the same. It is evident however that the rigid $\Delta\omega_{R}$ simulation shows some scattering of the order $\pm$10\% at large chain-lengths. Likewise, $\Delta\omega_{T}$ remains roughly the same since the density of rigid and flexible fluids are the same, implying that the $^1$H spin density is roughly the same. 

The correlation times $\tau_{R,T}$ computed using Eq. \ref{eq:TauRT} for rigid and flexible molecules are plotted in Fig. \ref{fg:TauAlkane} for (a) in{\it tra}molecular and (b) in{\it ter}molecular interactions. The rotational correlation-time $\tau_{R}$ for flexible benzene agrees well with previous estimates from NMR measurements and MD simulations \cite{laaksonen:jcp1998,witt:jpca2000}. In the case of rigid molecules, the maximum autocorrelation time $t = 150$ ps did not fully capture the decay in $G_{R,T}(t)$, implying that $\tau_{R,T}$ is underestimated for rigid $n$C9 and rigid $n$C10, and $T_{1,2}$ is correspondingly overestimated for rigid $n$C9 and rigid $n$C10. Nevertheless, Fig. \ref{fg:TauAlkane}(a) shows that the estimated in{\it tra}molecular $\tau_{R}$ is a factor $\simeq$ 2 greater for rigid $n$C5 than flexible $n$C5, and a factor $\simeq$ 12 greater for rigid $n$C10 than flexible $n$C10. Meanwhile, Fig. \ref{fg:TauAlkane}(b) shows that the estimated in{\it ter}molecular $\tau_{T}$ is roughly the same between rigid $n$C5 and flexible $n$C5, but a factor $\simeq$ 5 greater for rigid $n$C10 than flexible $n$C10. These findings clearly imply that internal motion effects become more prominent with increasing chain-length.

The next step is to compute the corresponding relaxation times from the $\Delta\omega_{R,T}$ and $\tau_{R,T}$ data using Eq. \ref{eq:T12RTmotional}. Fig. \ref{fg:T1Alkanes}(a) shows the ratio in relaxation times $T_{1,2,T}/T_{1,2,R}$, where $T_{1,2,T}/T_{1,2,R} \gg 1$ indicates that in{\it tra}molecular relaxation dominates over in{\it ter}molecular relaxation, while $T_{1,2,T}/T_{1,2,R} \ll 1$ indicates that in{\it ter}molecular dominates over in{\it tra}molecular instead. The data indicates that internal motions decrease the in{\it tra}molecular contribution relative to the in{\it ter}molecular contribution by a factor $\simeq$ 2 on average, although some scattering exists in the $T_{1,2,T}/T_{1,2,R}$ data for rigid $n$C9 and rigid $n$C10. This is consistent with the predictions in \cite{woessner:jcp1965}. These results can also be viewed as the ratio $\tau_D/\tau_R$ (where $\tau_D= \frac{5}{2}\tau_T$) in Fig. \ref{fg:TauRatio}, which shows that $\tau_D/\tau_R$ for rigid molecules is on average by a factor $\simeq$ 2 further away from the hard-sphere model. This indicates that rigid $n$-alkanes are less ``spherical" than flexible $n$-alkanes, as might be expected. 

Finally, Fig. \ref{fg:T1Alkanes}(b) presents the total relaxation time $T_{1,2}$ computed using Eq. \ref{eq:T12motional}, showing that $T_{1,2}$ is a factor $\simeq$ 1.4 shorter for rigid $n$C5 than flexible $n$C5, while it is a factor $\simeq$ 13 shorter for rigid $n$C10 than flexible $n$C10. Again, these findings clearly imply that internal motions effects become more prominent with increasing chain-length, in line with predictions from \cite{woessner:jcp1965}.

It should be noted in passing that for the flexible isomers discussed in Section \ref{ssc:Sphere}, Fig. \ref{fg:T1Alkanes} shows that while neopentane and isooctane show results consistent with their corresponding $n$-alkane, benzene and cyclohexane clearly do not. Specifically, Fig. \ref{fg:T1Alkanes}(a) shows that for all cases $T_{1,2}$ relaxation is dominated by in{\it tra}molecular interactions, i.e. $T_{1,2,T}/T_{1,2,R} \gg 1$, {\it except} for benzene and cyclohexane which show that $T_{1,2}$ is dominated by in{\it ter}molecular interactions instead, i.e. $T_{1,2,T}/T_{1,2,R} \ll 1$. In the case of benzene, this is a result of a lower $\Delta\omega_{R}$ (Fig. \ref{fg:Dipolar}(a)) compared to all the other $n$-alkanes. In the case of cyclohexane, this is a result of a larger $\tau_T$ (Fig. \ref{fg:TauAlkane}(b)) compared to $n$-hexane. These difference result in the spread of $T_{1,2}$ (Fig. \ref{fg:T1Alkanes}(b)) for benzene and cyclohexane compared with $n$-hexane. 

\subsection{Site-by-site simulations and distribution in correlation times} \label{ssc:Site}

The above results show that internal motions are not the primary cause of the stretched decay. Our next task is therefore to determine whether the multi-exponential decay in $G_{R}(t)$ (and the deviations in $G_{T}(t)$ from the Torrey model) are a result of underlying variations in molecular dynamics across the chain-length. This scenario was postulated previously, where the fast rotation (i.e. short $\tau_R$) of the methyl groups act as relaxation sinks for the macro-molecule \cite{kalk:jmr1976}. For convenience, all data in this section are listed in the supplementary material.

\begin{figure}[!ht]
	\begin{center}
		\includegraphics[width=0.9\columnwidth]{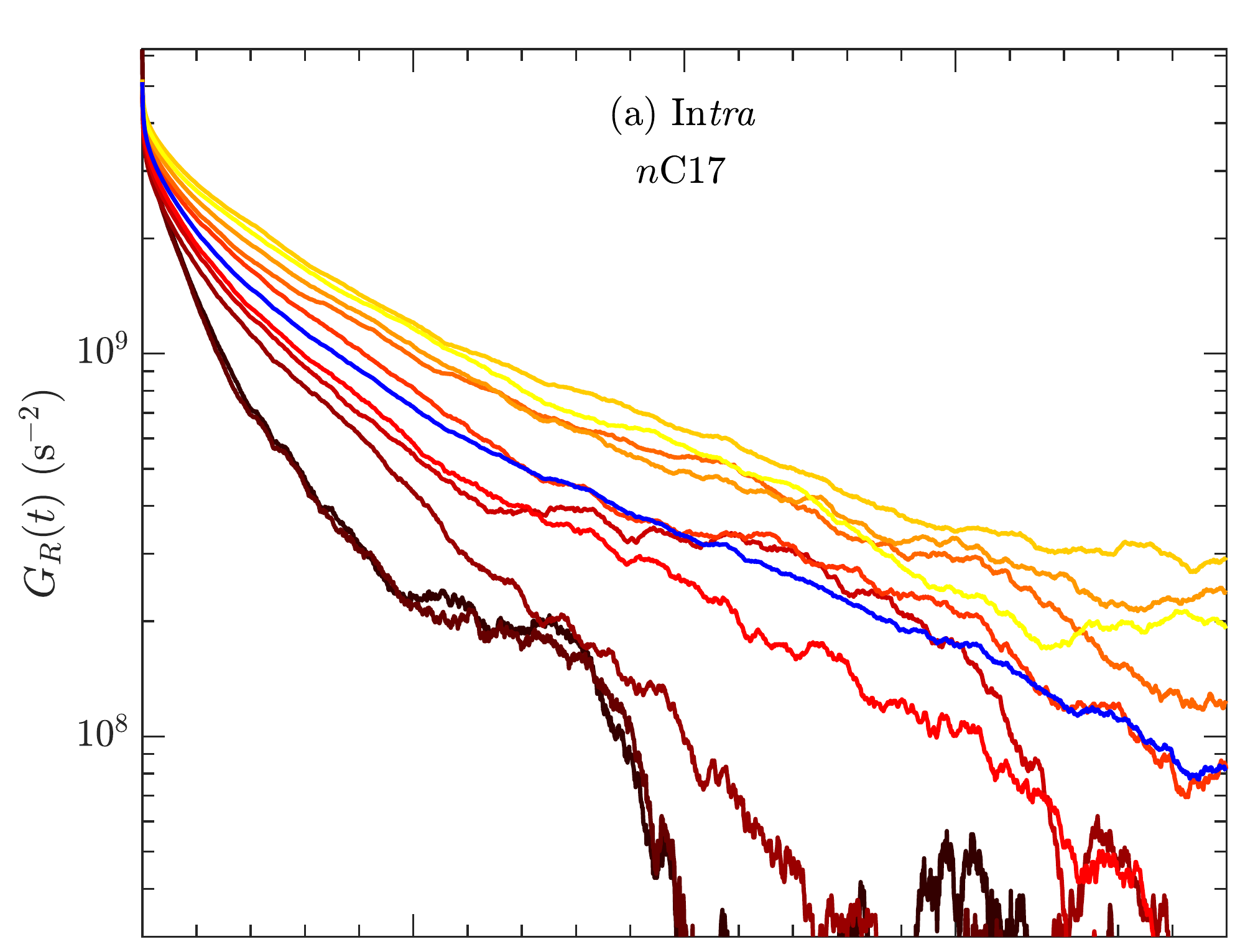}
		\includegraphics[width=0.9\columnwidth]{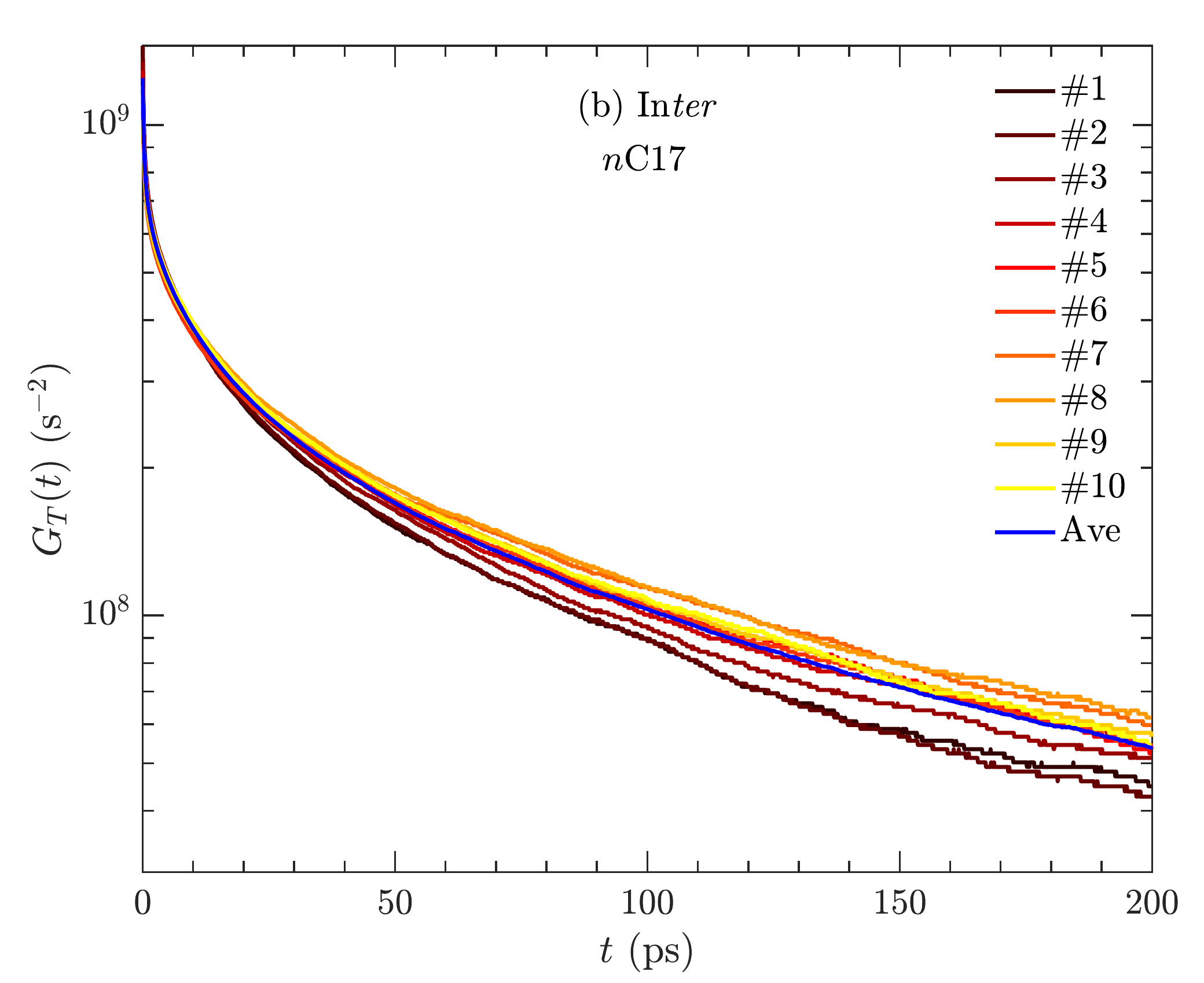}
	\end{center}
	\caption{MD simulation results of the autocorrelation function for (a) in{\it tra}molecular $G_R(t)$ and (b) in{\it ter}molecular $G_T(t)$ interactions using Eq. \ref{eq:GmRT}, for $n$-heptadecane ($n$C17) on a site-by-site basis with labels defined in Fig. \ref{fg:Molecules}. ``Ave" indicates weighted average over the site-by-site results.} 
	\label{fg:GtC17}
\end{figure}

\begin{figure}[!ht]
	\begin{center}
		\includegraphics[width=0.9\columnwidth]{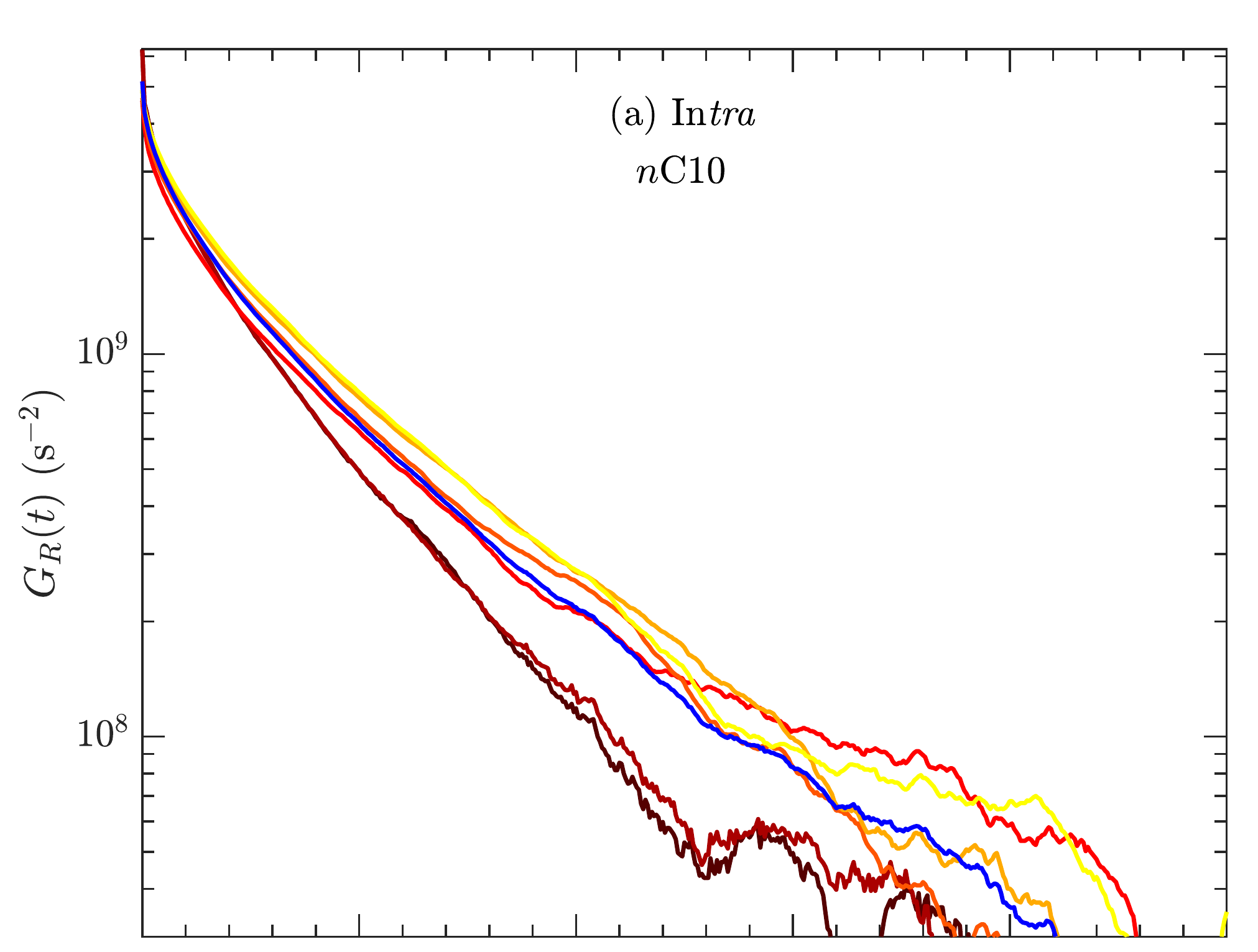}
		\includegraphics[width=0.9\columnwidth]{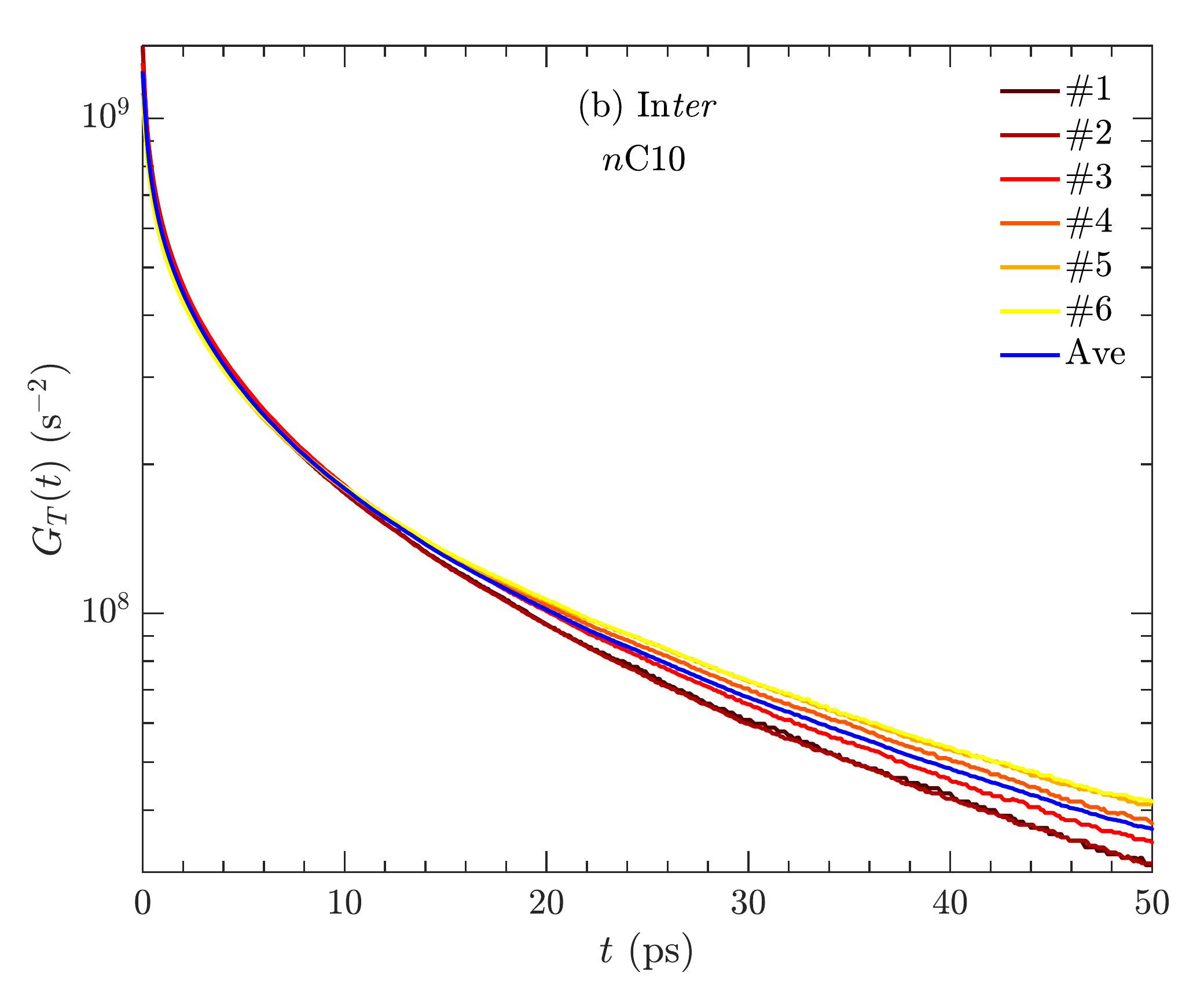}
	\end{center}
	\caption{MD simulation results of the autocorrelation function for (a) in{\it tra}molecular $G_R(t)$ and (b) in{\it ter}molecular $G_T(t)$ interactions using Eq. \ref{eq:GmRT}, for $n$-decane ($n$C10) on a site-by-site basis with labels defined in Fig. \ref{fg:Molecules}. ``Ave" indicates weighted average over the site-by-site results.} 
	\label{fg:GtC10}
\end{figure}

In order to investigate variations across the chain, we perform the MD simulations for each $^1$H across the chain, and then compute the correlation times $\tau_{R,T}$ and relaxation times $T_{1,2}$ on a site-by-site basis across the chain. The $^1$H sites are labeled in Fig. \ref{fg:Molecules} for $n$-decane and $n$-heptadecane, where the $\#1$ $^1$H is on the methyl end-group, and the largest number is in the middle of the chain (more details in Section \ref{ssc:Molecules}). Intramolecular $G_{R}(t)$ for $n$-heptadecane in Fig. \ref{fg:GtC17}(a) shows a large distribution in correlation times, with the $\#1$ and $\#$2 $^1$H's on the methyl clearly having the steepest decay (i.e. shortest  $\tau_{R}$), as expected from \cite{kalk:jmr1976}. The average ``Ave" is the weighted average over all sites, and constitutes what was reported in the previous sections. Meanwhile, in{\it ter}molecular $G_{T}(t)$ for $n$-heptadecane in Fig. \ref{fg:GtC17}(b) shows less variation between sites than in{\it tra}molecular, which is intuitive since the distance of closest approach between molecules should be roughly independent of the location along the chain. The same site-by-site method is used for $n$-decane. Both the in{\it tra}molecular $G_{R}(t)$ for $n$-decane in Fig. \ref{fg:GtC10}(a) and the in{\it ter}molecular $G_{T}(t)$ for $n$-decane in Fig. \ref{fg:GtC10}(b) show less variation between sites than $n$-heptadecane. This is intuitive since one expects there to be more variation in the molecular dynamics across longer chains. 

\begin{figure}[!ht]
	\begin{center}
		\includegraphics[width=0.9\columnwidth]{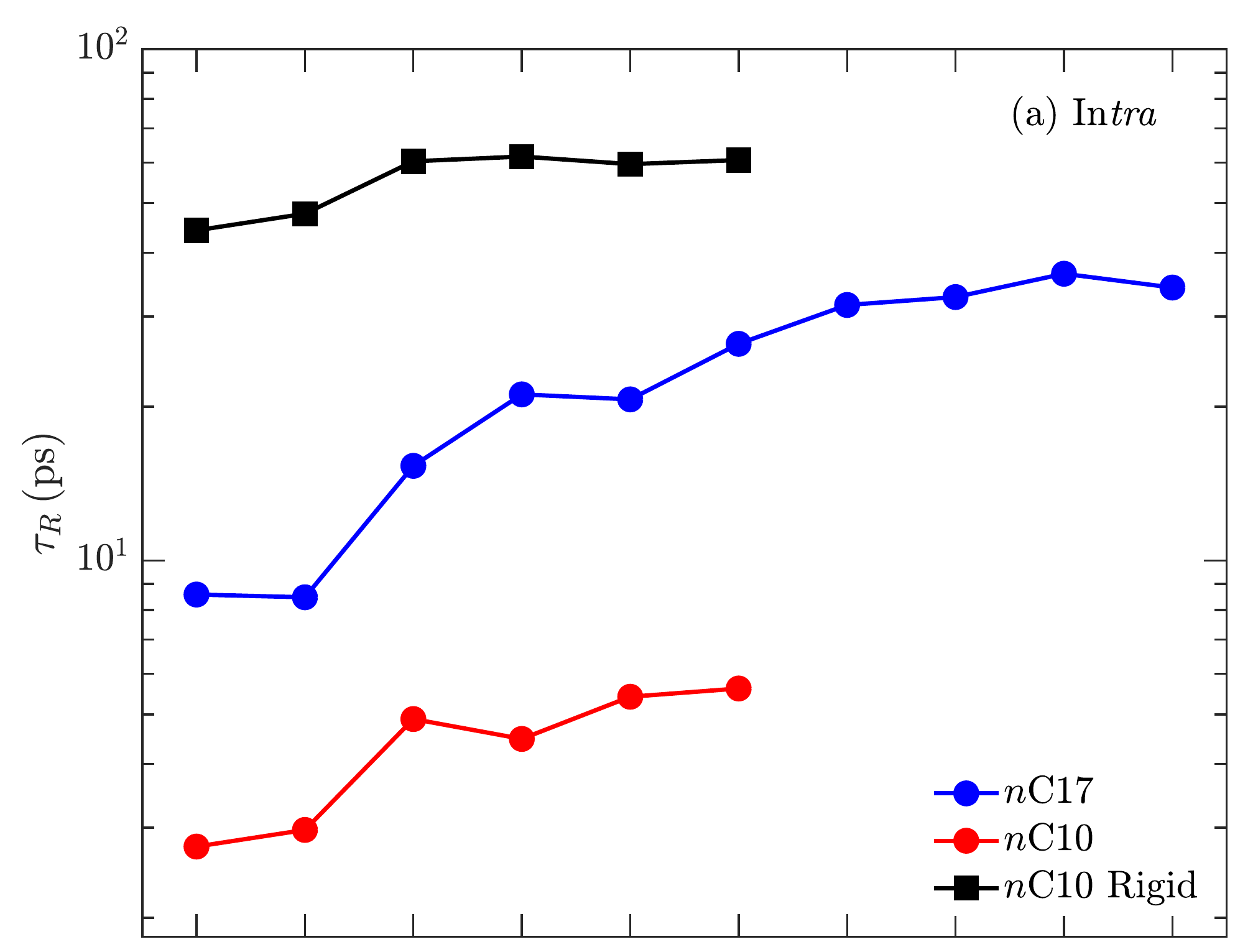} 
		\includegraphics[width=0.9\columnwidth]{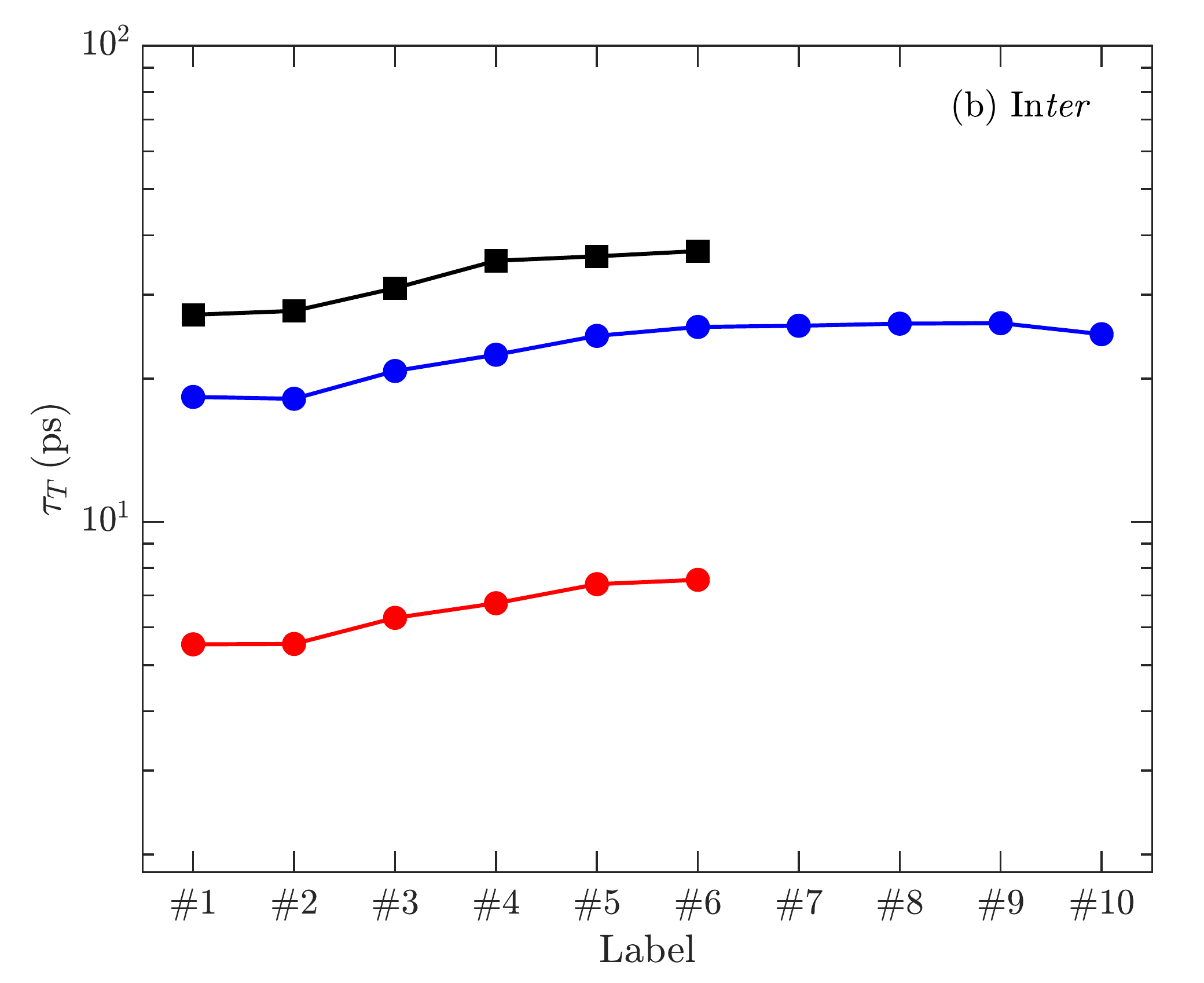} 
	\end{center}
	\caption{MD simulation results for correlation times from (a) in{\it tra}molecular $\tau_R$ and (b) in{\it ter}molecular $\tau_T$ interactions, using Eq. \ref{eq:TauRT}, on a site-by-site basis with site labels defined in Fig. \ref{fg:Molecules}. ``Rigid" refers to rigid hydrocarbons, while all others are flexible.} \label{fg:TauSite}
\end{figure}

A more quantitative summary of the site-by-site correlation times $\tau_{R,T}$ (using Eq. \ref{eq:TauRT}) is given in Fig. \ref{fg:TauSite}, for both $n$-decane and $n$-heptadecane. In the case of $n$-heptadecane, the $^1$H's labeled $\#$1 and $\#$2 (both on the methyl end-group) show a factor $\simeq$ 4 shorter in{\it tra}molecular $\tau_{R}$ than at the middle of the chain, and a factor $\simeq$ 1.4 shorter in{\it ter}molecular $\tau_{T}$. In the case of $n$-decane, $\#$1 and $\#$2 show a factor $\simeq$ 2 shorter in{\it tra}molecular $\tau_{R}$ than at the middle of the chain, and a factor $\simeq$ 1.4 shorter in{\it ter}molecular $\tau_{T}$. Also shown in Fig. \ref{fg:TauSite} are results from the site-by-site simulation for rigid $n$-decane. It is interesting to note that for rigid $n$-decane, $\#$1 and $\#$2 only show a factor $\simeq$ 1.4 shorter in{\it tra}molecular $\tau_{R}$ than the chain middle, which is somewhat less variation than for flexible $n$-decane. In other words, the internal motions have a tendency to enhance the site-by-site variation of in{\it tra}molecular $\tau_{R}$ across the chain. In the case of in{\it ter}molecular $\tau_{T}$, both rigid and flexible $n$-decane (as well as flexible $n$-heptadecane) show the same factor $\simeq$ 1.4 shorter $\tau_{T}$ for $\#$1 and $\#$2 versus the chain middle, which is reasonable given that the distance of closest approach between molecules should be the same across the chain, regardless of whether the molecule is rigid or not.

\begin{figure}[!ht]
	\begin{center}
		\includegraphics[width=0.9\columnwidth]{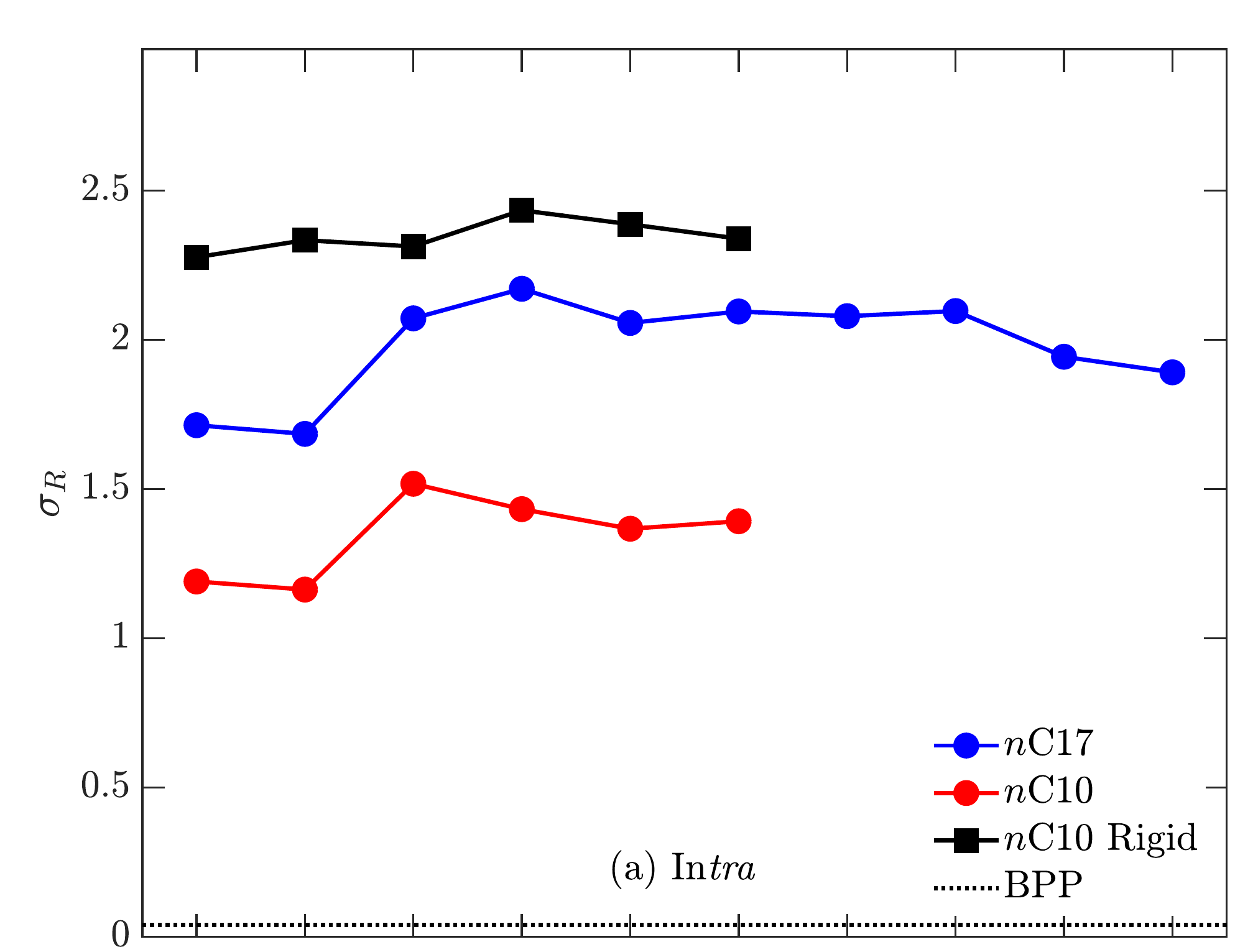}  
		\includegraphics[width=0.9\columnwidth]{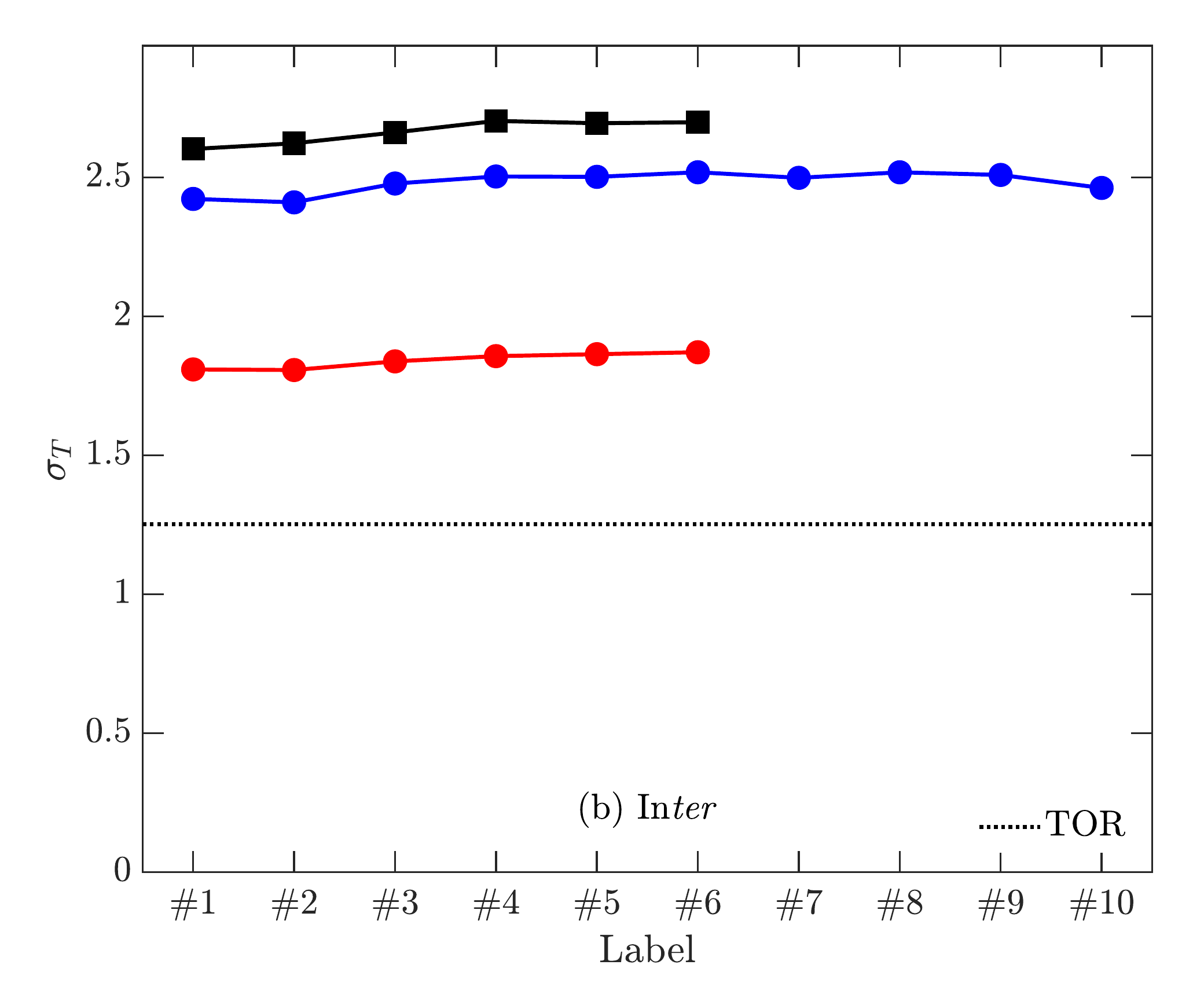} 
	\end{center}
	\caption{MD simulation results for the standard deviation of (a) in{\it tra}molecular $\sigma_R$ and (b) in{\it ter}molecular $\sigma_T$ interactions, using Eq. \ref{eq:CvRT} determined from the $P_{R,T}(\tau)$ distributions (Eq. \ref{eq:ILT}). Same hydrocarbon labels as Fig. \ref{fg:TauSite}. Also shown are the BPP model which predicts $\sigma_R = 0$ (or $\simeq$ 0.038 in our computation), and the Torrey (TOR) model which predicts $\sigma_T \simeq$ 1.25.} \label{fg:CvRT_Site}
\end{figure}

Given the large variation in $\tau_{R}$ across the chain (Fig. \ref{fg:TauSite}), one expects the average $G_{R}(t)$ to be a multi-exponential (i.e. stretched) decay, since the average $G_{R}(t)$ is the weighted sum of decays over all sites. Indeed $n$-decane (Fig. \ref{fg:GtC10}) and $n$-heptadecane (Fig. \ref{fg:GtC17}) show a stretched decay for the average. What is remarkable, however, is that $G_{R}(t)$ for both $n$-decane (Fig. \ref{fg:GtC10}) and $n$-heptadecane (Fig. \ref{fg:GtC17}) show a stretched decay {\it at every site}. The extent of the stretched (i.e. multi-exponential) decay can be quantified on a site-by-site basis using $\sigma_{R,T}$ (Eq. \ref{eq:CvRT}), in a similar fashion as Section \ref{ssc:Sphere}. As shown in \ref{fg:CvRT_Site}(a), $\sigma_R$ does not vary significantly across the chain, although the data may indicate a slightly lower value at the chain ends ($\#$1 and $\#$2), for both $n$C10 and $n$C17. The fact that $\sigma_R$ is roughly uniform across the chain implies that not only does each site show a stretched decay, but all the sites show roughly the {\it same} stretched functional-form as the average. Similar results are found for the case of site-by-site $G_{R}(t)$ for {\it rigid} $n$-decane, where Fig. \ref{fg:CvRT_Site}(a) shows a roughly uniform $\sigma_R$ across the rigid chain. All these findings imply that the overall molecular geometry is a crucial factor, perhaps even the {\it dominant} factor, behind the functional form in the decay of $G_{R}(t)$.
 
\begin{figure}[!ht]
	\begin{center}
		\includegraphics[width=0.9\columnwidth]{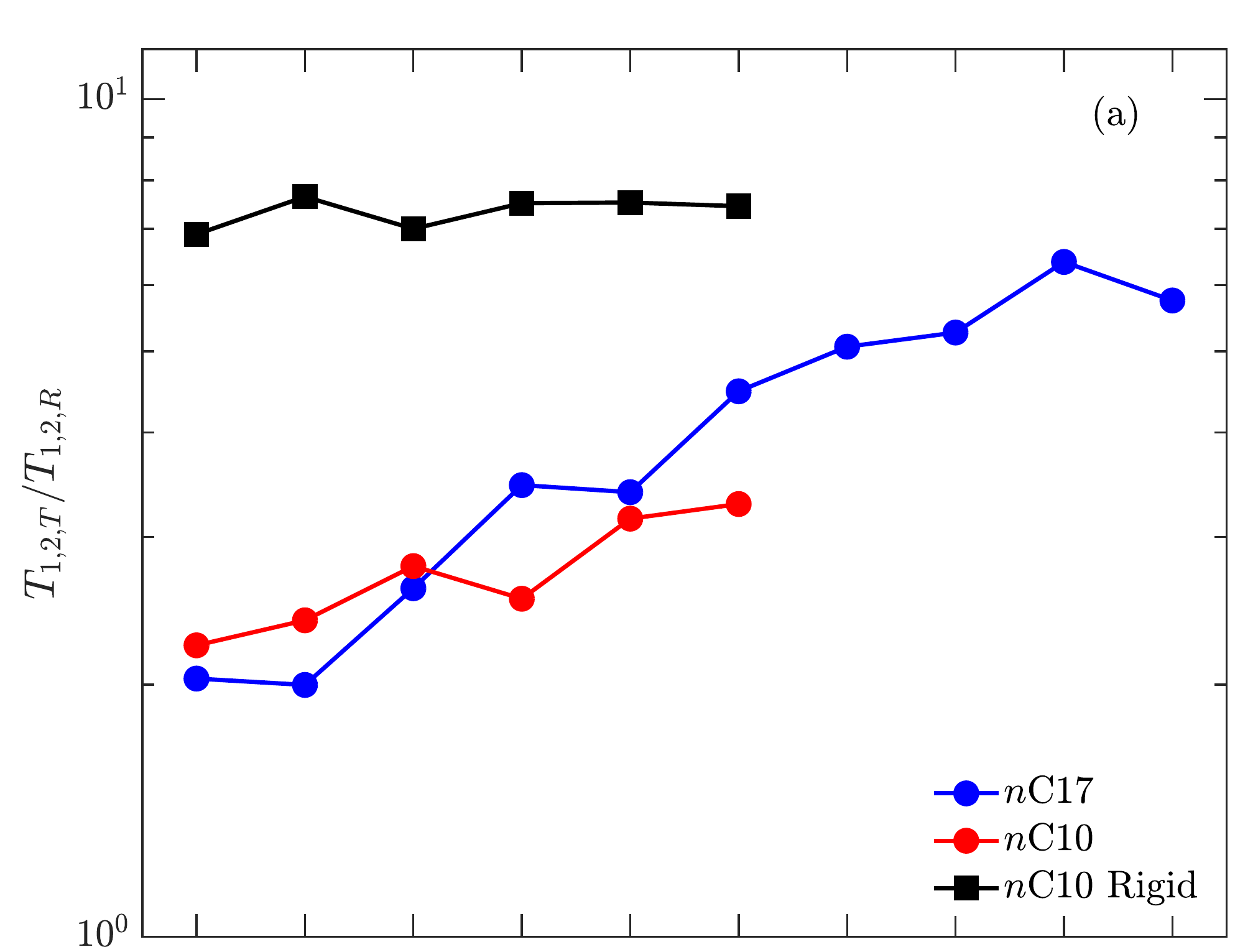} 
		\includegraphics[width=0.9\columnwidth]{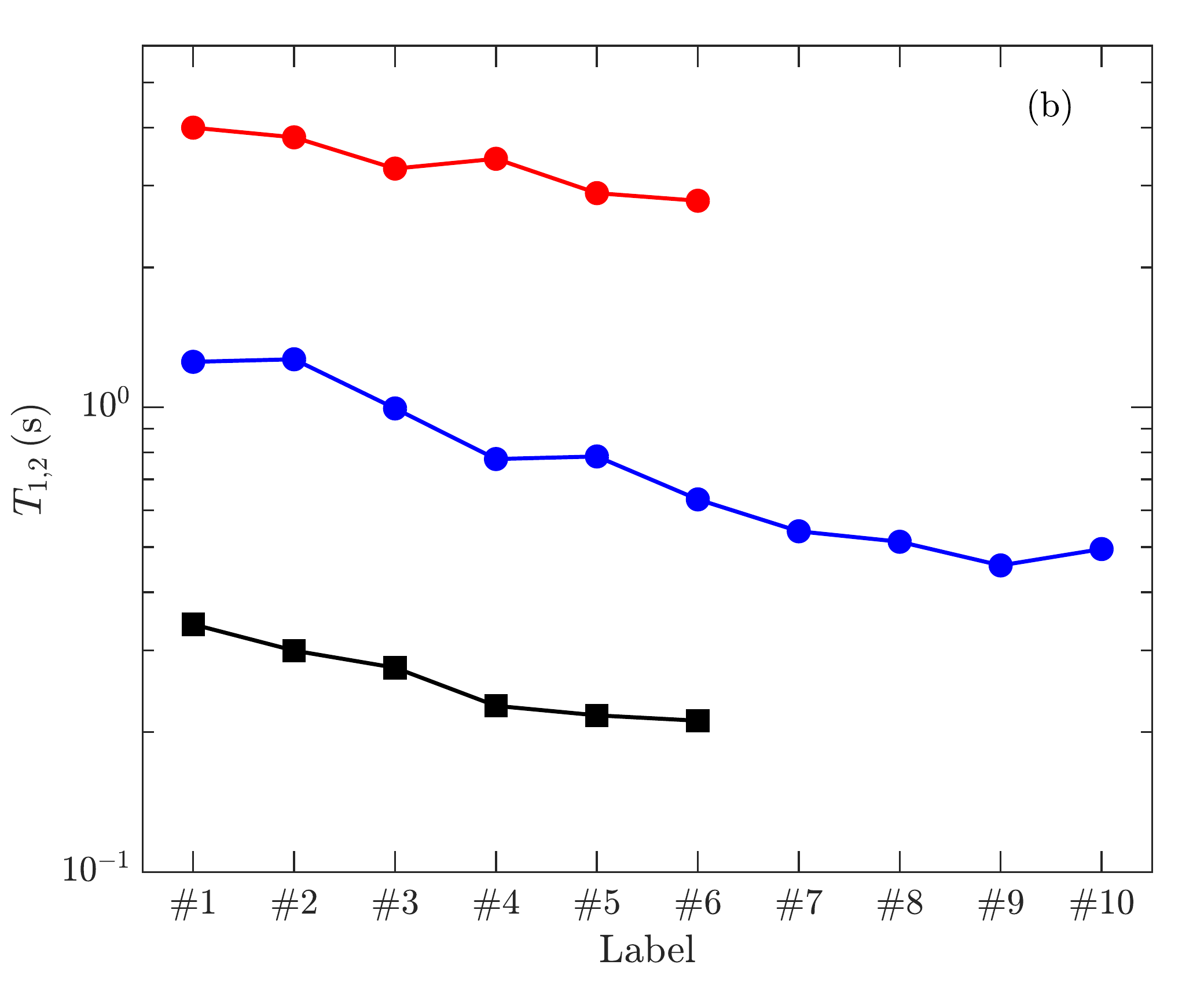}
	\end{center}
	\caption{MD simulation results for (a) the ratio of relaxation times $T_{1,2,T}/T_{1,2,R}$ using Eq. \ref{eq:T12RTmotional}, where $T_{1,2,T}/T_{1,2,R} \gg 1$ indicates that in{\it tra}molecular dominates over in{\it ter}molecular, and (b) the total relaxation time $T_{1,2,}$ using Eq. \ref{eq:T12motional}. Same hydrocarbon labels as Fig. \ref{fg:TauSite}.} \label{fg:T12Site}
\end{figure} 

Similar conclusions can also be made for in{\it ter}molecular $G_{T}(t)$ in Figs. \ref{fg:GtC10}(b) and \ref{fg:GtC17}(b), which clearly show that the site-by-site decays overlap with the average. The same can be inferred from the $\sigma_T$ data in Fig. \ref{fg:CvRT_Site}(b), with even more certainty since $\sigma_T$ shows less scattering than $\sigma_R$ .

It is also interesting to track the strength of in{\it tra}molecular versus in{\it ter}molecular relaxation on a site-by-site basis across the chain. Fig. \ref{fg:T12Site}(a) presents the ratio $T_{1,2,T}/T_{1,2,R}$ across the chain sites, which shows that the in{\it tra}molecular relaxation dominates in the chain middle (i.e. $T_{1,2,T}/T_{1,2,R} \gg 1$), while in{\it tra}molecular relaxation is a factor of $\simeq$ 2 stronger than in{\it ter}molecular at the chain ends (i.e. $\#$1 and $\#$2). In the case of $n$-heptadecane, the chain middle shows a factor $\simeq$ 3 larger value of $T_{1,2,T}/T_{1,2,R}$ than the chain ends. In the case of $n$-decane, the chain middle shows only a factor $\simeq$ 1.5 larger value of $T_{1,2,T}/T_{1,2,R}$ than the chain ends, which is reasonable given the smaller variation in correlation times $\tau_{R,T}$ for $n$-decane (Fig. \ref{fg:TauSite}). Rigid $n$-decane shows almost no variation in $T_{1,2,T}/T_{1,2,R} \simeq$ 7, and clearly in{\it tra}molecular relaxation dominates across the entire chain.

Finally, Fig. \ref{fg:T12Site}(b) presents the total relaxation time $T_{1,2}$ on a site-by-site basis. In the case of $n$-heptadecane, the chain middle shows a factor $\simeq$ 3 shorter value of $T_{1,2}$ than the chain ends. In the case of rigid and flexible $n$-decane, the chain middle shows only a factor $\simeq$ 1.5 shorter value of $T_{1,2}$ than the chain ends, which is reasonable given the smaller variation in correlation times $\tau_{R,T}$ in Fig. \ref{fg:TauSite}.

\subsubsection{Cross-relaxation effects and comparison with measurements}\label{sssc:Cross}

Despite the site-by-site variation in correlation times $\tau_{R,T}$ (Fig. \ref{fg:TauSite}) and subsequent variation in relaxation times $T_{1,2}$ (Fig. \ref{fg:T12Site}(b)) across the chain, the measured distribution in $T_{1,2}$ values for liquids is never as large. This is a result of strong cross-relaxation effects, a.k.a. ``spin diffusion" effects, which tend to average out  (i.e. ``wash out") any such variations across the chain \cite{solomon:pr1955,kalk:jmr1976,edzes:nature1977,lambert:book}. 
\begin{figure}[!ht]
	\begin{center}
		\includegraphics[width=0.9\columnwidth]{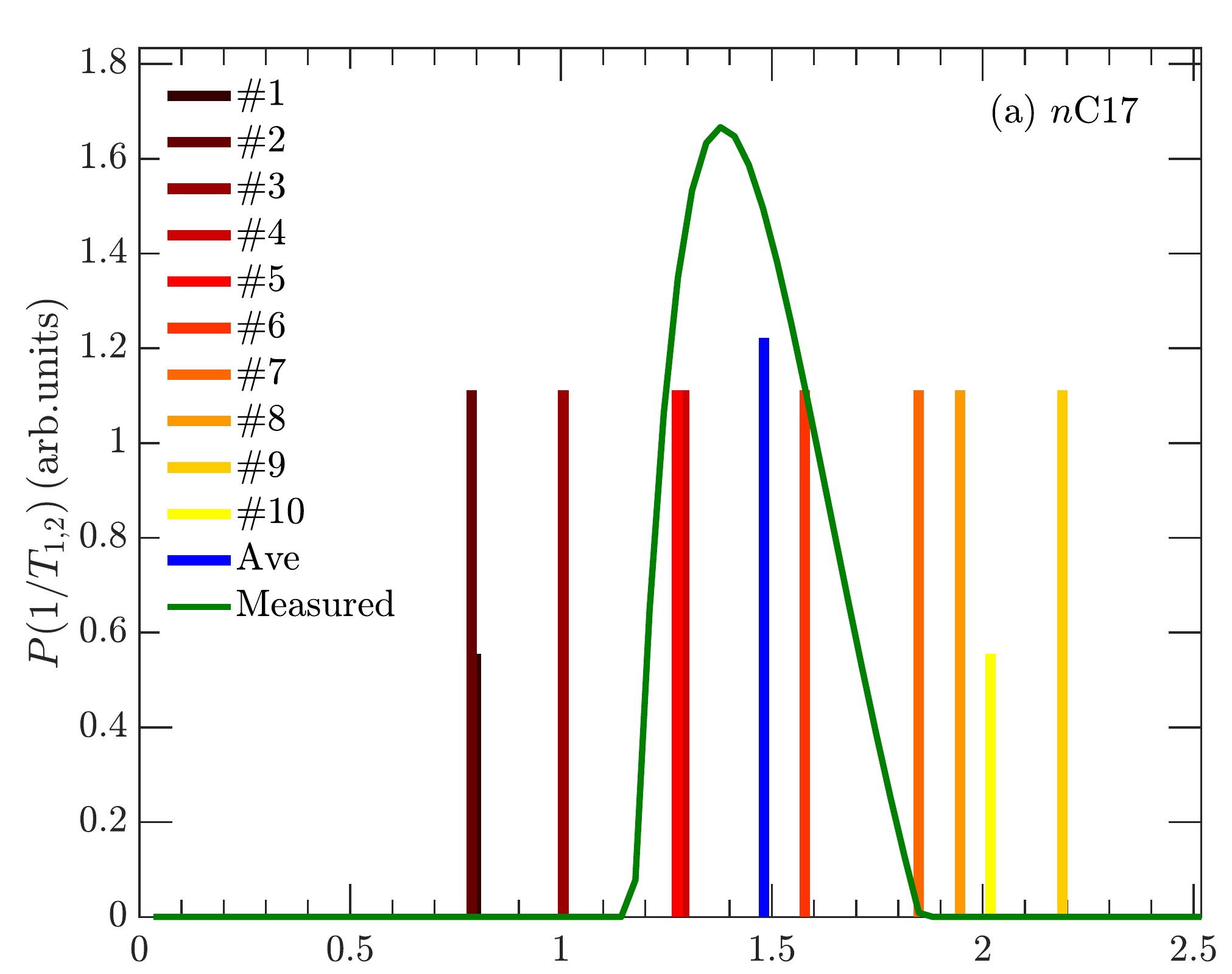}
		\includegraphics[width=0.9\columnwidth]{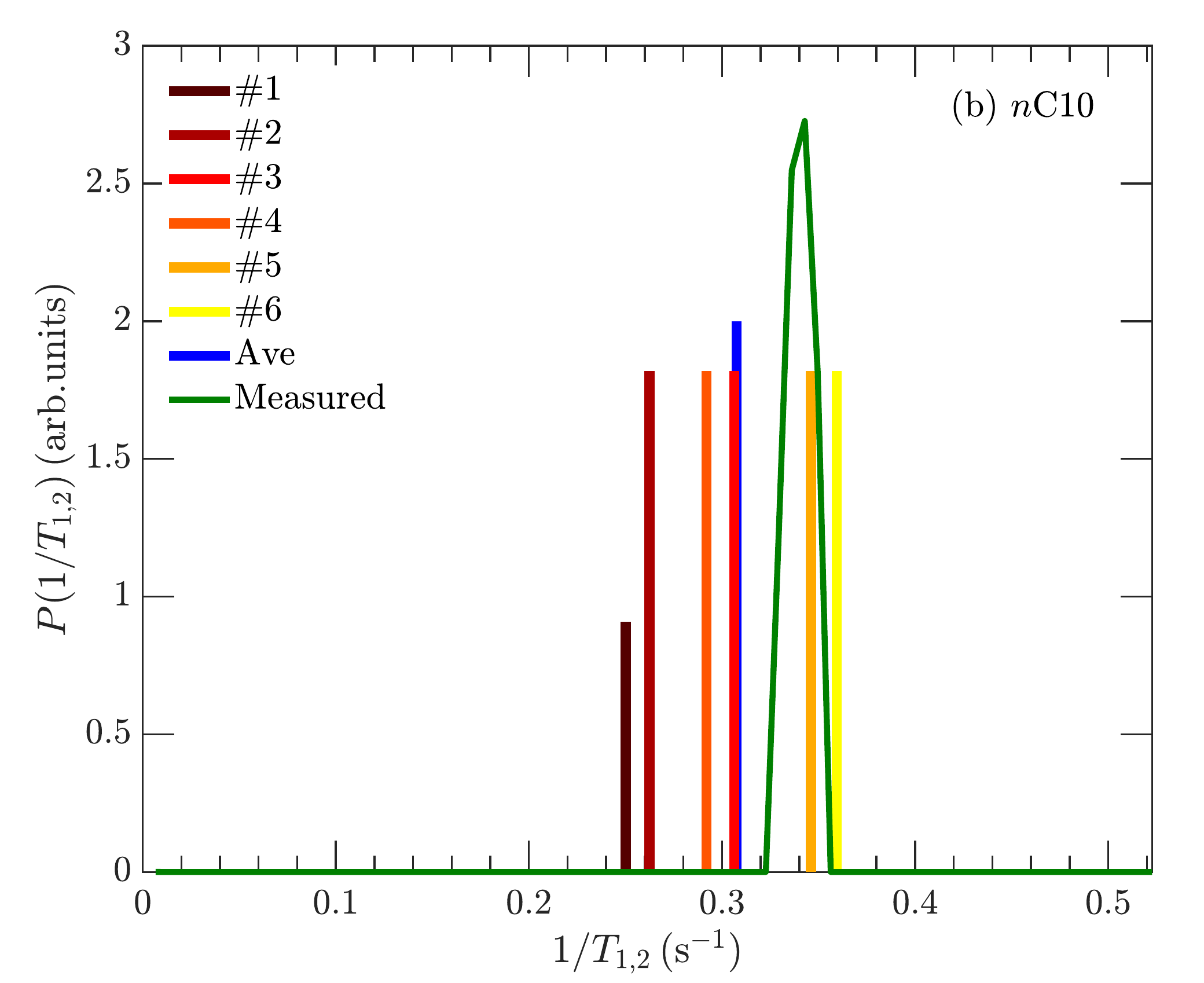}  
	\end{center}
	\caption{(a) MD simulation result for distribution in site-by-site values of $1/T_{1,2,i}$ taken from Fig. \ref{fg:T12Site}(b), for (a) $n$C17 and (b) $n$C10, with labels defined in Fig. \ref{fg:Molecules} and with relative heights according to degeneracy. ``Ave" indicates weighted average over the site-by-site results Eq. \ref{eq:Cross}, with arbitrary height. Simulations results are compared with measured distributions of $1/T_{2}$, with arbitrary height.} \label{fg:T1dist}
\end{figure} 

The condition for either strong, intermediate, or weak cross-relaxation is determined by the relative strength of the cross-relaxation rate $\sigma_{1,2,ij}$ between spin-pairs $i$ and $j$ compared with the difference in individual relaxation rates $|1/T_{1,2,i} - 1/T_{1,2,j}|$ (see supplementary material for more details). More specifically, the different cross-relaxation regimes can be defined as such \cite{kalk:jmr1976,kowalewski:book}:
\begin{eqnarray}
\begin{aligned}
\sigma_{1,2,ij} &\gg \frac{1}{2}|1/T_{1,2,i} - 1/T_{1,2,j}| \,\,\,\, ({\rm strong}),\\
\sigma_{1,2,ij} &\simeq \frac{1}{2}|1/T_{1,2,i} - 1/T_{1,2,j}| \,\,\,\, ({\rm intermediate}),\\
\sigma_{1,2,ij} &\ll \frac{1}{2}|1/T_{1,2,i} - 1/T_{1,2,j}| \,\,\,\, ({\rm weak}).
\label{eq:Regimes}
\end{aligned}
\end{eqnarray}
For strong cross-relaxation, which is generally the case for low-viscosity liquids, one expects the measured $T_{1,2}$ to be single-valued and given by the average rate \cite{kalk:jmr1976,woessner:jcp1962b}:
\begin{equation}
\frac{1}{T_{1,2}} = \frac{1}{N} \sum_{i} \frac{1}{T_{1,2,i}},
\label{eq:Cross}
\end{equation}
where $N$ is the number of $^1$H's in the chain, and $1/T_{1,2,i}$ is the relaxation rate of the $i$'th $^1$H in the chain shown in Fig. \ref{fg:T12Site}(b). One consequence of averaging the relaxation {\it rates} in Eq. \ref{eq:Cross} is that the average $1/T_{1,2}$ is equivalent to computing the weighted average $G_{R,T}(t)$ in Figs. \ref{fg:GtC17} and \ref{fg:GtC10}, and {\it then} using Eqs. \ref{eq:TauRT}$-$\ref{eq:T12motional}.

The comparison between simulation and measurements are shown in Fig. \ref{fg:T1dist}. In the case of $n$-heptadecane, the simulation predicts an average value of $1/T_{1,2} \simeq$ 1.48 s$^{-1}$, which is close to the measured value of $1/T_{1,2} \simeq$ 1.45 s$^{-1}$. In the case of $n$-decane, the simulation predicts an average value of $1/T_{1,2} \simeq$ 0.308 s$^{-1}$, which is also close to the measured value of $1/T_{1,2} \simeq$ 0.328 s$^{-1}$. Such agreement was previously reported in \cite{singer:jmr2017}, which validates our MD simulation methodology. 

However, while the mean values agree, Fig. \ref{fg:T1dist} clearly shows that the widths of simulation versus measurement do not agree. In the case of $n$-heptadecane (Fig. \ref{fg:T1dist}(a)), the simulation predicts a width of $\Delta_{1,2} \simeq$ 1.40 s$^{-1}$ (defined as the difference between maximum and minimum values in the distribution), which is a factor $\simeq$ 3 times larger than the measured full-width at half-maximum $W_{1,2} \simeq$ 0.42 s$^{-1}$ of the distribution. The $n$-heptadecane results therefore indicate that cross-relaxation partially averages out the site-by-site distribution in relaxation. In other words, in the case of weak cross-relaxation $\sigma_{1,2,ij} \ll \frac{1}{2}|1/T_{1,2,i} - 1/T_{1,2,j}|$, the measured $W_{1,2}$ would agree with $\Delta_{1,2} \simeq$ 1.40 s$^{-1}$ from simulation. In the case of strong cross-relaxation $\sigma_{1,2,ij} \gg \frac{1}{2}|1/T_{1,2,i} - 1/T_{1,2,j}|$, the measured $W_{1,2}$ would tend towards zero (i.e. the $1/T_{1,2}$ distribution would tend towards a delta function), which in practice will be limited by the experimental resolution of $\simeq$  $0.06/T_{1,2}$ (as determined on a water sample). Fig. \ref{fg:T1dist}(a) indicates that $\sigma_{1,2,ij}$ is somewhere in-between the strong and weak cross-relaxation regime, i.e. in the intermediate regime, suggesting that $\sigma_{1,2,ij}$ could in principle be calculated from the observed difference between simulation and measurements at low magnetic-fields ($\omega_0/2\pi\lesssim 2.3$ MHz).

In the case of $n$-decane (Fig. \ref{fg:T1dist}(b)), the simulation predicts a width of $\Delta_{1,2} \simeq$ 0.11 s$^{-1}$, which is much larger than the measured $W_{1,2} \simeq$ 0.02 s$^{-1}$. Given that the experimental resolution limit has been reached, the true measured width for $n$-decane must satisfy $W_{1,2} <$ 0.02 s$^{-1}$. This implies that $\Delta_{1,2}$ is at least a factor $\gtrsim$ 6 times larger than $W_{1,2}$, indicating that cross-relaxation is more efficient for $n$-decane than for $n$-heptadecane.

\section{Conclusions}\label{sc:Conc}

The traditional hard-sphere models developed by Bloembergen, Purcell, Pound \cite{bloembergen:pr1948} and Torrey \cite{torrey:pr1953} continue to be the building blocks in describing NMR relaxation from rotational ($R$) and translational ($T$) diffusion of molecules, respectively. The BPP hard-sphere model predicts a single-exponential decay in the autocorrelation function $G_R(t)$, however $G_R(t)$ for the $n$-alkanes becomes increasingly ``stretched" (i.e. multi-exponential) with increasing chain-length \cite{singer:jmr2017}. Likewise, $G_T(t)$ shows greater departure from the Torrey hard-sphere model with increasing chain-length \cite{singer:jmr2017}. We quantify the departure of $G_R(t)$ from the single-exponential model by fitting $G_{R}(t)$ to a sum of exponential decays using an inverse Laplace transform, and determine the standard deviation $\sigma_R$ (i.e. the normalized standard deviation) of the underlying distribution $P_{R}(\tau)$ in correlation times $\tau$. We find that hydrocarbons of greater molecular symmetry such as neopentane and isooctane have $\simeq $ 25 $\%$ lower $\sigma_R$ than their corresponding linear $n$-alkanes. Furthermore, in the case of the spherically-symmetric neopentane, we find that the ratio of translational-diffusion to rotational correlation times is found to be $\tau_D/\tau_R = 8.76$ (where $\tau_D = \frac{5}{2}\tau_T$), which agrees well with the Stokes-Einstein prediction of $\tau_D/\tau_R = 9$. 

By comparing the relaxation of rigid and flexible $n$-alkanes, we find a factor $\simeq$ 12 increase in the rotational correlation-time $\tau_R$ for rigid $n$-decane compared with flexible $n$-decane, together with a factor $\simeq$ 5 increase in the translational correlation-time $\tau_T$, thereby revealing the strong influence of internal motions on $\tau_{R,T}$ and $T_{1,2}$ for long-chain $n$-alkanes. The extent of the stretched (i.e. multi-exponential) decay of $G_{R,T}(t)$ is greater for rigid $n$-alkanes, namely the $\sigma_{R,T}$ values are a factor $\simeq$ 2 larger for rigid compared to flexible $n$-alkanes, indicating that internal motions tend to narrow the underlying distribution $P_{R,T}(\tau)$ (Eq. \ref{eq:ILT}) in correlation times $\tau$. 

We find that $T_{1,2}$ relaxation from in{\it tra}molecular interactions dominates over in{\it ter}molecular interactions (i.e. $T_{1,2,T}/T_{1,2,R} \gg 1$) for all hydrocarbons investigated, {\it except} for benzene and cyclohexane where in{\it ter}molecular interactions dominate (i.e. $T_{1,2,T}/T_{1,2,R} \ll 1$). The rigid $n$-alkanes show a factor $\simeq$ 2 larger ratio $T_{1,2,T}/T_{1,2,R}$ than flexible $n$-alkanes, indicating that internal motions somewhat diminish the influence of  in{\it tra}molecular interactions.

Site-by-site simulations of $G_{R,T}(t)$ for the $^1$H's across the chain indicate that $\tau_R$ decreases by a factor $\simeq$ 4 towards the chain-ends of $n$-heptadecane, together with a factor $\simeq$ 1.4 decrease in $\tau_T$, thereby revealing variations in $\tau_{R,T}$ and $T_{1,2}$ across the chain for long-chain $n$-alkanes. Moreover, the simulations indicate that the stretched functional-forms of site-by-site $G_{R,T}(t)$ are approximately the {\it same} as the chain averages, namely the $\sigma_{R,T}$ values are roughly the same across the chain, indicating that the overall molecular geometry plays a crucial (if not dominant) role in the functional-form of the decay in $G_{R,T}(t)$.

Corresponding $T_{1,2}$ measurements of $n$-heptadecane indicate a narrower distribution in $T_{1,2}$ than site-by-site simulations, implying that cross-relaxation (partially) averages-out the variations in $\tau_{R,T}$ and $T_{1,2}$ across the chain of long-chain $n$-alkanes. Such $T_{1,2}$ comparisons between site-by-site simulations and measurements at low magnetic-field ($\omega_0/2\pi \lesssim $ 2.3 MHz) could in principle be used to compute the cross-relaxation rates $\sigma_{1,2,ij}$ for long-chain $n$-alkanes. 

This work informs our on-going work in understanding the NMR relaxation and diffusion of hydrocarbons (and other fluids) in nano-confined pores, such as the light hydrocarbons found in the organic-matter pores of kerogen and bitumen \cite{singer:petro2016,chen:petro2017,singer:EF2018} typically found in organic-shale reservoirs.

\section{Acknowledgments} \label{sc:Acknow}

This work was funded by the Rice University Consortium on Processes in Porous Media, and the American Chemical Society Petroleum Research Fund [ACS-PRF-58859-ND6]. We gratefully acknowledge the National Energy Research Scientific Computing Center, which is supported by the Office of Science of the U.S. Department of Energy [DE-AC02-05CH11231], for HPC time and support. We also gratefully acknowledge the Texas Advanced Computing Center (TACC) at The University of Texas at Austin (URL: http://www.tacc.utexas.edu) for providing HPC resources.


%

\vfill\eject
\section*{Appendix A} \label{app:Rigid}

\begin{figure}[!ht]
	\begin{center}
		\includegraphics[width=0.9\columnwidth]{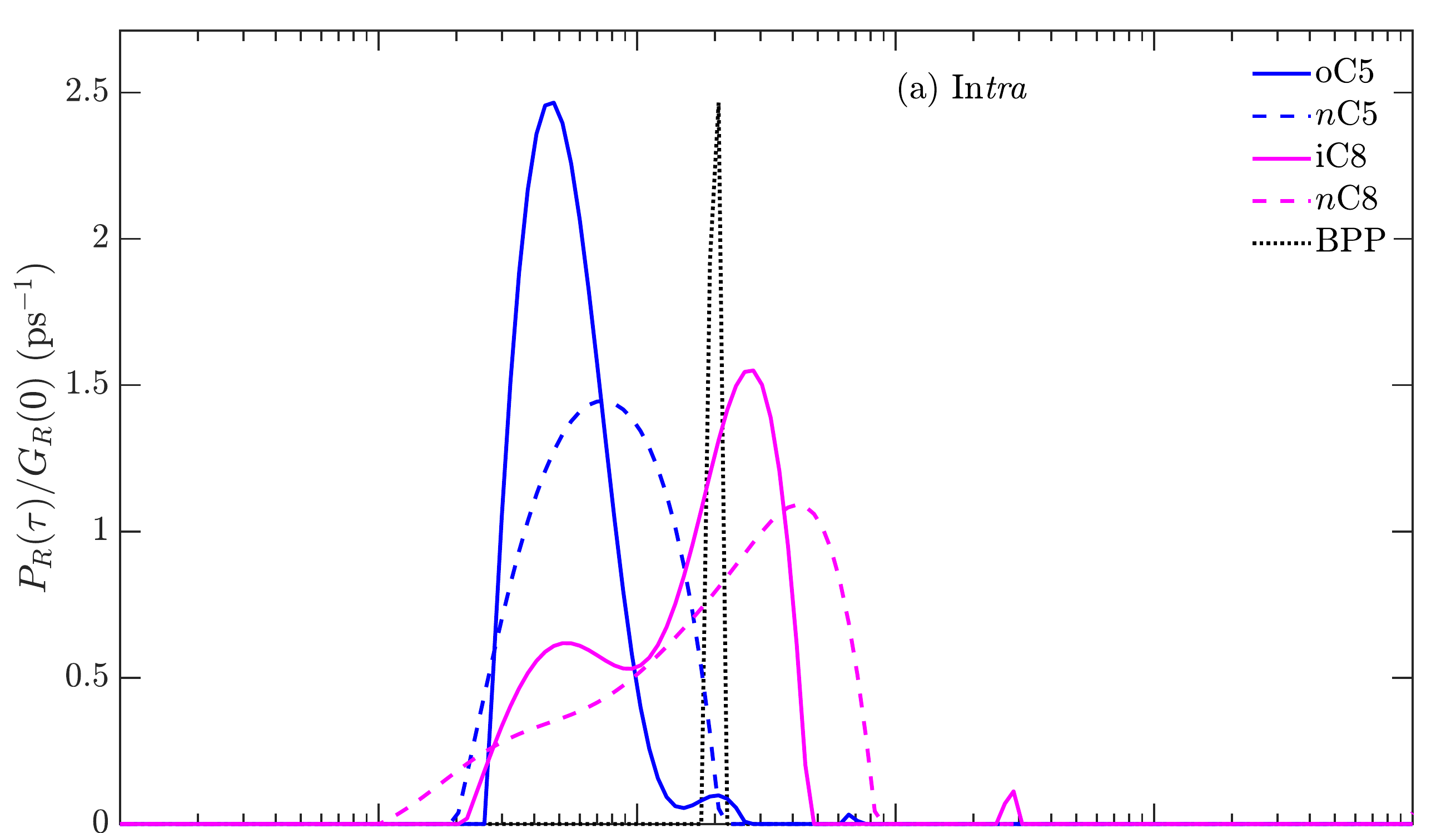}  
		\includegraphics[width=0.9\columnwidth]{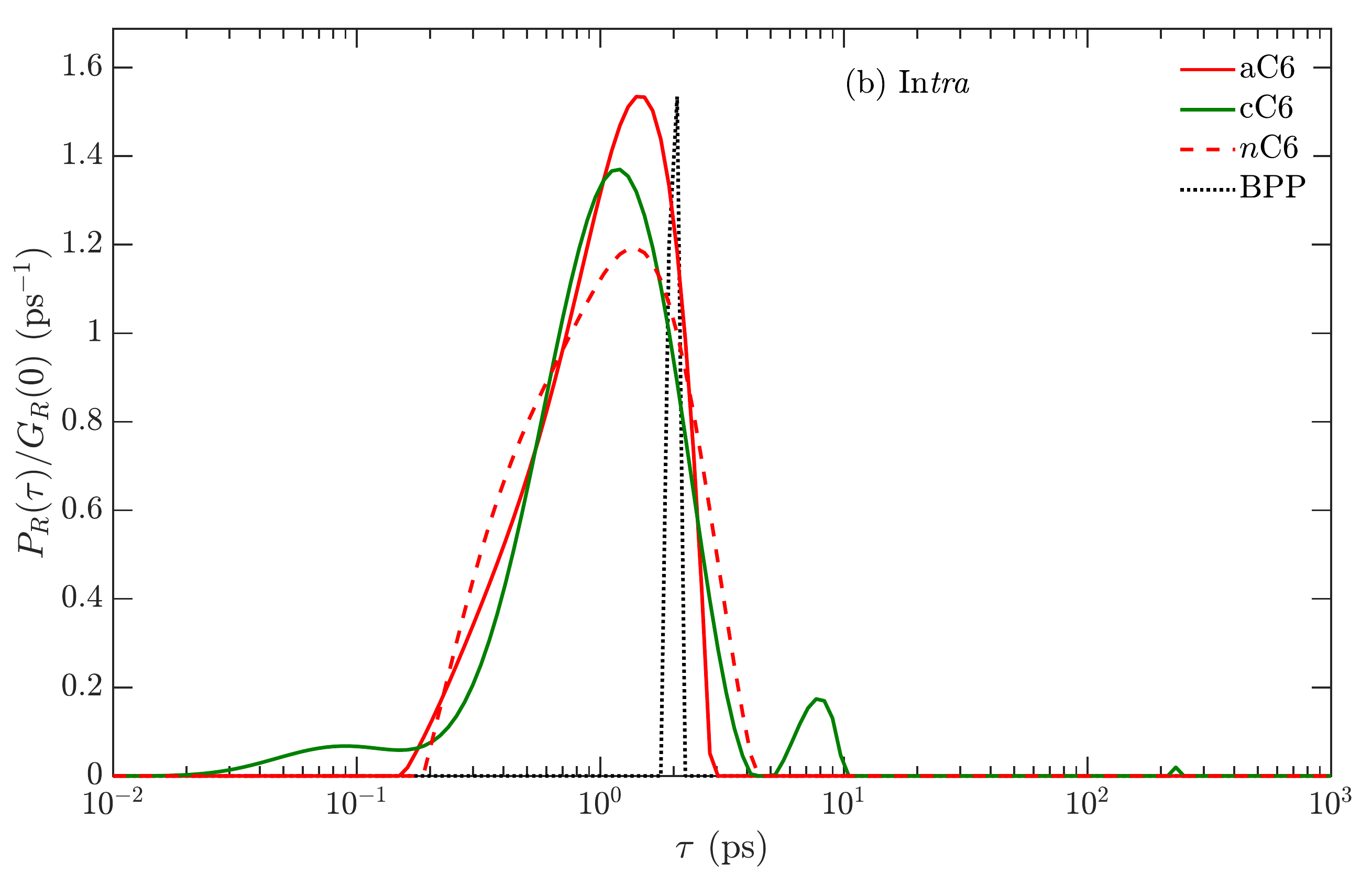} 
	\end{center}
	\caption{Probability distribution function $P_R(\tau)$ of rotational correlation time $\tau$ derived from the inverse Laplace transform (Eq. \ref{eq:ILT}) of the $G_R(t)$ simulations in Fig. \ref{fg:GtIntra}(a) and (b), respectively. Also shown is the BBP prediction $G_R(t)$ from Eq. \ref{eq:Gmodel}, generated with an arbitrarily chosen value of $\tau_R = $ 2 ps. The $y$-axis has been divided by $G_R(0)$, which normalizes the area of the distributions to unity (except for the BPP model).} 
	\label{fg:FEstIntra}
\end{figure}

\begin{figure}[!ht]
	\begin{center}
		\includegraphics[width=0.9\columnwidth]{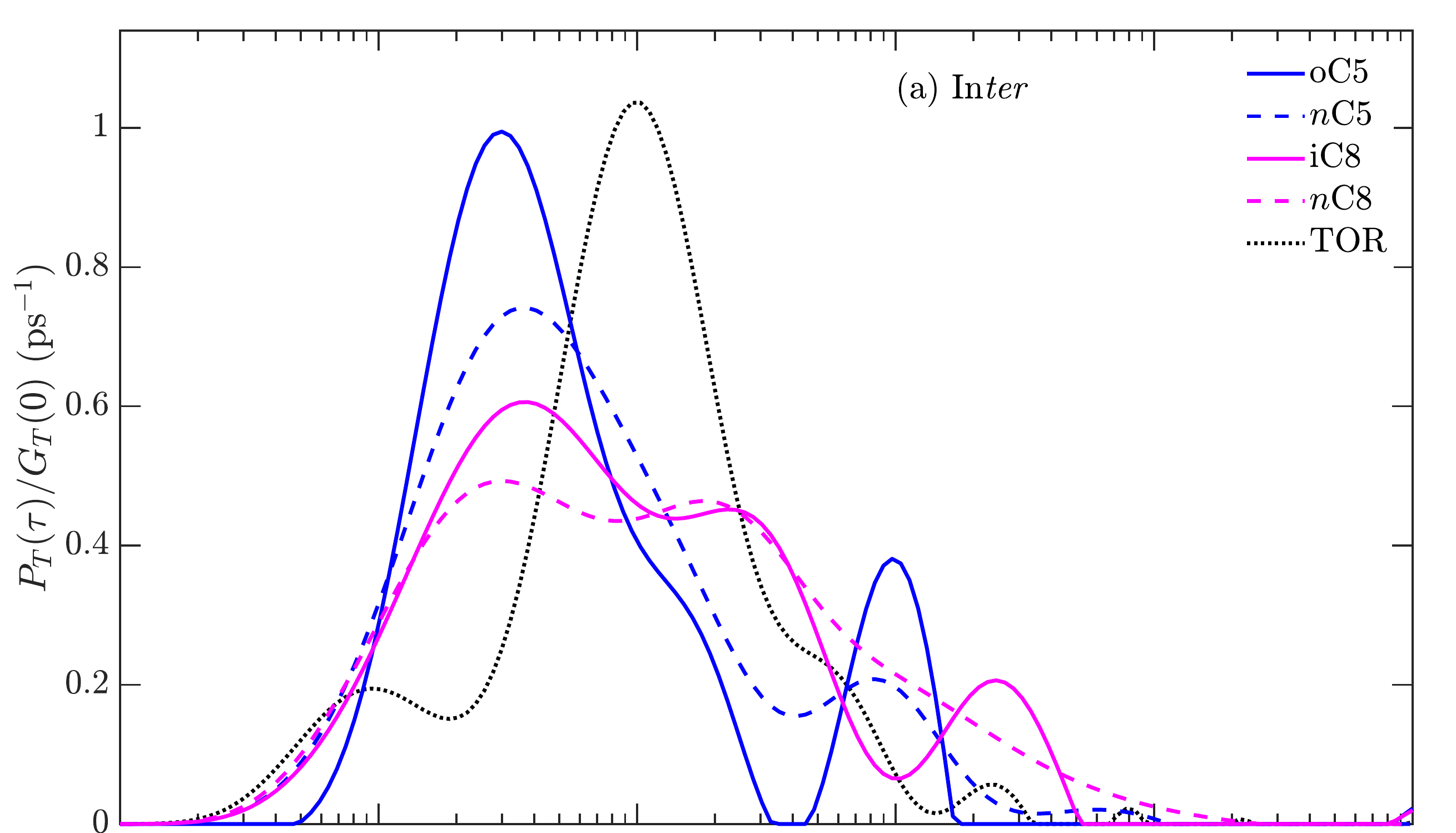}  
		\includegraphics[width=0.9\columnwidth]{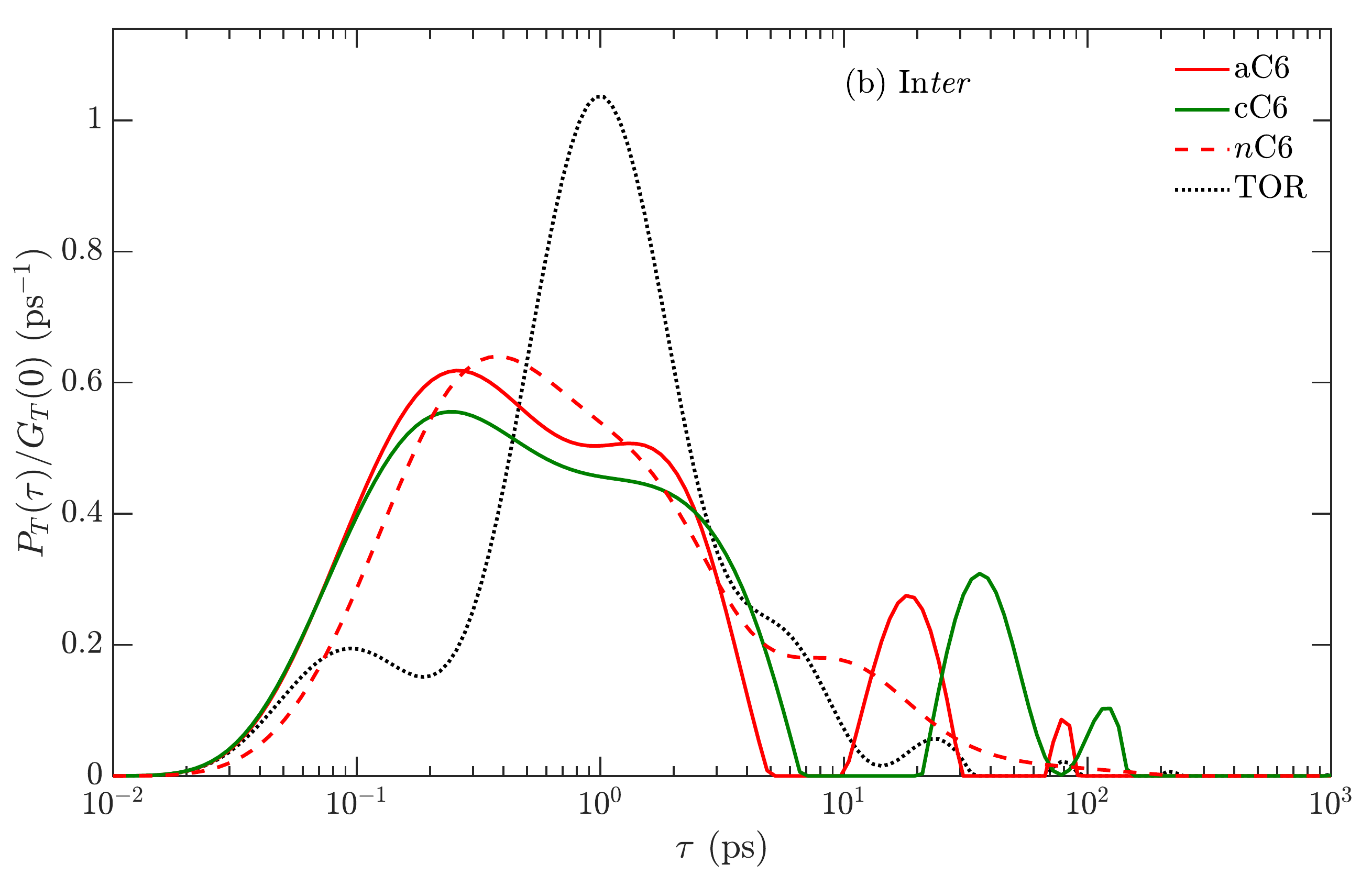} 
	\end{center}
	\caption{Probability distribution function $P_T(\tau)$ of translational correlation time $\tau$ derived from the inverse Laplace transform (Eq. \ref{eq:ILT}) of the $G_T(t)$ simulations in Fig. \ref{fg:GtInter}(a) and (b), respectively. Also shown is the Torrey (TOR) prediction $G_T(t)$ from Eq. \ref{eq:Gmodel}, generated with an arbitrarily chosen value of $\tau_T = $ 2 ps. The $y$-axis has been divided by $G_T(0)$, which normalizes the area of the distributions to unity.} 
	\label{fg:FEstInter}
\end{figure}

As mentioned in Section \ref{ssc:Sphere}, the probability distribution functions $P_{R,T}(\tau)$ of correlation time $\tau$ are shown in Figs. \ref{fg:FEstIntra} and \ref{fg:FEstInter}. The $P_{R,T}(\tau)$ are derived from the inverse Laplace transforms in Eq. \ref{eq:ILT} of the $G_{R,T}(t)$ simulations in Figs. \ref{fg:GtIntra} and \ref{fg:GtInter}, respectively. The $P_{R,T}(\tau)$ are then used to derive the coefficients of variance $\sigma_{R,T}$ using Eq. \ref{eq:CvRT}. For instance, it is clear from Figs. \ref{fg:FEstIntra} and \ref{fg:FEstInter} that $P_{R,T}(\tau)$ are narrower for neopentane, benzene and isooctane (i.e. have lower $\sigma_{R,T}$) compared to their corresponding straight-chain $n$-alkane, respectively. 

\begin{figure}[!ht]
	\begin{center}
		\includegraphics[width=0.9\columnwidth]{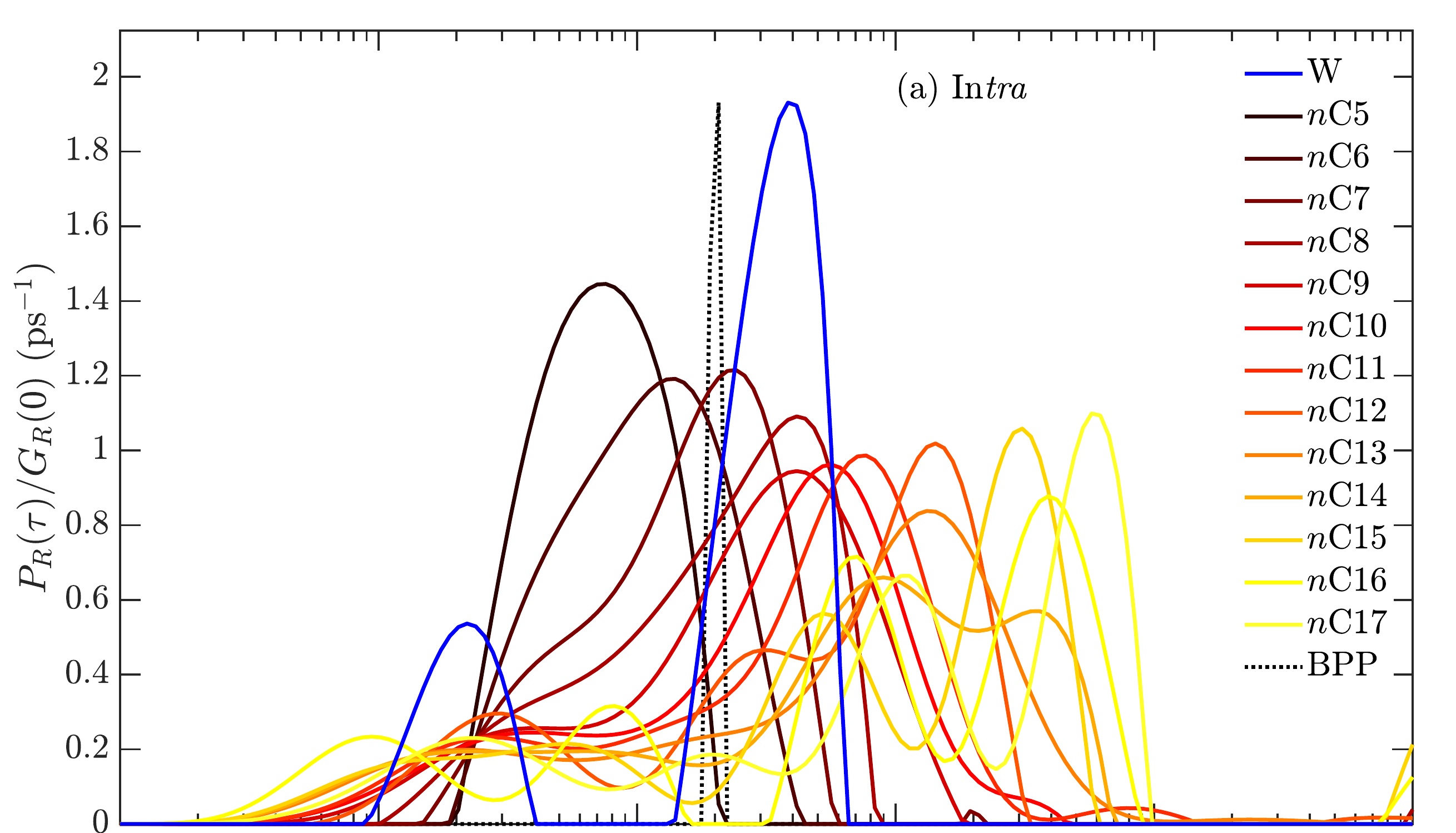}  
		\includegraphics[width=0.9\columnwidth]{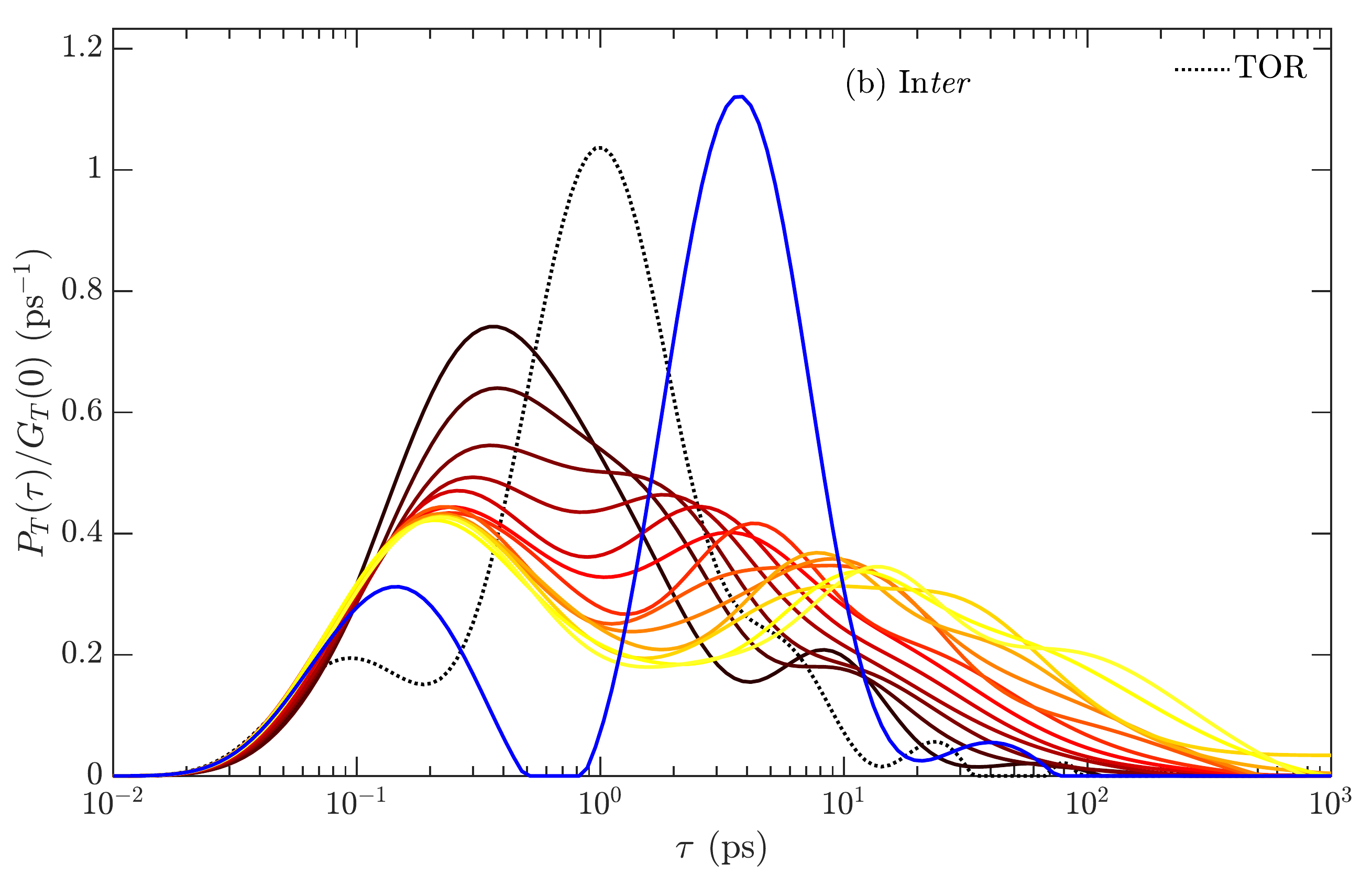} 
	\end{center}
	\caption{Probability distribution function for (a) rotational $P_R(\tau)$ and (b) translational $P_T(\tau)$ correlation time $\tau$ derived from the inverse Laplace transform (Eq. \ref{eq:ILT}) of the $G_{R,T}(t)$ simulations taken from Ref. \cite{singer:jmr2017}, including water (W). Also shown are the BBP and Torrey (TOR) predictions $G_{R,T}(t)$ from Eq. \ref{eq:Gmodel}, generated with an arbitrarily chosen value of $\tau_{R,T} = $ 2 ps. The $y$-axis has been divided by $G_{R,T}(0)$, which normalizes the area of the distributions to unity (except for the BPP model).} 
	\label{fg:FEstWalk}
\end{figure}

As a consistency check of the inverse Laplace algorithm and the resulting $P_{R,T}(\tau)$ distributions, it was verified that the predicted $\tau_{R,T}$ defined as: 
\begin{equation}
\tau_{R,T} = \frac{1}{G_{R,T}(0) }\int\! P_{R,T}(\tau) \, \tau \,d\tau,
\label{eq:TauP}
\end{equation}
gave similar values as $\tau_{R,T}$ from Eq. \ref{eq:TauRT}. Indeed, the $\tau_{R,T}$ predicted from Eq. \ref{eq:TauP} was within $\pm$ 5 $\%$ of Eq. \ref{eq:TauRT}, for all the flexible molecules. In the case of the rigid $n$-alkanes however, Eq. \ref{eq:TauP} predicted (on average) a factor $\simeq $ 1.5 longer $\tau_{R,T}$ compared to Eq. \ref{eq:TauRT}. For the rigid $n$-alkanes, this discrepancy is a result of the incomplete decay in $G_{R,T}(t)$ at $t = 150 $ ps (Figs. \ref{fg:GtRigidIntra} and \ref{fg:GtRigidInter}), which tends to underestimate $\tau_{R,T}$ according to Eq. \ref{eq:TauRT}.

For completeness, Fig. \ref{fg:FEstWalk} shows the $P_{R,T}(\tau)$ distributions for the complete set of flexible $n$-alkanes and water, where the $G_{R,T}(t)$ were previously reported in Ref. \cite{singer:jmr2017}. The case of water shows two distinct peaks in rotational distribution $P_{R}(\tau)$, which agrees well with previous reports \cite{calero:jpcb2015,madhavi:jpcb2017}. According to Ref. \cite{madhavi:jpcb2017} for water, the larger peak ($\simeq$ 79 $\%$ relative intensity) with longer rotational correlation-time ($\tau$ = 3.8 ps) is interpreted as Debye continuous-time rotational diffusion, while the smaller peak ($\simeq$ 21 $\%$ relative intensity) with shorter rotational correlation-time  ($\tau$ = 0.22 ps) is interpreted as large-amplitude discrete jumps.

\clearpage

\setcounter{page}{1}
\section*{Supplementary material} \label{app:Cross}

Here we provide some background about cross-relaxation effects in liquids, and the definition of the $T_1$ and $T_2$ cross-relaxation rates $\sigma_{1,2,ij}$ between spin-pairs $i$ and $j$, respectively. Cross-relaxation plays an important role in $^1$H relaxation in liquids, and exact theories of cross-relaxation exist for the simple case of two-spin systems \cite{solomon:pr1955,kalk:jmr1976,kowalewski:book}. The case of $n$-heptadecane ($n$-C$_{17}$H$_{36}$) reported here consists of 36 $^1$H spins for in{\it tra}molecular relaxation, and therefore presents a much more complex scenario than the two-spin case. However, we use the two-spin case to briefly explain the principles of cross-relaxation, and we neglect any possible cross-correlation effects with other relaxation mechanisms.

\subsubsection{Longitudinal cross-relaxation}
The time evolution of longitudinal magnetization for a two-spin system is solved using the well known Solomon equations \cite{solomon:pr1955}:
\begin{widetext}
\renewcommand{\arraystretch}{1.2}
\begin{align}
\frac{d}{dt} \!
\begin{bmatrix} 
\left<\vcenter{\hbox{\(\hat{I}_z\)}}\right> \\ 
\left<\vcenter{\hbox{\(\hat{S}_z\)}}\right>  
\end{bmatrix} &= -
\begin{bmatrix} 
1/T_{1,I} -  \sigma_{1,IS}& \sigma_{1,IS} \\ 
\sigma_{1,IS} & 1/T_{1,S} -  \sigma_{1,IS}
\end{bmatrix} 
\begin{bmatrix} 
\left<\vcenter{\hbox{\(\hat{I}_z\)}}\right> -I^{eq}_z  \\ 
\left<\vcenter{\hbox{\(\hat{S}_z\)}}\right>-S^{eq}_z
\end{bmatrix} \label{eq:T1sol}.
\end{align}

\begin{align}
\frac{d}{dt} \!
\begin{bmatrix} 
\left<\vcenter{\hbox{\(\hat{I}_{\scriptscriptstyle \pm}\)}}\right> \\ 
\left<\vcenter{\hbox{\(\hat{S}_{\scriptscriptstyle \pm}\)}}\right> 
\end{bmatrix} &= -
\begin{bmatrix} 
1/T_{2,I} -  \sigma_{2,IS} \mp i\delta_{IS}/2& \sigma_{2,IS} \\ 
\sigma_{2,IS} &1/T_{2,S} -  \sigma_{2,IS}  \pm i\delta_{IS}/2
\end{bmatrix} 
\begin{bmatrix} 
\left<\vcenter{\hbox{\(\hat{I}_{\scriptscriptstyle \pm}\)}}\right>  \\ 
\left<\vcenter{\hbox{\(\hat{S}_{\scriptscriptstyle \pm}\)}}\right> 
\end{bmatrix} \label{eq:T2sol}.
\end{align}

\end{widetext}
$\left<\vcenter{\hbox{\(\hat{I}_z\)}}\right>$ and $\left<\vcenter{\hbox{\(\hat{S}_z\)}}\right>$ are the time-dependent expectation values of the operators for longitudinal (i.e. $z$) magnetization for $^1$H spins $I$ and $S$ along the chain, respectively. $I_z^{eq}$ and $S_z^{eq}$ are the equilibrium values of the longitudinal magnetization for $^1$H spins $I$ and $S$ along the chain, respectively. $1/T_{1,I}$ and $1/T_{1,S}$ are the longitudinal relaxation rates for $^1$H spins $I$ and $S$, respectively. $\sigma_{1,IS}$ is the longitudinal cross-relaxation rate between $^1$H spins $I$ and $S$. According to Ref. \cite{kalk:jmr1976}, in the case of $\sigma_{1,IS} \ll \frac{1}{2}|1/T_{1,I} - 1/T_{1,S}|$, spins $I$ and $S$ relax at their own distinct rates \cite{kalk:jmr1976}, as would be predicted by the site-by-site simulations (e.g. $\#$1, $\#$2, etc.) in Fig. \ref{fg:T1dist}. In the opposite case of $\sigma_{1,IS} \gg \frac{1}{2}|1/T_{1,I} - 1/T_{1,S}|$, the two spins relax at the same rate given by the average $1/T_1 = \frac{1}{2}(1/T_{1,I} + 1/T_{1,S})$ defined in Eq. \ref{eq:Cross}, as would be predicted by the ``Ave"  in Fig. \ref{fg:T1dist}. 

When $^1$H spins $I$ and $S$ are not identical (i.e. when their Larmor frequencies are not exactly equal $\omega_I \neq \omega_S$), one can manipulate the two spins independently and measure $T_{1,I}$, $T_{1,S}$ and $\sigma_{1,IS}$ using NOESY (nuclear Overhauser effect spectroscopy), and/or $T_{2,I}$, $T_{2,S}$ and $\sigma_{2,IS}$ (see below) using ROESY (rotating frame Overhauser effect spectroscopy), both at high magnetic-fields (typically $\omega_0/2\pi \gtrsim 100$ MHz) \cite{kowalewski:book}. 

\subsubsection{Transverse cross-relaxation}
A slightly more complicated expression exists for the time evolution of transverse magnetization for a two-spin system \cite{solomon:pr1955} is given in Eq. \ref{eq:T2sol}, where $\left<\vcenter{\hbox{\(\hat{I}_{\scriptscriptstyle \pm}\)}}\right>$ and $\left<\vcenter{\hbox{\(\hat{S}_{\scriptscriptstyle \pm}\)}}\right>$ are the time-dependent expectation values of the operators for transverse (i.e. $x,y$) magnetization for $^1$H spins $I$ and $S$ along the chain, respectively, where step-up $\hat{I}_{+}$ and step-down $\hat{I}_{-}$ operators are used ($\hat{I}_{\pm} =\hat{I}_x \pm i \hat{I}_y$). $1/T_{2,I}$ and $1/T_{2,S}$ are the transverse relaxation rates for $^1$H spins $I$ and $S$, respectively. $\sigma_{2,IS}$ is the transverse cross-relaxation rate between $^1$H spins $I$ and $S$. The additional term for transverse relaxation involves the frequency splitting $\delta_{IS} = | \omega_I - \omega_S |$ between the two $^1$H sites, which for example a methyl (spin $I$) versus a methylene (spin $S$) is $\delta_{IS} \simeq$ 0.4 ppm $\cdot \omega_{0}$ \cite{lambert:book}, or, $\delta_{IS}/2\pi \simeq $ 1 Hz at $\omega_{0}/2\pi = $ 2.3 MHz. Eq. \ref{eq:T2sol} for $T_2$ is equivalent to Eq. \ref{eq:T1sol} for $T_1$, provided $\delta_{IS}/2 \lesssim \sigma_{2,IS}$ \cite{kowalewski:book}, which is the case here given that the measured $T_1$ and $T_2$ distributions are found to be roughly the same.

\subsubsection{Fast-motion regime}

Further insight can be obtained about the cross-relaxation rate $\sigma_{1,2,IS}$, given that the present case is in the fast-motion regime, i.e. $\omega_{0}\, \tau_{R,T} \ll 1$. In the fast-motion regime, the following equalities hold: $T_{1,I} = T_{2,I}$, $T_{1,S} = T_{2,S}$, $\sigma_{1,IS} = \sigma_{2,IS}$, plus none of these quantities are dispersive (i.e. none depend on $\omega_{0}$). One additional equality holds in the fast-motion regime, namely $\sigma_{1,2,IS} = \frac{1}{3} \, 1/T_{1,2}$ \cite{kalk:jmr1976}, with the average defined as $1/T_{1,2} = \frac{1}{2}(1/T_{1,2,I} + 1/T_{1,2,S})$ from Eq. \ref{eq:Cross}. 

One can then loosely apply the above two-spin case in the fast-motion regime to the more complex case of $n$-heptadecane in Fig. \ref{fg:T1dist}(a); though this is a gross oversimplification, it can check whether our interpretation is at all reasonable. According to Fig. \ref{fg:T1dist}(a), the average $1/T_{1,2} \simeq $ 1.5 s$^{-1}$, which loosely implies that  $\sigma_{1,2,IS} \simeq \frac{1}{3} \,1/T_{1,2} \simeq$ 0.5 s$^{-1}$. The simulations further indicate that the spread is given by $\Delta_{1,2} \simeq |1/T_{1,I} - 1/T_{1,S}| \simeq$ 1.4 s$^{-1}$. Putting these two estimates together loosely indicates that $\sigma_{1,2,IS} \simeq \frac{1}{2}|1/T_{1,I} - 1/T_{1,S}| \simeq$ 0.5-0.7 s$^{-1}$, implying an intermediate cross-relaxation regime, which is qualitatively consistent with the findings in Fig. \ref{fg:T1dist}(a). Such consistency is motivation for a more thorough analysis of $\sigma_{1,2,ij}$ for all spin-pairs, provided extensions of Eqs. \ref{eq:T1sol} and \ref{eq:T2sol} are developed which (for instance) break down the 36 $^1$H spin system into magnetization modes \cite{kowalewski:book}. This is the first instance we are aware of where cross-relaxation effects can be studied by comparing simulations with measurements at low magnetic-fields ($\omega_0/2\pi \lesssim 2.3$ MHz).
%
%

\begin{turnpage}
	\renewcommand{\arraystretch}{1.2}
	\begin{table*}
		\begin{adjustwidth}{0in}{0in} 
			\begin{tabularx}{23cm}{CCCCCCCCCCCCC}
				\hline
				
				\multicolumn{2}{c}{$^{{\strut}{}}$Name} & Label &  Rigid &$\Delta\omega_R/2\pi$ & $\Delta\omega_T/2\pi$  & $\tau_R$ & $\tau_T$ & $\tau_D/\tau_R$ & $\sigma_R$ & $\sigma_T$ & $T_{1,2,T}/T_{1,2,R}$ & $T_{1,2}$  \\ 
				\multicolumn{2}{c}{$^{{\strut}{}}$}& &  & (kHz) &  (kHz) &  (ps) &  (ps) & & & & &  (s)\\ 
				\hline

		\multicolumn{2}{c}{neopentane}	&	oC5	&		&	21.62	&	9.11	&	0.55	&	1.93	&	8.76	&	0.39	&	1.35	&	1.61	&	18.17	\\
		\multicolumn{2}{c}{benzene}	&	aC6	&		&	7.93	&	8.83	&	1.11	&	3.21	&	7.22	&	0.61	&	1.57	&	0.28	&	23.71	\\
		\multicolumn{2}{c}{cyclohexane}	&	cC6	&		&	17.86	&	10.55	&	1.43	&	7.40	&	12.90	&	0.86	&	1.87	&	0.56	&	5.93	\\
		\multicolumn{2}{c}{isooctane}	&	iC8	&		&	21.44	&	9.36	&	1.75	&	3.87	&	5.52	&	0.80	&	1.62	&	2.38	&	6.64	\\
		\multicolumn{2}{c}{$n$-pentane}	&	$n$C5	&		&	20.17	&	9.66	&	0.72	&	2.27	&	7.92	&	0.53	&	1.44	&	1.38	&	15.10	\\
		\multicolumn{2}{c}{$n$-hexane}	&	$n$C6	&		&	20.05	&	9.72	&	1.10	&	2.84	&	6.45	&	0.68	&	1.52	&	1.65	&	10.69	\\
		\multicolumn{2}{c}{$n$-heptane}	&	$n$C7	&		&	19.99	&	9.73	&	1.66	&	3.50	&	5.28	&	0.82	&	1.62	&	2.00	&	7.65	\\
		\multicolumn{2}{c}{$n$-octane}	&	$n$C8	&		&	19.93	&	9.73	&	2.45	&	4.30	&	4.38	&	1.02	&	1.72	&	2.40	&	5.50	\\
		\multicolumn{2}{c}{$n$-nonane}	&	$n$C9	&		&	19.87	&	9.76	&	3.34	&	5.19	&	3.88	&	1.21	&	1.84	&	2.67	&	4.19	\\
		\multicolumn{2}{c}{$n$-decane}	&	$n$C10	&		&	19.83	&	9.74	&	4.61	&	6.25	&	3.39	&	1.38	&	1.95	&	3.06	&	3.16	\\
		\multicolumn{2}{c}{$n$-undecane}	&	$n$C11	&		&	19.82	&	9.74	&	6.00	&	7.40	&	3.08	&	1.51	&	2.05	&	3.36	&	2.49	\\
		\multicolumn{2}{c}{$n$-dodecane}	&	$n$C12	&		&	19.80	&	9.73	&	8.22	&	10.57	&	3.22	&	1.59	&	2.15	&	3.22	&	1.80	\\
		\multicolumn{2}{c}{$n$-tridecane}	&	$n$C13	&		&	19.76	&	9.70	&	10.74	&	12.37	&	2.88	&	1.82	&	2.23	&	3.60	&	1.42	\\
		\multicolumn{2}{c}{$n$-tetradecane}	&	$n$C14	&		&	19.71	&	9.66	&	12.34	&	13.61	&	2.76	&	1.90	&	2.29	&	3.77	&	1.25	\\
		\multicolumn{2}{c}{$n$-pentadecane}	&	$n$C15	&		&	19.75	&	9.69	&	16.24	&	16.31	&	2.51	&	2.07	&	2.41	&	4.13	&	0.97	\\
		\multicolumn{2}{c}{$n$-hexadecane}	&	$n$C16	&		&	19.66	&	9.69	&	18.90	&	18.19	&	2.41	&	2.26	&	2.48	&	4.28	&	0.84	\\
		\multicolumn{2}{c}{$n$-heptadecane}	&	$n$C17	&		&	19.79	&	9.59	&	22.35	&	20.00	&	2.24	&	2.08	&	2.53	&	4.76	&	0.72	\\
		\multicolumn{2}{c}{$n$-pentane}	&	$n$C5	&	Rigid	&	19.77	&	9.83	&	1.31	&	2.13	&	4.07	&	1.01	&	1.38	&	2.48	&	10.58	\\
		\multicolumn{2}{c}{$n$-hexane}	&	$n$C6	&	Rigid	&	19.80	&	9.93	&	2.65	&	3.37	&	3.18	&	1.13	&	1.52	&	3.13	&	5.54	\\
		\multicolumn{2}{c}{$n$-heptane}	&	$n$C7	&	Rigid	&	19.53	&	10.05	&	6.77	&	5.81	&	2.14	&	1.40	&	1.68	&	4.40	&	2.40	\\
		\multicolumn{2}{c}{$n$-octane}	&	$n$C8	&	Rigid	&	19.59	&	10.13	&	16.42	&	8.19	&	1.25	&	1.85	&	1.82	&	7.50	&	1.06	\\
		\multicolumn{2}{c}{$n$-nonane}	&	$n$C9	&	Rigid	&	17.71	&	10.16	&	31.95	&	24.49	&	1.92	&	2.56	&	2.66	&	3.96	&	0.61	\\
		\multicolumn{2}{c}{$n$-decane}	&	$n$C10	&	Rigid	&	21.78	&	10.51	&	56.87	&	33.01	&	1.45	&	2.37	&	2.68	&	7.39	&	0.25	\\

				\hline
			\end{tabularx}
			\caption{MD simulations results, including: fluid name, fluid label (chemical formula given in Fig. \ref{fg:Molecules} caption), rigid or not, in{\it tra}molecular dipolar strength $\Delta\omega_R/2\pi$ (Eq. \ref{eq:Dipolar}), in{\it ter}molecular dipolar strength $\Delta\omega_T/2\pi$ (Eq. \ref{eq:Dipolar}), in{\it tra}molecular correlation time $\tau_R$ (Eq. \ref{eq:TauRT}), in{\it ter}molecular correlation time $\tau_T$ (Eq. \ref{eq:TauRT}), ratio of translational-diffusion to rotational correlation times $\tau_D/\tau_R$ (where $\tau_D=\frac{5}{2}\tau_T$), in{\it tra}molecular standard deviation $\sigma_R$ (Eq. \ref{eq:CvRT}), in{\it ter}molecular standard deviation $\sigma_T$ (Eq. \ref{eq:CvRT}), ratio of in{\it ter}molecular to in{\it tra}molecular relaxation times $T_{1,2,T}/T_{1,2,R}$ (Eq. \ref{eq:T12RTmotional}), and total relaxation time $T_{1,2}$ (Eq. \ref{eq:T12motional}). \label{tb:1} }
		\end{adjustwidth}
	\end{table*}
\end{turnpage}

\begin{turnpage}
	\renewcommand{\arraystretch}{1.2}
	\begin{table*}
		\begin{adjustwidth}{0in}{0in}
			\begin{tabularx}{23cm}{CCCCCCCCCCCCC}
				\hline
				
				$^{{\strut}{}}$Label & Site & Degen. &  Rigid &$\Delta\omega_R/2\pi$ & $\Delta\omega_T/2\pi$  & $\tau_R$ & $\tau_T$ & $\tau_D/\tau_R$ &  $\sigma_R$ & $\sigma_T$ & $T_{1,2,T}/T_{1,2,R}$ & $T_{1,2}$  \\ 
				$^{{\strut}{}}$& & &  & (kHz) &  (kHz) &  (ps) &  (ps) & & & &  &  (s)\\ 
				\hline
				
		$n$C17	&	$\#$1	&	2	&		&	21.77	&	10.46	&	8.58	&	18.30	&	5.33	&	1.71	&	2.42	&	2.03	&	1.25	\\
		$n$C17	&	$\#$2	&	4	&		&	21.71	&	10.50	&	8.47	&	18.13	&	5.35	&	1.68	&	2.41	&	2.00	&	1.27	\\
		$n$C17	&	$\#$3	&	4	&		&	18.99	&	10.11	&	15.31	&	20.74	&	3.39	&	2.07	&	2.48	&	2.61	&	0.99	\\
		$n$C17	&	$\#$4	&	4	&		&	18.99	&	9.91	&	21.13	&	22.44	&	2.66	&	2.17	&	2.50	&	3.46	&	0.77	\\
		$n$C17	&	$\#$5	&	4	&		&	19.04	&	9.47	&	20.66	&	24.58	&	2.98	&	2.06	&	2.50	&	3.39	&	0.78	\\
		$n$C17	&	$\#$6	&	4	&		&	19.22	&	9.24	&	26.53	&	25.67	&	2.42	&	2.10	&	2.52	&	4.48	&	0.63	\\
		$n$C17	&	$\#$7	&	4	&		&	19.27	&	9.47	&	31.60	&	25.81	&	2.04	&	2.08	&	2.50	&	5.06	&	0.54	\\
		$n$C17	&	$\#$8	&	4	&		&	19.48	&	9.52	&	32.75	&	26.08	&	1.99	&	2.10	&	2.52	&	5.26	&	0.51	\\
		$n$C17	&	$\#$9	&	4	&		&	19.89	&	9.28	&	36.38	&	26.14	&	1.80	&	1.94	&	2.51	&	6.39	&	0.46	\\
		$n$C17	&	$\#$10	&	2	&		&	19.55	&	9.58	&	34.18	&	24.78	&	1.81	&	1.89	&	2.46	&	5.75	&	0.50	\\
		$n$C10	&	$\#$1	&	2	&		&	21.82	&	10.32	&	2.76	&	5.53	&	5.01	&	1.19	&	1.81	&	2.23	&	4.00	\\
		$n$C10	&	$\#$2	&	4	&		&	21.75	&	10.31	&	2.97	&	5.54	&	4.66	&	1.16	&	1.81	&	2.39	&	3.81	\\
		$n$C10	&	$\#$3	&	4	&		&	18.69	&	9.91	&	4.90	&	6.28	&	3.21	&	1.52	&	1.84	&	2.77	&	3.26	\\
		$n$C10	&	$\#$4	&	4	&		&	18.85	&	9.65	&	4.48	&	6.74	&	3.77	&	1.43	&	1.86	&	2.53	&	3.43	\\
		$n$C10	&	$\#$5	&	4	&		&	19.21	&	9.25	&	5.41	&	7.40	&	3.42	&	1.37	&	1.86	&	3.16	&	2.89	\\
		$n$C10	&	$\#$6	&	4	&		&	19.30	&	9.19	&	5.62	&	7.55	&	3.36	&	1.39	&	1.87	&	3.29	&	2.78	\\
		$n$C10	&	$\#$1	&	2	&	Rigid	&	20.97	&	10.18	&	44.23	&	27.23	&	1.54	&	2.28	&	2.60	&	6.90	&	0.34	\\
		$n$C10	&	$\#$2	&	4	&	Rigid	&	21.69	&	10.28	&	47.67	&	27.74	&	1.45	&	2.33	&	2.62	&	7.65	&	0.30	\\
		$n$C10	&	$\#$3	&	4	&	Rigid	&	20.01	&	10.56	&	60.37	&	30.97	&	1.28	&	2.31	&	2.66	&	7.00	&	0.28	\\
		$n$C10	&	$\#$4	&	4	&	Rigid	&	21.85	&	10.53	&	61.67	&	35.35	&	1.43	&	2.43	&	2.70	&	7.51	&	0.23	\\
		$n$C10	&	$\#$5	&	4	&	Rigid	&	22.76	&	10.66	&	59.60	&	36.13	&	1.52	&	2.39	&	2.70	&	7.53	&	0.22	\\
		$n$C10	&	$\#$6	&	4	&	Rigid	&	22.84	&	10.71	&	60.70	&	37.05	&	1.53	&	2.34	&	2.70	&	7.45	&	0.21	\\

				\hline
			\end{tabularx}
			\caption{Site-by-site MD simulations results, including: fluid label (chemical formula given in Fig. \ref{fg:Molecules} caption), site number (Fig. \ref{fg:Molecules}), degeneracy, rigid or not, in{\it tra}molecular dipolar strength $\Delta\omega_R/2\pi$ (Eq. \ref{eq:Dipolar}), in{\it ter}molecular dipolar strength $\Delta\omega_T/2\pi$ (Eq. \ref{eq:Dipolar}), in{\it tra}molecular correlation time $\tau_R$ (Eq. \ref{eq:TauRT}), in{\it ter}molecular correlation time $\tau_T$ (Eq. \ref{eq:TauRT}), ratio of translational-diffusion to rotational correlation times $\tau_D/\tau_R$ (where $\tau_D=\frac{5}{2}\tau_T$), in{\it tra}molecular standard deviation $\sigma_R$ (Eq. \ref{eq:CvRT}), in{\it ter}molecular standard deviation $\sigma_T$ (Eq. \ref{eq:CvRT}), ratio of in{\it ter}molecular to in{\it tra}molecular relaxation times $T_{1,2,T}/T_{1,2,R}$ (Eq. \ref{eq:T12RTmotional}), and total relaxation time $T_{1,2}$ (Eq. \ref{eq:T12motional}). \label{tb:2} }
		\end{adjustwidth}
	\end{table*}
\end{turnpage}

\end{document}